\documentclass[10pt, a4paper]{article}

\usepackage[english]{babel}
\usepackage[utf8]{inputenc}
\usepackage{amsmath, amssymb} 
\usepackage{dsfont} 
\usepackage[top = 1in, left = 1in, bottom = 1in, right = 1in]{geometry}
\usepackage{indentfirst} 
\usepackage[round]{natbib} 
\usepackage{color}
\usepackage{multirow}
\usepackage{graphicx} 
\usepackage{multirow} 
\usepackage{diagbox} 
\usepackage{makecell} 
\usepackage{arydshln} 
\usepackage{xcolor}
\usepackage{subcaption}
\usepackage{url}
\usepackage{enumitem}
\usepackage{caption}
\usepackage[hidelinks]{hyperref} 
\usepackage{setspace}
\usepackage{xurl}

\usepackage[mathlines]{lineno}

\usepackage{etoolbox}

\usepackage[colorinlistoftodos]{todonotes}

\usepackage{xr}

\newcommand{\change}[1]{{#1}}

\makeatletter
\newcommand*{\addFileDependency}[1]{
  \typeout{(#1)}
  \@addtofilelist{#1}
  \IfFileExists{#1}{}{\typeout{No file #1.}}
}
\makeatother

\newcommand*{\myexternaldocument}[1]{
    \externaldocument{#1}
    \addFileDependency{#1.tex}
    \addFileDependency{#1.aux}
}

\myexternaldocument{./supporting}

\newcommand*\linenomathpatch[1]{%
	\cspreto{#1}{\linenomath}%
	\cspreto{#1*}{\linenomath}%
	\csappto{end#1}{\endlinenomath}%
	\csappto{end#1*}{\endlinenomath}%
}
\newcommand*\linenomathpatchAMS[1]{%
	\cspreto{#1}{\linenomathAMS}%
	\cspreto{#1*}{\linenomathAMS}%
	\csappto{end#1}{\endlinenomath}%
	\csappto{end#1*}{\endlinenomath}%
}

\expandafter\ifx\linenomath\linenomathWithnumbers
\let\linenomathAMS\linenomathWithnumbers
\patchcmd\linenomathAMS{\advance\postdisplaypenalty\linenopenalty}{}{}{}
\else
\let\linenomathAMS\linenomathNonumbers
\fi

\linenomathpatch{equation}
\linenomathpatchAMS{gather}
\linenomathpatchAMS{multline}
\linenomathpatchAMS{align}
\linenomathpatchAMS{alignat}
\linenomathpatchAMS{flalign}

\begin{document}
	\pagenumbering{gobble}
	
	\begin{center}
		{\Large\textbf{Extended Excess Hazard Models for Spatially Dependent Survival Data}} \\
		\vspace{30pt}
		{\large Andr\'e Victor Ribeiro Amaral ${}^{\dagger, *}$, Francisco Javier Rubio ${}^{\ddagger}$, Manuela Quaresma ${}^{\mathsection}$, Francisco J. Rodr\'iguez-Cort\'es ${}^{\mathparagraph}$ and Paula Moraga ${}^{\dagger}$}\\
	    \vspace{24pt}
	    ${}^{\dagger}$\hspace{2pt}CEMSE Division, King Abdullah University of Science and Technology. Thuwal, Saudi Arabia. \\
	    ${}^{\ddagger}$\hspace{2pt}Department of Statistical Science, University College London. London, UK. \\
	    ${}^{\mathsection}$\hspace{2pt}Department of Non-Communicable Disease Epidemiology, London School of Hygiene \& Tropical Medicine. London, UK. \\
	    ${}^{\mathparagraph}$\hspace{2pt}Escuela de Estad\'istica, Universidad Nacional de Colombia.
Medell\'in, Colombia. \\
		${}^{*}$ Corresponding author. E-mail: \texttt{\href{mailto:andre.ribeiroamaral@kaust.edu.sa}{andre.ribeiroamaral@kaust.edu.sa}}
		\vspace{18pt}
	\end{center}

	\begin{center}
		\textbf{Abstract}
	\end{center} \vspace{-6pt}
	
	\noindent
    Relative survival represents the preferred framework for the analysis of population cancer survival data. The aim is to model the survival probability associated to cancer in the absence of information about the cause of death. Recent data linkage developments have allowed for incorporating the place of residence into the population cancer data bases; however, modeling this spatial information has received little attention in the relative survival setting. We propose a flexible parametric class of spatial excess hazard models (along with inference tools), named ``Relative Survival Spatial General Hazard'' (RS-SGH), that allows for the inclusion of fixed and spatial effects in both time-level and hazard-level components. We illustrate the performance of the proposed model using an extensive simulation study, and provide guidelines about the interplay of sample size, censoring, and model misspecification. \change{We present a case study using real data from colon cancer patients in England.} This case study illustrates how a spatial model can be used to identify geographical areas with low cancer survival, as well as how to summarize such a model through marginal survival quantities and spatial effects.
    \vspace{12pt}
	
	
	\newpage
	\pagenumbering{arabic}
	\setcounter{footnote}{0} 
	
\section{Introduction} \label{sec:introduction}

Survival analysis represents one of the main branches in Statistics, which concerns the study of times-to-event, potentially subject to censoring. The main quantity of interest in survival analysis is the probability of survival beyond a specific time point, associated with either the entire population under study or for subgroups of such a population. The most relevant approaches for analyzing survival data are: (i) the overall survival framework, which aims at analyzing all-cause mortality; (ii) the cause-specific survival framework, which incorporates information about the cause of death; and (iii) the relative survival framework, which aims at quantifying the survival associated to a cause of death of interest (such as cancer) in the absence of information about the cause of death. In the context of cancer epidemiology, national and international health agencies are interested in monitoring the survival probability of cancer patients at the population level \citep{allemani:2015}. The preferred approach for population-based cancer survival analysis is the ``relative survival'' framework \citep{esteve:1990, perme:2012}. 

The relative survival approach aims at estimating the survival (or hazard) function associated to cancer, in the absence of reliable information about the cause of death for the whole population (since information about the cause of death is typically unreliable at the population level). The main idea is to assume an additive decomposition of the hazard function $h(\cdot)$ into two components, namely the hazard associated to other causes of death $h_{\text{O}}(\cdot)$, and the hazard associated to cancer $h_{\text{E}}(\cdot)$. The latter is typically referred to as the ``excess hazard.'' That is,
\begin{align} \label{eq:main-haz}
    h(t; \mathbf{x}) = h_{\text{O}}(\text{age} + t) + h_{\text{E}}(t; \mathbf{x}),~ t \geq 0,
\end{align}
where $t$ is the time measured from the date of diagnosis, ``age'' is age at diagnosis of cancer, and $\mathbf{x} \in {\mathbb R}^p$ is the vector of available covariates. Since $h_{\text{O}}(\text{age} + t)$ is unknown in practice, it is usually estimated using the population hazard $h_{\text{P}}(\text{age} + t; \mathbf{z})$, which is obtained from the life tables based on the characteristics $\mathbf{z} \in {\mathbb R}^q \subseteq \mathbf{x}$. Depending on the country, the available life tables may be stratified by age, calendar year, sex, deprivation level, et cetera. Several excess hazard models have been proposed using both parametric and nonparametric estimation approaches (see \cite{eletti:2022} for a recent review). The main quantity of interest in the relative survival framework is the ``net survival,'' which is the survival associated to the excess hazard $S_{\text{N}}(t;\mathbf{x}) = \exp\left\{-\int_0^t h_{\text{E}}({r};\mathbf{x})d{r}\right\}$. The net survival only depends on the excess hazard function. Thus, it is a useful quantity for comparing the performance of cancer management between different populations since it is not affected by differences in population mortality hazards. For that reason, comparisons between different countries, regions, or periods of time are based on the marginal net survival
\begin{equation*}
S_{\text{N}}(t) = \dfrac{1}{m}\sum_{i=1}^m S_\text{N}(t;\mathbf{x}_i),    
\end{equation*}
where $\{\mathbf{x}_i\}_{i=1}^m$ represents the covariates associated to the (sub-)population of interest, \change{such that $m$ denotes the population size.}

\change{The utilization of spatial information regarding the residence of cancer-diagnosed patients may enable the identification of regional inequalities in cancer survival \citep{GOV,exarchakou:2018,quaresma:2022}.} Furthermore, such information may facilitate the identification of areas with low cancer survival, which can be  be used to inform policymaking aiming at improving cancer survival. Indeed, cancer registry data may present a spatially dependent structure, as individuals from adjacent neighborhoods are likely to share environmental and socio-economical factors \citep{li:2002}. In that case, the individuals' locations would act like a proxy for non-observed regional characteristics \citep{zhou:2018}. 

Spatial survival modeling has received a great deal of attention in the overall survival framework. The main idea is usually to incorporate a spatial term into a survival regression model (see \cite{klein:2014} for a book-length review on classical survival models). For instance, \cite{li:2002} propose adding a spatial frailty, modeled as a zero-mean Gaussian process (GP), into a semiparametric Proportional Hazard (PH) model. \cite{banerjee:2003} fit a PH model with spatially dependent random effects for geostatistical and lattice data. \cite{carlin:2003} propose a Bayesian semiparametric survival model for spatio-temporal correlated data based on including generalized multivariate conditionally autoregressive (MCAR) region-specific frailties into a hazard regression model. \cite{li:2015} propose a normal transformation model of the General Hazard (GH) model (\citealp{chen:2001}, also known as Extended Hazard (EH) model). The spatial variability is modeled through the covariance matrix of the normal transformation. \cite{zhou:2018} propose a framework for fitting PH, Proportional Odds (PO), and accelerated failure time (AFT) models, accounting for different types of censoring, including random effects with intrinsic conditionally autoregressive (ICAR) priors. \cite{basak:2022} propose a semiprametric model for clustered interval-censored survival data. In that case, the hazard function is written as a product of the baseline hazard component and a non-parametric component modeled as a soft Bayesian additive regression tree (sBART) that is used to incorporate the possible clustering effects. \cite{rubio:2022} consider the GH structure with random effects at the cluster level; however, they do not account for the spatial structure and limit their proposal to modeling the clustering components. 

In contrast, spatial survival models in the relative survival framework have received less attention. For instance, \cite{charvat:2016} propose a parametric frailty model for the excess hazard function using different types of splines or parametric baseline hazards. However, the frailties are assumed to be independent, thus only accounting for clustering but not for the spatial dependence. \cite{cramb:2016} propose a Bayesian spatial frailty approach based on modeling the cumulative excess hazard using splines, thus requiring a different interpretation for the estimated effects. Their proposal does not include time-scale effects, and the frailties are modeled using an ICAR normal distribution. This method was later applied in \cite{cramb:2017}. Finally, \cite{eletti:2022} propose a link-based additive excess hazard model that allows for the inclusion of non-linear effects, temporal-dependent effects, and spatial effects via Markov random fields. In a slightly different vein, \cite{yu:2012} studied cure mixture models in the relative survival framework. They adopted a mixture of three accelerated failure time models for the excess hazard, with spatial frailties modeled using multivariate conditionally autoregressive (MCAR) distributions. 

In this paper, we introduce a general class of parametric spatial frailty models for survival data under the relative survival framework. The basic idea consists of adding spatial effects, at two levels (time and hazard), into the  General Hazard (GH) model \citep{etezadi:1987, chen:2001}, which is a hazard structure that generalizes the PH model, the accelerated failure time (AFT) model \citep{buckley:1979}, the accelerated hazard (AH) model \citep{chen:2000}, and others, as we will describe in Section \ref{sec:model}.

To do so, we extend the existing approaches by modeling the dependence structure through spatial smoothing methods, namely Intrinsic Conditional Autoregressive (ICAR) and Besag-York-Mollié (BYM2) model priors. It also allows for incorporating fixed and spatial effects at the time-scale and at the hazard scale without requiring numerical integration.  By taking such an approach, we can easily compute credible intervals as a measure of uncertainty, and further investigate other quantities of interest. For instance, similar to the discussion in \cite{moraga:2019}, we can compute the \textit{relative} exceedance probabilities, which are useful for assessing unusual elevation in any function of the linear predictor terms, such as the excess hazard, net survival, among others. The term ``relative'' is important, since we are extending the concept of exceedance probabilities to the relative survival framework. This quantity also helps detecting high-risk areas based on the analysis of the spatial random effects, as the possibly non-observed spatial heterogeneity is captured by such components. The \texttt{R} \citep{CRAN} and STAN \citep{carpenter:2017} scripts containing the implementation of the examples presented here, as well as additional examples using real data are available at \href{https://github.com/avramaral/relative_survival}{\url{https://github.com/avramaral/relative\_survival}}.

The remainder of this paper is organized as follows. Section \ref{sec:model} introduces notation and presents the proposed modeling approach. We also discuss some particular sub-models of interest. We introduce two spatial smoothing methods that account for non-observed spatial characteristics and list all implemented models. In Section \ref{sec:inference}, we detail the inference procedure and present a brief discussion on the prior distributions specification for our class of models. In Section \ref{sec:simulations}, we provide a simulation study that illustrate the performance of our model under different scenarios, and present guidelines about the interplay of sample size, censoring, and model misspecification. \change{Section \ref{sec:case_study} presents a case study that analyzes the variation of colon cancer survival for different geographic regions in England.} Finally, in Section \ref{sec:discussion}, we present a general discussion, and comment on the limitations and possible extensions of our work.

\section{Spatial models} \label{sec:model}

In this section, we introduce the proposed general model structure, and discuss the particular models that can be derived from it. Also, we discuss different spatial smoothing methods that can be used with our approach and list all possible modeling scenarios.

\subsection{Excess hazard model} \label{ssec:hazard-model}

Let us first introduce some notation. Let $o_{ij} \in \mathbb{R}_+$ be a sample of times-to-event, where $i = 1, \ldots, r$ indicates the region and $j = 1, \ldots, n_i$ denotes the individuals. Also, let $c_{ij} \in \mathbb{R}_+$ be the right-censoring times, and $t_{ij} = \min\{o_{ij}, c_{ij}\}$ be the observed survival times. Let $\delta_{ij} = \mathds{1}(o_{ij} < c_{ij})$ be the vital status indicators (that is, $\delta_{ij} = 1$, if dead, and $\delta_{ij} = 0$, if right-censored or alive), and $n = \sum_{i = 1}^r n_i$ be the total sample size across the $r$ regions. Let $\mathbf{x}_{ij}\in \mathbb{R}^{p}$ be the vector of available covariates. Similar to the mixed effects survival regression model (for overall survival) proposed in \cite{rubio:2022}, we consider the excess hazard model
    \begin{align} \label{eq:exc-haz}
        h_{\text{E}}(t; \mathbf{x}_{ij} \mid \boldsymbol{\theta}, \boldsymbol{\alpha}, \boldsymbol{\beta}, \boldsymbol{\gamma}, \tilde{u}_i, u_i) = h_0(t \exp\{\tilde{\mathbf{x}}_{ij}^{\top} \boldsymbol{\alpha} + \tilde{u}_i\} \mid  \boldsymbol{\theta}) \exp\{\mathbf{s}_{ij}^{\top} \boldsymbol{\gamma} + \mathbf{x}_{ij}^{\top} \boldsymbol{\beta} + u_i\},
    \end{align}
where $h_0(\cdot \mid \boldsymbol{\theta})$ is the baseline excess hazard function, defined through a flexible parametric distribution, $\boldsymbol{\theta}$ represents the corresponding distribution parameters, $\mathbf{x}_{ij}$ play the role of hazard-level effects, $\tilde{\mathbf{x}}_{ij} \subseteq \mathbf{x}_{ij}$ represent the time-level effects, where $\tilde{\mathbf{x}}_{ij} \in \mathbb{R}^{\tilde{p}}$, and $\boldsymbol{\alpha} = (\alpha_1, \ldots, \alpha_{\tilde{p}})^{\top}$ and $\boldsymbol{\beta} = (\beta_1, \ldots, \beta_{p})^{\top}$ are the regression coefficients associated to  $\tilde{\mathbf{x}}_{ij}$ and $\mathbf{x}_{ij}$, respectively. Additionally, $\mathbf{s}_{ij} = ({\mathbf{s}_{ij1}}^{\top}, \ldots, {\mathbf{s}_{ijk}}^{\top})^{\top} \in \mathbb{R}^{q}$ and $\boldsymbol{\gamma} = (\gamma_1, \ldots, \gamma_{q})^{\top}$, where $q = \sum_{l = 1}^{k} q_l$, such that $q_l$ is the dimension of $\mathbf{s}_{ijl}$, and $\mathbf{s}_{ijl}$ is the spline expansion of a (continuous) covariate $x_{ijl}$. Lastly, we assume that $\tilde{\mathbf{u}}$ and $\mathbf{u}$ are independent, with  $\tilde{\mathbf{u}} = (\tilde{u}_1, \ldots, \tilde{u}_r)^{\top} \sim \tilde{G}$ and $\mathbf{u} = (u_1, \ldots, u_r)^{\top} \sim G$, such that $\tilde{G}$ and $G$ are multivariate distributions that account the spatial dependence among regions. The spatial models used to define $\tilde{G}$ and $G$ will be introduced in Section \ref{ssec:spatial-model}. Thus, our proposal can be seen as an extension of the MEGH model proposed in \cite{rubio:2022} to the relative survival framework, but also with the incorporation of spatial effects.

We will denote Model \eqref{eq:exc-haz} as the RS-SGH (Relative Survival Spatial General Hazard) model, and we will also consider eight particular sub-models that might be useful for researchers and practitioners when fitting this class of models. These alternative modeling approaches are described in Table \ref{tab:models} (Appendix \ref{appendix:submodels}). 

Let $\boldsymbol{\xi} = (\boldsymbol{\theta}^\top,\boldsymbol{\alpha}^\top,\boldsymbol{\beta}^\top,\boldsymbol{\gamma}^\top)^\top$, then the cumulative hazard function $H(\cdot \mid  \mathbf{x}_{ij}, \boldsymbol{\xi}, \tilde{u}_i, u_i)$ associated with Model \eqref{eq:main-haz} can be written in the following manner
\begin{align*}
    H(t; \mathbf{x}_{ij} \mid  \boldsymbol{\xi}, \tilde{u}_i, u_i) &= \int_{0}^{t} h({\zeta}; \mathbf{x}_{ij} \mid  \boldsymbol{\xi}, \tilde{u}_i, u_i) d{\zeta} \\
    &= H_{\text{P}}(\text{age}_{ij} + t; \mathbf{z}_{ij}) - H_{\text{P}}(\text{age}_{ij}; \mathbf{z}_{ij}) + H_{\text{E}}(t; \mathbf{x}_{ij} \mid  \boldsymbol{\xi}, \tilde{u}_i, u_i),
\end{align*}
where $H_{\text{P}}(\cdot; \mathbf{z}_{ij})$ is the cumulative population hazard, and $H_{\text{E}}(\cdot ; \mathbf{x}_{ij} \mid \boldsymbol{\xi}, \tilde{u}_i, u_i)$ is the cumulative excess hazard function. Moreover, the cumulative excess hazard function can be written in closed-form as
\begin{align*}
    H_{\text{E}}(t; \textbf{x}_{ij} \mid  \boldsymbol{\xi}, \tilde{u}_i, u_i) & = \int_{0}^{t} h_{\text{E}}({\zeta}; \textbf{x}_{ij} \mid  \boldsymbol{\xi}, \tilde{u}_i, u_i) d{\zeta}\\
    & = H_0(t \exp\{\tilde{\mathbf{x}}_{ij}^{\top} \boldsymbol{\alpha} + \tilde{u}_i\} \mid  \boldsymbol{\theta}) \exp\{\mathbf{x}_{ij}^{\top} \boldsymbol{\beta} - \tilde{\mathbf{x}}_{ij}^{\top} \boldsymbol{\alpha} + u_i - \tilde{u}_i\},
\end{align*}
where $H_0(\cdot \mid \boldsymbol{\theta})$ is the cumulative baseline excess hazard. 

We can now adapt the concept of individual net survival based on the proposed spatial excess hazard model. 
The net survival, for a specific covariate and conditional on model parameters, and random effects, can be defined as
\begin{align} \label{eq:net-surv}
    S_{\text{N}}(t; \textbf{x}_{ij} \mid  \boldsymbol{\xi}, \tilde{u}_i, u_i) = \exp\{-H_{\text{E}}(t; \textbf{x}_{ij} \mid  \boldsymbol{\xi}, \tilde{u}_i, u_i)\}.
\end{align}
Consequently, the region-specific net survival associated to the $i$-th region is defined as follows
\begin{align} \label{eq:netsur-strat}
    S_{\text{N},i}(t \mid  \boldsymbol{\xi}) = \frac{1}{n_i} \sum_{j = 1}^{n_i} \int_{\mathbb{R}^2} S_{\text{N}}(t; \textbf{x}_{ij} \mid  \boldsymbol{\xi}, \tilde{u}_i, u_i) d\tilde{G}(u_i)d G(u_i).
\end{align}

Let us now discuss some specific choices for modeling the parametric baseline hazard function $h_0(\cdot \mid  \boldsymbol{\theta})$. Since the Weibull baseline hazard is the only choice that leads to a non-identifiable model \citep{chen:2001}, we will adopt distributions that do not belong to the Weibull family. We will focus on 2-parameter and 3-parameter distribution that can account for a variety of shapes. These include the Log-normal (LN), Log-logistic (LL), Power Generalized Weibull (PGW), Gamma (GAM), and Generalized Gamma (GG) distributions. In Web Appendix 1 (Supporting Information), we specify all possible distributions for such a baseline component. 

\change{Lastly, notice that, by setting $h_{\text{P}}(\text{age}_{ij} + t; \mathbf{z}_{ij}) = 0$, for all individuals in all regions, we shift to the overall survival framework. Therefore, the RS-SGH model generalizes several well-known modeling approaches in different directions and under different frameworks.}

\subsection{Spatial effects} \label{ssec:spatial-model}

We aim at incorporating spatial effects in the excess hazard Model \eqref{eq:exc-haz} by incorporating the neighborhood structure into the distribution of the random effects $\tilde{\mathbf{u}} = (\tilde{u}_1, \ldots, \tilde{u}_r)^{\top}$ and $\mathbf{u} = (u_1, \ldots, u_r)^{\top}$. To this end, we will define them based on two approaches: the Intrinsic Conditional Autoregressive (ICAR) and Besag-York-Mollié (BYM2) models. To formulate these models, we need to introduce the concept of adjacency matrix. Briefly, given two regions $k$ and $l$, we will say that $k$ and $l$ are neighbors (written $k \sim l$, with $k \neq l$) if those regions share any boundary. Notice that if $k \sim l$, then $l \sim k$. However, a region will not be its own neighbor. Based on this ``neighbor operator'' ($\sim$), we can define an adjacency matrix $\mathbf{A}$, such that $a_{k l} = 1$, if $k \sim l$, and $a_{k l} = 0$, otherwise. The diagonal of $\mathbf{A}$ is defined as zero, that is $\text{diag}(\mathbf{A}) = \mathbf{0}$. \change{As a remark, \cite{freni:2018} present guidelines on how to adapt these models if the corresponding spatial graph is disconnected. Additionally, \cite{morris:2019} (Section 3.5) comment on how to implement these extensions using STAN.}

\subsubsection{Intrinsic Conditional Autoregressive (ICAR) model} \label{sssec:icar}

For the Intrinsic Conditional Autoregressive (ICAR) model, the conditional distribution of $u_k$ given all other random effects $u_l$, such that $l \neq k$ (written $\mathbf{u}_{-k}$), is
\begin{align*} 
    \pi(u_k \mid \mathbf{u}_{-k}) = \text{Normal}\left(\frac{\sum\limits_{s \in \Lambda_{k}} u_s}{\lambda_{k}}, \frac{1}{\lambda_k\tau_u}\right),
\end{align*}
where $\Lambda_k$ and $\lambda_k$ correspond to the neighbors and the number of neighbors of region $k$, respectively, and $\tau_u$ is the precision term. \cite{besag:1974} proved that the corresponding joint specification of $\mathbf{u}$ follows a multivariate normal distribution with mean $\mathbf{0}$ and precision matrix  $\mathbf{Q} = \tau_u(\mathbf{D} - \mathbf{A})$, where $\mathbf{D}$ is a ($r \times r$) diagonal matrix with $d_{k k}$ containing the number of neighbors of $k$, and $d_{k l} = 0$, $\forall k \neq l$. Moreover, as shown \change{in \cite{besag:1991, besag:1995},}
the joint distribution of $\mathbf{u}$ as specified above can be further simplified to the following pairwise difference
\begin{align} \label{eq:pairwise}
	\pi(\mathbf{u}) \propto \tau_u^{\frac{r - 1}{2}}\exp\left\{-\frac{\tau_u}{2}\sum_{k \sim l} (u_k - u_l) ^ 2\right\}.
\end{align}
However, from Equation \eqref{eq:pairwise}, one can notice that the joint distribution of $\mathbf{u}$ is non-identifiable (adding any constant to all elements of $\mathbf{u}$ does not change the joint distribution). To overcome this issue, it suffices to impose the constraint $\sum_{k = 1}^{r} u_k = 0$. From a practical point of view, and under the Bayesian framework, the approximate condition $\sum_{k = 1}^{r} u_k \approx 0$ is implemented instead, using a ``soft sum-to-zero constraint''. That is, when implementing the model, we assign a zero-mean prior distribution to the mean of $\mathbf{u}$ with very small variance. Such an approach is recommended by \cite{morris:2019}, as the STAN's Hamiltonian Monte Carlo sampler runs faster under this setting. Finally, the same modeling procedure will be adopted for $\tilde{\mathbf{u}}$. 

\subsubsection{Besag-York-Molli\'e (BYM2) model} \label{sssec:bym2}

Alternatively, unstructured (or non-spatial) random effects could be added, along with the structured ICAR components, to the excess hazard Model \eqref{eq:exc-haz}. This approach is known as a Besag-York-Mollié (BYM)-type model \citep{besag:1991}. However, as commented in \cite{mahmood:2022}, such a parameterization might present some shortcomings. For instance, a model expressed based on such a convolution of the structured and unstructured random effects may fail to fit, as one of the two components can account for almost all observed variance \citep{morris:2019}. Also, it might be difficult to set reasonable priors for the corresponding scale parameters \citep{banerjee:2003a}. Aiming to avoid these issues, instead, the BYM2 model is often used \citep{riebler:2016}. 

To formulate the BYM2 model, the unstructured and structured random effects ($\mathbf{v} = (v_1, \ldots, v_r)^{\top}$ and $\mathbf{s} = (s_1, \ldots, s_r)^{\top}$, respectively), can be written as
\begin{align*} 
\mathbf{u} = \mathbf{v} + \mathbf{s} = \sigma(\sqrt{1 - \rho} \mathbf{v}^{\star} + \sqrt{\rho}\mathbf{s}^{\star}),
\end{align*}
where $\sigma$ is the overall standard deviation, $\rho \in [0, 1]$ determines the proportion of the variance that comes from the structured random effects, $\mathbf{v}^{\star} \sim \text{Normal}(\mathbf{0}, \textbf{I}_r)$, such that $\textbf{I}_r$ is an $(r \times r)$ identity matrix, and $\mathbf{s}^{\star}$ is the scaled ICAR model \change{\citep{sorbye:2017}}, such that $\text{Var}(s_i) \approx 1$, $\forall i$. As before, similar reasoning is applied to define $\tilde{\mathbf{u}}$ in terms $\tilde{\mathbf{v}}$ and $\tilde{\mathbf{s}}$.

\subsubsection{IID model} \label{sssec:iid}

One last alternative would be defining $\tilde{\mathbf{u}}$ and $\textbf{u}$ purely based on an ``independent and identically distributed'' (i.i.d.) model; that is, $\textbf{u} \sim \text{Normal} (\mathbf{0}, \sigma_u^2 \textbf{I}_r)$ and $\tilde{\textbf{u}} \sim \text{Normal} (\mathbf{0}, \sigma_{\tilde{u}}^2 \textbf{I}_r)$. This would be the same including a clustering effect per region. Under the overall survival framework, this idea has been explored by \cite{rubio:2022} using likelihood inference, and we will also implement such a model in a Bayesian setting. All implemented models for the possible baseline hazard distributions, spatial random effects, and overall model structure are detailed in Table \ref{tab:all-models} in Appendix \ref{appendix:implemented-models}.

\subsubsection{Point data model} \label{sssec:pointdata}

\change{Although our focus is on employing areal data to model spatial dependence, one might also be interested in using latitude-longitude coordinates (if available) to determine a patient's location. In such cases, the spatial structure may be accounted for by a model for \textit{point} data. For instance, ``penalized spline regression'' \citep{fahrmeir:2013} is a popular method for spatial smoothing. Alternatively, \cite{diggle:2007} proposed a geostatistical framework to model the spatial correlation structure in the point data while enabling rigorous statistical inference. For the latter, additional assumptions about the sampling scheme can be made, e.g., ``preferential sampling'' \citep{diggle:2010}---with a non-stationary extension proposed by \cite{amaral:2023}. Throughout this paper, we will employ the methods described in Sections \ref{sssec:icar}--\ref{sssec:iid}.}

\section{Inference} \label{sec:inference}

In this section, we introduce the inference procedure used for fitting Model \eqref{eq:main-haz} with excess hazard given by Model \eqref{eq:exc-haz}. Also, we present some guidelines for setting the prior distributions, and define a model selection measure.

\subsection{Likelihood function} \label{ssec:likelihood}

Let $\mathcal{D} = \{(t_{ij}, \delta_{ij}, \mathbf{x}_{ij}, \mathbf{z}_{ij}); \, i = 1, \ldots, r, \text{ and } j = 1, \ldots, n_i\}$ be the observed data. In that case, the likelihood function for the vector of unknown parameters can be written as proportional to
\begin{align} \label{eq:like}
    \prod_{i = 1}^{r}\prod_{j = 1}^{n_i} \left\{h_{\text{P}}(\text{age}_{ij} + t_{ij}; \mathbf{z}_{ij}) + h_{\text{E}}(t_{ij}; \mathbf{x}_{ij} \mid  \boldsymbol{\xi}, \tilde{u}_i, u_i)\right\}^{\delta_{ij}} \exp\{-H_{\text{E}}(t_{ij}; \textbf{x}_{ij} \mid  \boldsymbol{\xi}, \tilde{u}_i, u_i)\},
\end{align}
where $h_{\text{P}}(\text{age}_{ij} + t_{ij}; \mathbf{z}_{ij})$ is obtained from the life tables. From Equation \eqref{eq:like}, notice that the only term in the likelihood function that distinguishes an overall survival model from a relative survival model is $h_{\text{P}}(\text{age}_{ij} + t_{ij}; \mathbf{z}_{ij})$, therefore, by setting it to zero, we could also retrieve the overall survival framework. 

\change{Nevertheless, as proved by \cite{chen:2001}, the General Hazard model (and thus, the RS-SHG model, as it extends the GH approach) is non-identifiable if the baseline hazard in $h_{\text{E}}(t_{ij}; \mathbf{x}_{ij} \mid  \boldsymbol{\xi}, \tilde{u}_i, u_i)$ is Weibull.  However, this scenario is not of concern since, if the true model is Weibull, it means that a simpler model would fit the data well---see \cite{rubio:2019} for further details. Furthermore, our capability to simultaneously recover the two spatial structures in Equation \eqref{eq:like} is noteworthy. As briefly demonstrated in Sections \ref{ssec:simulation-spatial-effects} and \ref{sec:case_study}, we can estimate $\sigma_{u} = 1 / \sqrt{\tau_{u}}$ (and $\sigma_{\tilde{u}} = 1 / \sqrt{\tau_{\tilde{u}}}$) for all proposed models. To do so, in practice, we must have a certain number of \textit{uncensored} observations per cluster (in addition to avoiding the Weibull distribution when defining the baseline hazard). In this case, there is an interplay between the number of individuals in each region and the censoring rate in these areas.} 

The next section presents our prior elicitation strategy. \change{Inference is performed by sampling from the corresponding posterior distributions based on the \texttt{RStan}'s implementation \citep{rstanpackage} of the Hamiltonian Monte Carlo algorithm \citep{betancourt:2015}.}

\subsection{Prior distributions} \label{ssec:priors}

Although we acknowledge that other choices can be made, in this section, we recommend some weakly informative priors for the model parameters. For the linear fixed effects, we set $\alpha_{\tilde{p}} \sim \text{Normal}(0, \sigma^2_{\alpha_{\tilde{p}}})$, $\forall \tilde{p}$, and $\beta_{p} \sim \text{Normal}(0, \sigma^2_{\beta_{p}})$, such that $\sigma^2_{\alpha_{\tilde{p}}}$ and $\sigma^2_{\beta_{p}}$ are large enough to reflect the vague prior information. On the other hand, for the non-linear fixed effects, we adopted a novel choice of g-priors \citep{zellner:1986} that account for censoring; that is, letting $\textbf{S}_{k}$ be the design matrix associated with the spline basis expansion of the $k$-th covariate $\mathbf{x}_{k}$, and defining $\mathbf{M}_{k} = g_{\boldsymbol{\gamma}} (\textbf{S}_{k}^{\top} \textbf{S}_{k})^{-1}$, we set $\boldsymbol{\gamma}_{k} \sim \text{MVN}(\mathbf{0}, \sigma^2_{\boldsymbol{\gamma_{k}}}\textbf{M}_{k})$, where $g_{\boldsymbol{\gamma}} = (n - (0.5 \times (n - n_{\text{obs}}))) / q$, $n_{\text{obs}}$ corresponds to the number of uncensored observations, and $\sigma^2_{\boldsymbol{\gamma_{k}}} \sim \text{Half-Cauchy}(0, \tau_{\sigma_{\gamma}})$, such that $\tau_{\sigma_{\gamma}} > 0$. In that case, notice that $g_{\boldsymbol{\gamma}}$ accounts, to some extent, for the effective number of observations---as the rescaled number of censored patients is subtracted from the total number of collected data points.

\change{In our setting, the g-priors can be seen as a type of shrinkage prior, where the induced shrinkage is mild as we only include a few variables in the models. Alternative prior specifications could have been employed to induce higher levels of shrinkage in place of the selected g-priors. However, in the context of our problem, we do not aim to induce higher levels of shrinkage since there are only a few variables available at the population level, and all of these variables are typically relevant for cancer survival. As an alternative, in the Bayesian smoothing literature, it is also common to assign priors to the spline coefficients that enforce smoothness between adjacent spline coefficients (similarly to what the ICAR model does in the spatial setting). These priors typically take the form of random walks or intrinsic Gaussian Markov random fields \citep{fahrmeir:2013, rue:2005}.}

Regarding the spatial smoothing distributions, for the ICAR model, we set $\tau_u \sim \text{Gamma}(\theta_{{\tau}_u}, \theta_{{\tau}_u})$ (same for $\tau_{\tilde{u}}$), such that $\theta_{{\tau}_u} > 0$ is a small number. \change{Although the Gamma distribution with such scale and shape parameters is commonly found in the literature---mainly due to The BUGS (Bayesian inference Using Gibbs Sampling) project implementation \citep{lunn:2009}, we again emphasize that it is possible to use other types of priors. For instance, \cite{gelman:2006} suggests the usage of a distribution from the half-$t$ family for the variance parameter in hierarchical models. Alternatively, the penalized complexity (PC) priors \citep{simpson:2017} could also be explored in our setting. In Section \ref{sec:discussion}, we briefly discuss possible extensions of our work concerning prior elicitation.} For the BYM2 model, we set $\sigma \sim \text{Half-Normal}(0, 1)$ and $\rho \sim \text{Beta}(0.5, 0.5)$ (both when defining $\mathbf{u}$ and $\tilde{\mathbf{u}}$), such that the latter is based on the recommendations given by \cite{morris:2019}. Finally, for each of the baseline hazard distributions listed in Section \ref{ssec:hazard-model}, and based on the parameterization given in Web Appendix 1 (Supporting Information), we set the priors as follows
\begin{enumerate}[noitemsep, topsep = 0pt]
    \item Log-normal: $\mu \sim \text{Normal}(0, \sigma^{2}_{\mu})$, where $\sigma^2_{\mu}$ is a large number, and $\sigma \sim \text{Half-Cauchy}(0, \tau_{\sigma})$, with $\sigma^2_{\mu}, \tau_{\sigma} > 0$ \citep{rubio:2018}.
    \item Log-logistic: as for the LN model, $\mu \sim \text{Normal}(0, \sigma^{2}_{\mu})$, where $\sigma^2_{\mu}$ is a large number, and $\sigma \sim \text{Half-Cauchy}(0, \tau_{\sigma})$, with $\sigma^2_{\mu}, \tau_{\sigma} > 0$.
    \item Power Generalized Weibull:  $\eta \sim \text{Half-Cauchy}(0, \tau_{\eta})$, $\nu \sim \text{Half-Cauchy}(0, \tau_{\nu})$, and $\kappa \sim \text{Gamma}$($0.65$, $1.83$), 
    with scale parameters $\tau_{\eta}, \tau_{\nu} > 0$. \change{The prior specification for $\kappa$ has been proven to be weakly informative \citep{dette:2018, alvares:2021}.}
    \item Gamma: $\eta \sim \text{Half-Cauchy}(0, \tau_{\eta})$ and  $\nu \sim \text{Half-Cauchy}(0, \tau_{\nu})$, with scale parameters $\tau_{\eta}, \tau_{\nu} > 0$.
    \item Generalized Gamma: as for the PGW model, $\eta \sim \text{Half-Cauchy}(0, \tau_{\eta})$, $\nu \sim \text{Half-Cauchy}(0, \tau_{\nu})$, and  $\kappa \sim \text{Gamma}$($0.65$, $1.83$), with scale parameters $\tau_{\eta}, \tau_{\nu} > 0$. 
\end{enumerate}

\subsection{Model selection} \label{ssec:modelselection}

To compare the fitted models, we will use a leave-one-out cross validation (LOO CV) procedure; that is, we will use the likelihood evaluated at the parameters' posterior samples as a goodness-of-fit measure. In particular, we will use the Pareto-smoothed importance sampling (PSIS) implementation \citep{vehtari:2017} and compute the corresponding quantities using the \texttt{loo} package \citep{loopackage}. Under the Bayesian framework, the LOO estimate of out-of-sample predictive fit will be computed as
\begin{align*}
    \text{elpd}_{\text{LOO}} &= \sum_{i = 1}^{r}\sum_{j = 1}^{n_i} \log\left[\pi(t_{ij}\mid\mathbf{t}_{-{ij}})\right],
\end{align*}
where $\pi(t_{ij} \mid \mathbf{t}_{-{ij}})$ is the LOO predictive density given $\mathbf{t}_{-ij}$, such that $\mathbf{t}_{-ij}$ corresponds to the vector of all observed time points, except $t_{ij}$. However, instead of re-fitting the model $n = \sum_{i = 1}^r n_i$ times, $\pi(t_{ij} \mid \mathbf{t}_{-{ij}})$ will be approximately computed, $\forall i, j$, using the PSIS technique. For the details, the reader can refer to \cite{vehtari:2015} and \cite{vehtari:2017}. Throughout this paper, we will denote such an estimate as $\widehat{\text{elpd}}_{\text{PSIS-LOO}}$. As a final remark, assuming well-specified and -fitted models, when comparing different approaches, the larger $\widehat{\text{elpd}}_{\text{PSIS-LOO}}$, the better---as such a quantity sums over the posterior predictive model evaluated at a new observation $t_{ij}$, for each $i$ and $j$.

\section{Simulation study} \label{sec:simulations}

In this simulation section, we will assess the performance of our RS-SGH model in three directions. First, we will evaluate our fitted models with respect to their ability to recover the true net survival function, as in Equation \eqref{eq:net-surv}. For that case, we will analyze our models performance based on different sample sizes, different censoring rates, and misspecified distributions for the baseline hazard. Second, fixing all components but the random effect structures, we will compare and select models based on the $\widehat{\text{elpd}}_{\text{PSIS-LOO}}$ criterion, as in Section \ref{ssec:modelselection}. Third, we will use our fitted models to identify riskier areas based on the analysis of the spatial effects. To do so, we will simulate data from the RS-SGH model as described in Web Appendix 3 (Supporting Information).

\subsection{Marginal quantities} \label{ssec:simulation-marginal-quantities}

Our first analysis concerns the estimation of marginal quantities, such as the net survival in Equation \eqref{eq:net-surv}. From that equation, notice that we are integrating out the effects of the spatial components; therefore, provided the model is well fitted and for a sufficiently large sample size in all regions, all random effect structures are expected to produce similar results---as the random effects are assumed to be zero-mean for all models in Section \ref{ssec:spatial-model}. In that case, we will benchmark our fitted spatial model with respect to the true corresponding curves.

We will focus on analyzing the effect of (i) different sample sizes; (ii) different censoring rates; and (iii) misspecified distributions for the baseline hazard. In this regard, we will set the sample size to $200$, $500$, $1000$, and $2000$ patients, the censoring rate to $25\%$ and $50\%$, and we will simulate and fit Model \eqref{eq:exc-haz} with the baseline hazard component defined by the Log-normal (LN) and Power Generalized Weibull (PGW) distributions. The simulation details are given in Appendix \ref{appendix:simulation-details}. \change{From that section, note that the considered spatial structure is defined based on the map of England, split into 9 regions (see Figure \ref{fig:map_england}). This choice is not arbitrary, as it is based on genuine epidemiological questions about cancer survival and how the England territory is administrated. Furthermore, for our simulation study, the number of regions was set to $9$ to ensure that we can repeatedly fit the model based on Equation \eqref{eq:like} within ``reasonable'' computational time---the average fitting times are presented in Table ST2, Supporting Information. For comparison, the fitting times for Section \ref{sec:case_study}, where a larger number of regions is considered (with also more patients), are presented in Table ST10. Lastly, recall that there must exist an interplay between the number of regions, censoring rate, and the sample size. As shown by \cite{rubio:2022}, we must have a sufficiently large number of uncensored observations in each cluster to correctly estimate the model parameters.} As a final remark, all simulation and fitting scenarios are listed in Table \ref{tab:simulated_scenarios}.

\begin{table}[!ht]
	\caption{All simulated scenarios for Section \ref{ssec:simulation-marginal-quantities}. ``Data Generating model'' refers to the model assumed for the data generating procedure, and ``Fitted model'' is defined as per Table \ref{tab:all-models}.}
    \resizebox{\textwidth}{!}{%
    \centering
	\begin{tabular}{ c | c | c | c | l | c | c | c | c | l }
	\# & Data Generating model & Censoring rate & Sample size & Fitted model & \# & Data Generating model & Censoring rate & Sample size & Fitted model \\ \hline 
	01 & RS-SGH LN ICAR & 25\% & \phantom{1}200 & RS-SGH LN  ICAR & 13 & RS-SGH PGW ICAR & 25\% & \phantom{1}200 & RS-SGH LN  ICAR \\
	02 & RS-SGH LN ICAR & 25\% & \phantom{1}500 & RS-SGH LN  ICAR & 14 & RS-SGH PGW ICAR & 25\% & \phantom{1}500 & RS-SGH LN  ICAR \\
	03 & RS-SGH LN ICAR & 25\% &           1000 & RS-SGH LN  ICAR & 15 & RS-SGH PGW ICAR & 25\% &           1000 & RS-SGH LN  ICAR \\
	04 & RS-SGH LN ICAR & 25\% &           2000 & RS-SGH LN  ICAR & 16 & RS-SGH PGW ICAR & 25\% &           2000 & RS-SGH LN  ICAR \\
	05 & RS-SGH LN ICAR & 25\% & \phantom{1}200 & RS-SGH PGW ICAR & 17 & RS-SGH PGW ICAR & 25\% & \phantom{1}200 & RS-SGH PGW ICAR \\
	06 & RS-SGH LN ICAR & 25\% & \phantom{1}500 & RS-SGH PGW ICAR & 18 & RS-SGH PGW ICAR & 25\% & \phantom{1}500 & RS-SGH PGW ICAR \\
	07 & RS-SGH LN ICAR & 25\% &           1000 & RS-SGH PGW ICAR & 19 & RS-SGH PGW ICAR & 25\% &           1000 & RS-SGH PGW ICAR \\
	08 & RS-SGH LN ICAR & 25\% &           2000 & RS-SGH PGW ICAR & 20 & RS-SGH PGW ICAR & 25\% &           2000 & RS-SGH PGW ICAR \\
	09 & RS-SGH LN ICAR & 50\% & \phantom{1}200 & RS-SGH LN  ICAR & 21 & RS-SGH PGW ICAR & 50\% & \phantom{1}200 & RS-SGH LN  ICAR \\
	10 & RS-SGH LN ICAR & 50\% & \phantom{1}500 & RS-SGH LN  ICAR & 22 & RS-SGH PGW ICAR & 50\% & \phantom{1}500 & RS-SGH LN  ICAR \\
	11 & RS-SGH LN ICAR & 50\% &           1000 & RS-SGH LN  ICAR & 23 & RS-SGH PGW ICAR & 50\% &           1000 & RS-SGH LN  ICAR \\
	12 & RS-SGH LN ICAR & 50\% &           2000 & RS-SGH LN  ICAR & 24 & RS-SGH PGW ICAR & 50\% & \phantom{1}500 & RS-SGH LN  ICAR 
	\end{tabular}%
    }
	\label{tab:simulated_scenarios}
\end{table}

From Table \ref{tab:simulated_scenarios}, notice that we are not fitting the PGW distribution model with 50\% censoring rate. We did it in this way since we identified that, for 3-parameter distributions (e.g., Power Generalized Weibull), it might be difficult to obtain well-mixed posterior chains for models fitted based on highly censored data sets. When generating the data, we will simulate $100$ data sets for each combination of sample size, censoring rate, and baseline hazard distribution. Next, for the fitting step, we will write Model \eqref{eq:exc-haz} using the same covariates as the selected ones for the data generating scheme. Also, for the MCMC-based (Markov chain Monte Carlo) code (implemented using \texttt{RStan} in the background) from Web Appendix 2 (Supporting Information), we set the number of chains, the number of iterations and the burn-in size as 4, $4000$, and $2000$, respectively (after fitting the models, the chains for the posterior sampled values were observed to be well mixed in all cases). Then, to assess the fitted models, we will plot the estimated net survival curves (averaged over all regions) along with an error measure defined as
\begin{align} \label{eq:error_integral}
    \text{Error} = \int_{\mathcal{T}} |f(t) - \hat{f}(t)| dt,
\end{align}
where $f$ is the true function, $\hat{f}$ is the corresponding estimated function, and $\mathcal{T} = [t_1, t_2]$ is the analyzed time interval.

\begin{figure}[!ht]
	\centering
	\includegraphics[width = 1\textwidth]{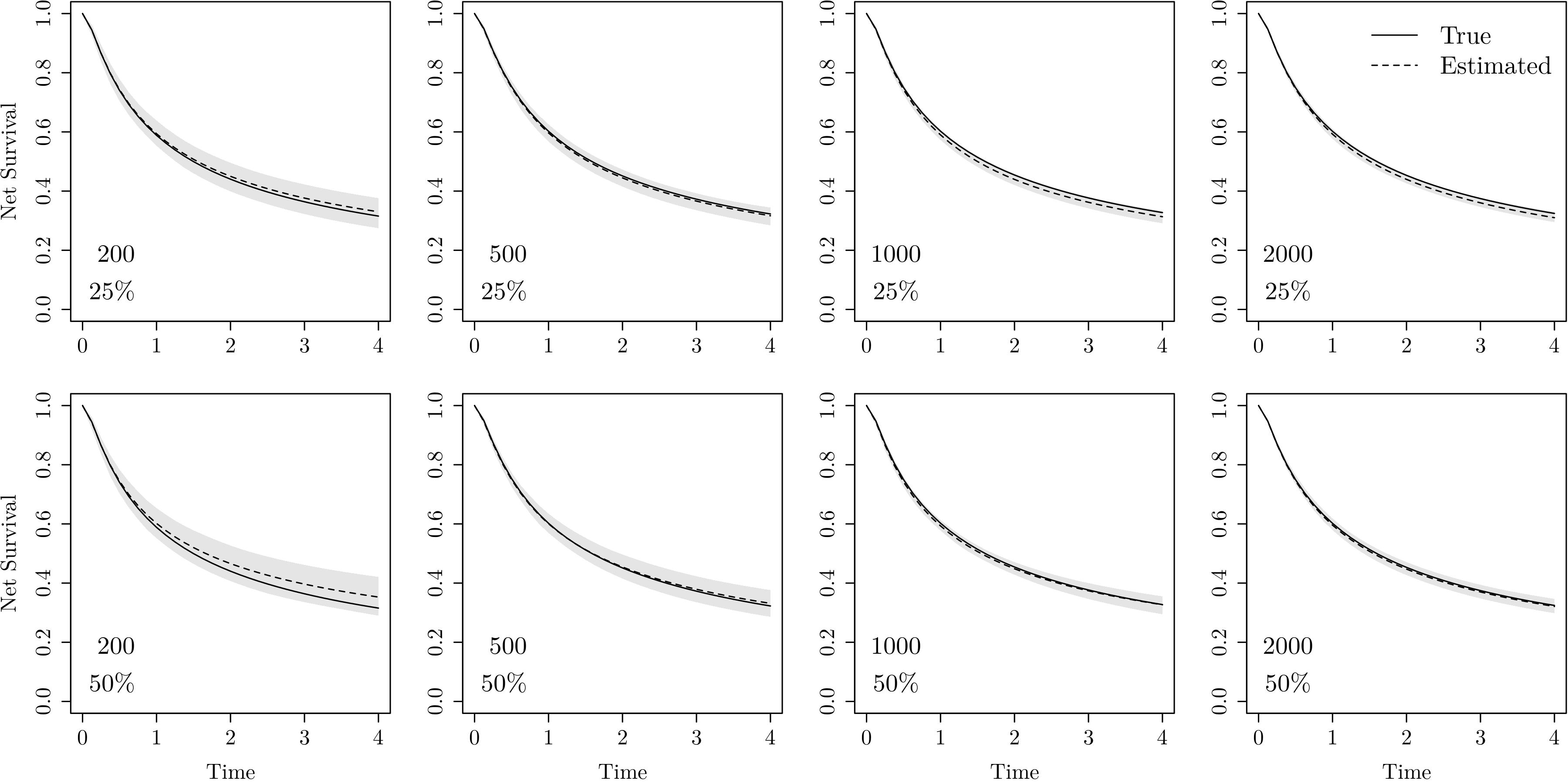}
    \caption{True and estimated (along with a 95\% equal-tailed credible interval) net survival curves based on the fitted RS-SGH LN ICAR model. The data were generated from the same model with 25\% and 50\% censoring rates and sample size set to $200$, $500$, $1000$, and $2000$ patients. Such estimates were obtained by averaging over the $100$ simulated data sets and all regions for each scenario (the corresponding uncertainty was computed based on the percentiles for the curves that average the regions' net survival).}
	\label{fig:first-block-netSur}
\end{figure}

\begin{figure}[!ht]
	\centering
	\includegraphics[width = 1\textwidth]{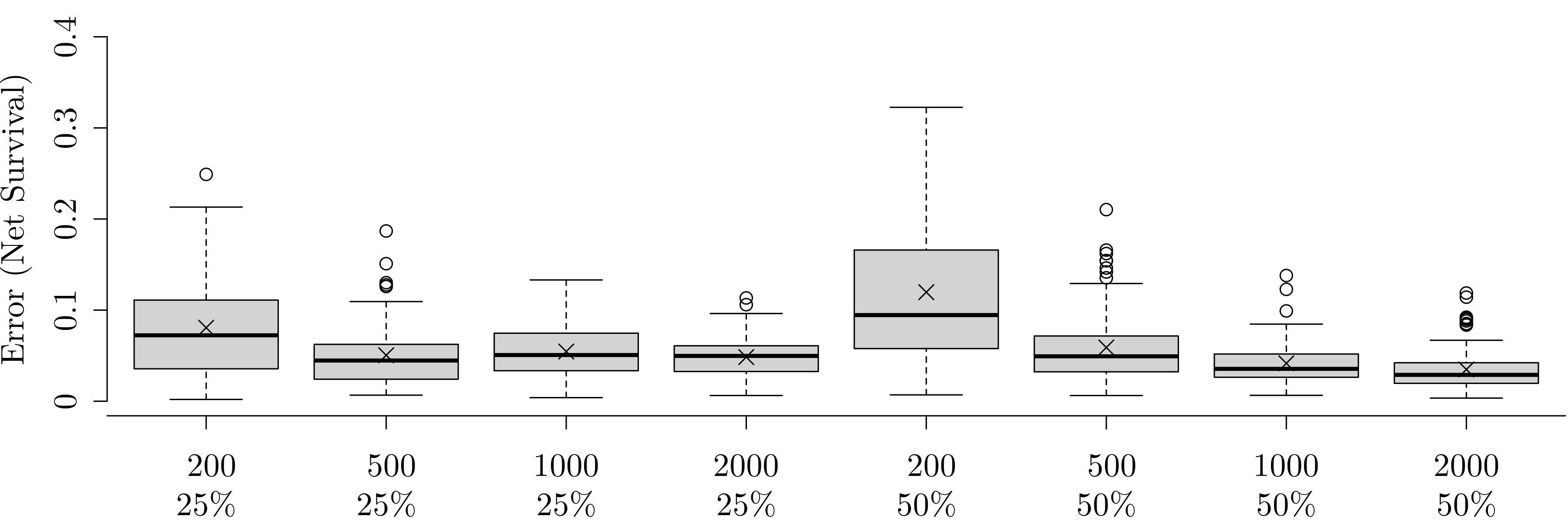}
    \caption{Error in estimating the true net survival function based on the fitted RS-SGH LN ICAR model. The data were generated from the same model with 25\% and 50\% censoring rates and sample size set to $200$, $500$, $1000$, and $2000$ patients. The computed errors aggregate the $100$ simulated data sets and all regions for each scenario. The crosses ($\times$) correspond to the boxplot values mean.}
	\label{fig:first-block-netSur-boxplot}
\end{figure}

First, Figures \ref{fig:first-block-netSur} and \ref{fig:first-block-netSur-boxplot} show the net survival curves and the corresponding errors, as per Equation \eqref{eq:error_integral}, in estimating the true functions for $\mathcal{T} = [0, 4]$, respectively, for data generated from the RS-SGH LN ICAR model with 25\% and 50\% censoring rates for all sample sizes. In that case, we fit the same model as the generating scheme; thus, here, we aim to assess the impact of the censoring rate and the sample size when employing such an approach. From these figures, we can see that our models recover well the true net survival functions for a 25\% censoring rate, with decreasing uncertainty as the sample size increases. In particular, we observe reasonable results for a sample size larger than $500$--$1000$ patients. In a similar manner, for scenarios with 50\% censoring rate, the estimates get better as we increase the number of patients; however, if the sample size is too small (e.g., $200$ patients), the observed bias (and the variability represented in Figure \ref{fig:first-block-netSur-boxplot}) when estimating the net survival curves is larger than before---although such a high censoring effect vanishes as the sample size gets larger.

\begin{figure}[!ht]
	\centering
	\includegraphics[width = 1\textwidth]{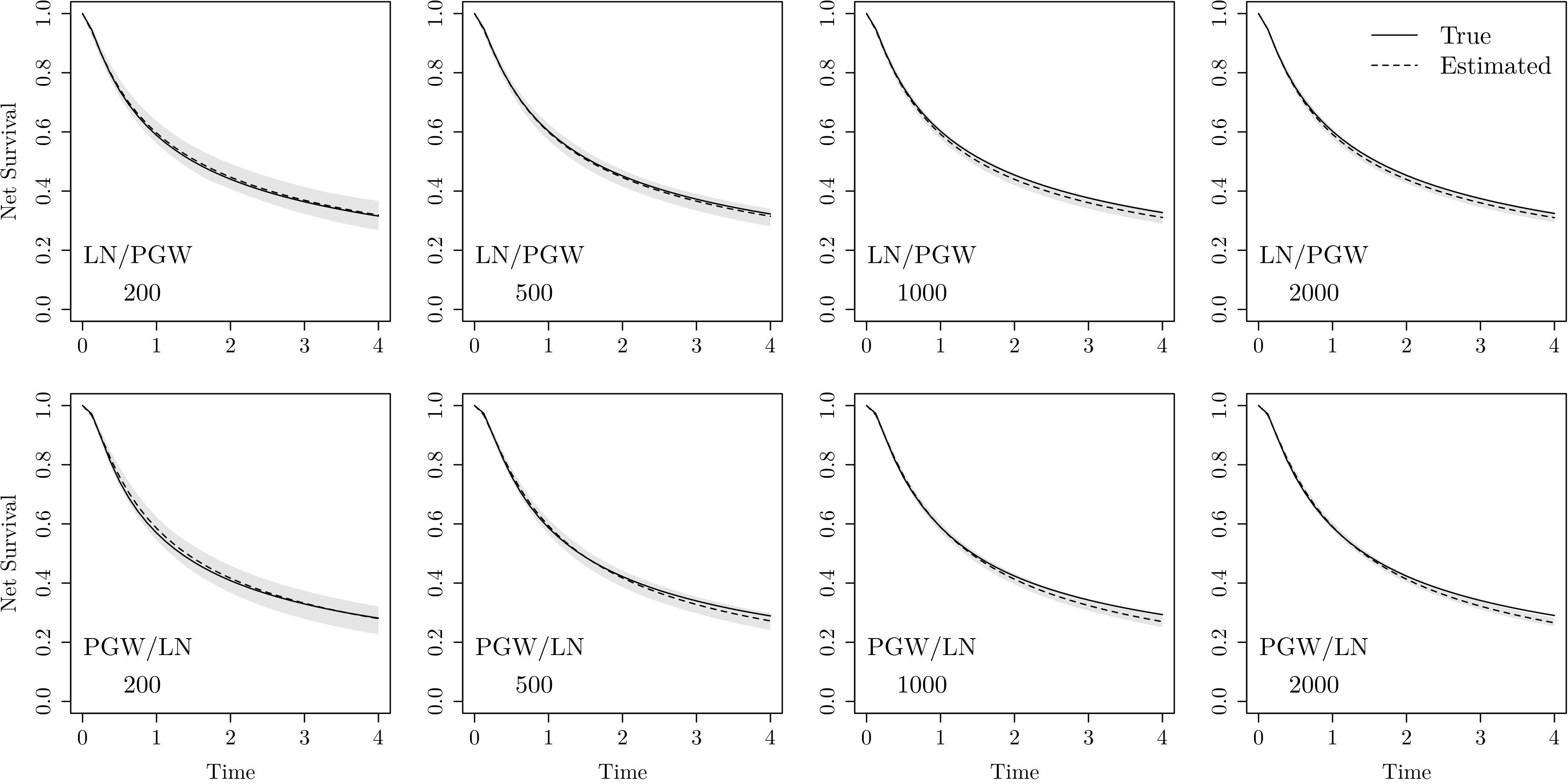}
    \caption{True and estimated (along with a 95\% equal-tailed credible interval) net survival curves based on the fitted RS-SGH PGW ICAR (first row) and RS-SGH LN (second row) models. In these two cases, the data were generated from models RS-SGH LN ICAR (first row) and RS-SGH PGW ICAR (second row), respectively, with 25\% censoring rate and sample size set to $200$, $500$, $1000$, and $2000$ patients. Such estimates were obtained by averaging over the $100$ simulated data sets and all regions for each scenario (the corresponding uncertainty was computed based on the percentiles for the curves that average the regions' net survival).}
	\label{fig:second-block-netSur}
\end{figure}

\begin{figure}[!ht]
	\centering
	\includegraphics[width = 1\textwidth]{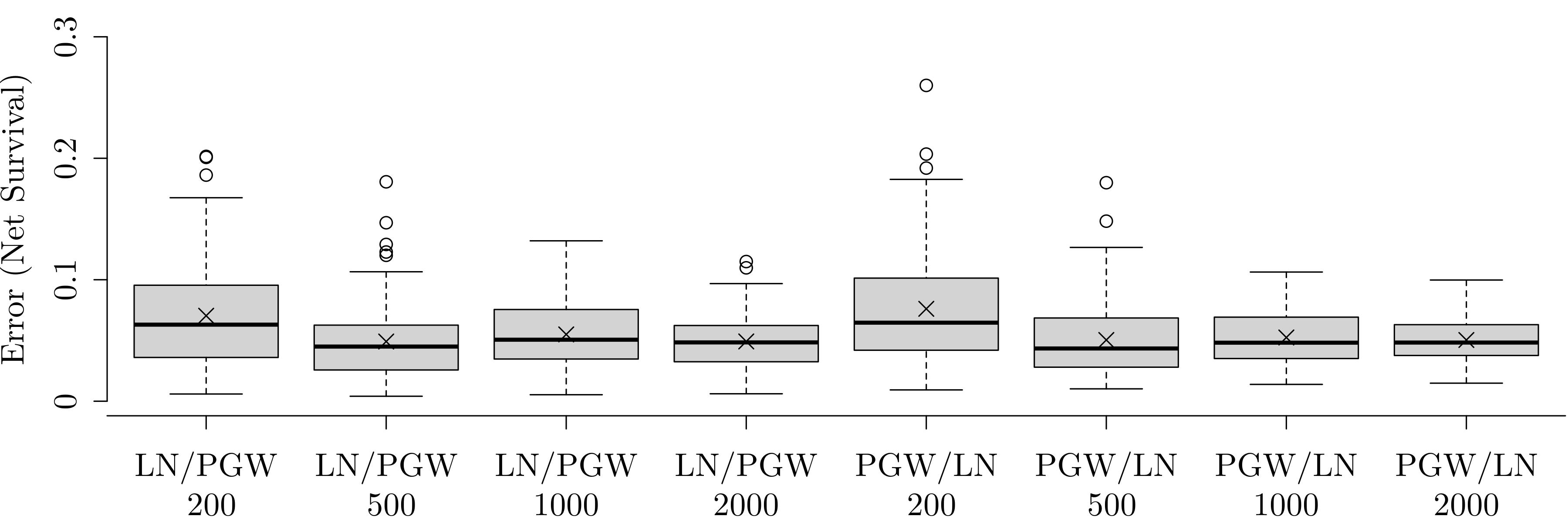}
    \caption{Error in estimating the true net survival function based on the fitted RS-SGH PGW ICAR and RS-SGH LN models. In these two cases, the data were generated from models RS-SGH LN ICAR (four first boxes) and RS-SGH PGW ICAR (four last boxes), respectively, with 25\% censoring rate and sample size set to $200$, $500$, $1000$, and $2000$ patients. The computed errors aggregate the $100$ simulated data sets and all regions for each scenario. The crosses ($\times$) correspond to the boxplot values mean.}
	\label{fig:second-block-netSur-boxplot}
\end{figure}

Second, Figures \ref{fig:second-block-netSur} and \ref{fig:second-block-netSur-boxplot} show similar plots to before, however, for data generated from the RS-SGH LN ICAR and RS-SGH PGW ICAR models with 25\% censoring rate for all sample sizes. In these two cases, the fitted models were RS-SGH PGW ICAR and RS-SGH LN ICAR, respectively; that is, we are fitting misspecified models for the baseline hazard component. From such figures, we can notice that misspecified baseline hazard distributions do not seem to be an important issue if one is solely interested in computing marginal quantities, as the net survival curves. However, as mentioned before, depending on the censoring rate, 3-parameter distributions (e.g., PGW) might require larger samples to fit. Lastly, Figures SF1, SF2, SF3, and SF4 (Supporting Information) show the corresponding results for the remaining scenarios from Table \ref{tab:simulated_scenarios}.

\subsection{Random effects selection} \label{ssec:simulation-model-selection}

In this section, we will analyze the role of the spatial effects in model selection. That is, fixing all components but the random effects, we will select the most appropriate model according the the estimated $\widehat{\text{elpd}}_{\text{PSIS-LOO}}$. To do so, we will, for the same data generating scenarios from Table \ref{tab:simulated_scenarios} with 25\% censoring rate, fit models with no random effects (RS-SGH), IID random effects (RS-SGH IID), ICAR random effects (RS-SGH ICAR), and BYM2 random effects (RS-SGH BYM2) with different distributions for the baseline hazard component. Table ST3 (Supporting Information) lists all considered combinations for data generation and model fitting. Then, similarly to Section \ref{ssec:simulation-marginal-quantities}, we will rank the models for the 100 different simulated data sets in all scenarios. 

After fitting all models (the ones that were not fitted yet in Section \ref{ssec:simulation-marginal-quantities}), we compute $\widehat{\text{elpd}}_{\text{PSIS-LOO}}$ (using the \texttt{loo} package) and compare such estimated quantities for all equivalent scenarios. As a remark, when fitting the models, all posterior chains were well mixed. Table \ref{tab:winning-model-selection-2000} reports the ``best-model proportions'' (i.e., the number of times, out of the 100 data sets, that a model was selected based on the estimated $\widehat{\text{elpd}}_{\text{PSIS-LOO}}$) for all scenarios with sample size set to $2000$ patients. Tables ST4, ST5, and ST6 (Supporting Information) report similar results, but based on the data sets containing $200$, $500$, and $1000$ patients, respectively.

\begin{table}[!ht]
	\caption{``Best-model proportions'' for model selection based on the estimated $\widehat{\text{elpd}}_{\text{PSIS-LOO}}$. In all scenarios, we assumed a 25\% censoring rate and set the sample size to $2000$ patients.}
    \resizebox{\textwidth}{!}{%
    \centering
	\begin{tabular}{ c | c | l | c | c | c | l | c }
	\# & Data Generating model & Fitted model & Best-model proportions & \# & Data Generating model & Fitted model & Best-model proportions \\ \hline 
	01 & RS-SGH LN ICAR & RS-SGH LN  ---  &  \phantom{0}2\% & 09 & RS-SGH PGW ICAR & RS-SGH LN  ---  &  \phantom{0}1\% \\
	02 & RS-SGH LN ICAR & RS-SGH LN  IID  &  \phantom{0}6\% & 10 & RS-SGH PGW ICAR & RS-SGH LN  IID  &            79\% \\
	03 & RS-SGH LN ICAR & RS-SGH LN  ICAR &            51\% & 11 & RS-SGH PGW ICAR & RS-SGH LN  ICAR &  \phantom{0}9\% \\
	04 & RS-SGH LN ICAR & RS-SGH LN  BYM2 &            41\% & 12 & RS-SGH PGW ICAR & RS-SGH LN  BYM2 &            11\% \\ \hline
	05 & RS-SGH LN ICAR & RS-SGH PGW ---  &            15\% & 13 & RS-SGH PGW ICAR & RS-SGH PGW ---  &  \phantom{0}7\% \\
	06 & RS-SGH LN ICAR & RS-SGH PGW IID  &            32\% & 14 & RS-SGH PGW ICAR & RS-SGH PGW IID  &            26\% \\
	07 & RS-SGH LN ICAR & RS-SGH PGW ICAR &            28\% & 15 & RS-SGH PGW ICAR & RS-SGH PGW ICAR &            41\% \\
	08 & RS-SGH LN ICAR & RS-SGH PGW BYM2 &            25\% & 16 & RS-SGH PGW ICAR & RS-SGH PGW BYM2 &            26\% 
	\end{tabular}%
    }
	\label{tab:winning-model-selection-2000}
\end{table}

\change{From Table \ref{tab:winning-model-selection-2000} (and Tables ST4--ST6, Supporting Information), we can see that models that account for some random effects structure were selected more often in all scenarios (except for 200 patients, as in Table ST4). Also, as the sample size increases (500 patients or more), not only the models with spatial effects were selected with higher proportions, but also the correct model (RS-SGH ICAR) was the most frequently selected approach for some of the specified settings with generating model based on the Log-normal distribution for the baseline hazard. On the contrary, for the misspecified scenarios with generating scheme based on the Power Generalized Weibull distribution, the model with clustering effects seemed to perform better than the competing approaches---as a reason for this to happen, recall that the Log-normal model might fail to recover the PGW hazard shape; in that case, it is possible that the spatial structure for the random effects gets suppressed by the error from the poorly fitted fixed components and the IID model performs better. In that way, under the assumption that the baseline hazard distribution can capture the corresponding hazard shape from the data, the employed model selection approach seems to work well when selecting an appropriate random structure, provided that we have a minimum of $500$--$1000$ data points (as we also identified in Section \ref{ssec:simulation-marginal-quantities}).}

\subsection{Spatial effects analysis} \label{ssec:simulation-spatial-effects}

For the following analysis, we showcase the insights we can obtain from the estimated spatial structure. For that, we will generate data from a model with manually set spatial effects. In particular, we will simulate data from the RS-SGH model with a Log-normal distribution for the baseline hazard, such that $\mu = 0.65$ and $\sigma = 1.15$, as in Section \ref{ssec:simulation-marginal-quantities}. Also, we will choose the covariates and set the corresponding coefficients as in Appendix \ref{appendix:simulation-details}. Lastly, we will set $\tilde{\mathbf{u}} = \mathbf{u} = (2.0$, $1.5$, $1.0$, $0.5$, $0$, $-0.5$, $-1.0$, $-1.5$, $-2.0)^{\top}$ for the 1--9 regions in England (as per Figure \ref{fig:map_england}), respectively.  Then, for a data set with $10000$ individuals and censoring rate of 25\%, we will fit the RS-SGH LN model, with no random effects and with IID, ICAR, and BYM2 random structures, once, such that the MCMC setting parameters will be defined as in Section \ref{ssec:simulation-marginal-quantities}, and analyze the estimated spatial effects (if any) based on the corresponding posterior distributions.

\begin{figure}[!ht]
	\centering
	\includegraphics[width = 0.9\textwidth]{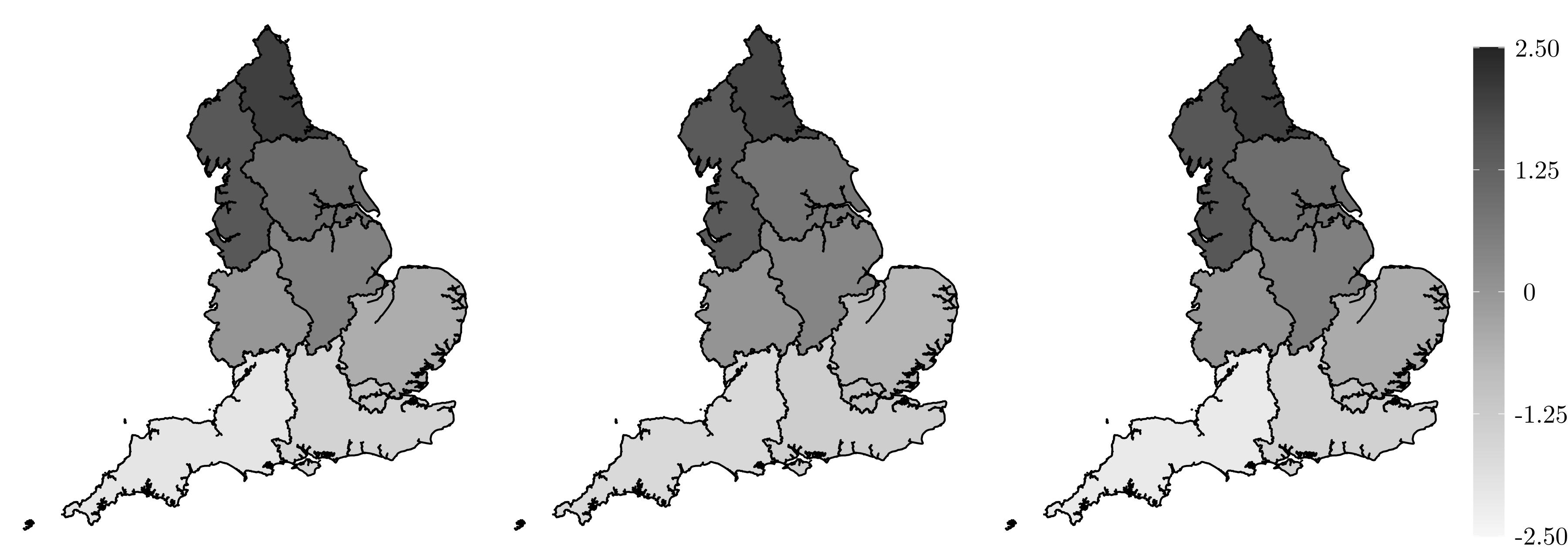}
    \caption{Spatial effects for the RS-SGH LN ICAR model. {Left panel:} True spatial effects $\tilde{\mathbf{u}} = \mathbf{u} = (2.0$, $1.5$, $1.0$, $0.5$, $0$, $-0.5$, $-1.0$, $-1.5$, $-2.0)^{\top}$. {Middle panel:} Estimated time-level spatial effects $\tilde{\mathbf{u}} = (1.86, 1.45, 0.81, 0.41, 0.04, -0.69, -0.90, -1.33, -1.65)^{\top}$. {Right panel:} Estimated hazard-level spatial effects $\mathbf{u} = (1.96$, $1.54$, $0.97$, $0.53$, $0.04$, $-0.45$, $-1.01$, $-1.43$, $-2.15)^{\top}$.}
	\label{fig:re_LN_ABST}
\end{figure}

Similar to previous simulations, when fitting the models, all posterior chains were well mixed. However, when comparing such approaches according to the $\widehat{\text{elpd}}_{\text{PSIS-LOO}}$ criterion, the results pointed out to the ICAR random structure as the most appropriate model---although the pairwise differences between the ICAR model and the IID and BYM2 models seem to be non-significant, as shown in Table ST7 (Supporting Information). The model with no spatial effects was ranked in the lowest position. This means that, although we do need to account for non-observed spatial heterogeneity, for such a large data set, all random structures captured well the spatial effects. Figure \ref{fig:re_LN_ABST} shows the true and estimated spatial effects (at both time- and hazard-levels) for the ICAR random structure (Figures SF5 and SF6, in Web Appendix 6 (Supporting Information), show the corresponding maps for the IID and BYM2 structures, respectively), such that the plotted estimates were computed based on the mean of the sample obtained from the corresponding random effect posterior distributions. The first thing we can observe from these maps is that we were able to recover the spatial effects reasonably well. Table ST8 (Supporting Information) shows the estimates for all RS-SGH LN ICAR model parameters along with a 95\% equal-tail credible interval for the same model (Table ST8 also presents similar results for models RS-SGH LN IID and RS-SGH LN BYM2). Second, based on these estimates only, we can study the geographical inequalities for different population groups. That is, fixing all terms but $\tilde{\mathbf{u}}$ and $\mathbf{u}$, the risk of dying is larger for patients who live in regions with positive estimates for the hazard-level spatial effects. The time-level spatial effects have a similar interpretation, but they have to be analyzed along with the baseline hazard shape; i.e., if $h_0(t; \boldsymbol{\theta})$ increases with $t$, then positive $\tilde{u}_i$'s imply in riskier areas; contrarily, if the baseline hazard is a decreasing function, a positive time-level random effect decreases the risk of dying in $i$, for patients with the same characteristics, in comparison to other regions with smaller effects of the same kind.

However, as suggested by \cite{taylor:2017}, analyzing the spatial effects as in Figure \ref{fig:re_LN_ABST} ignores the precision of the estimates---recall that we would be more precise in estimating $\tilde{\mathbf{u}}$ and $\mathbf{u}$ in regions with more patients. Alternatively, we could compute and analyze the relative exceedance probability $\mathbb{P}(u_i > c)$, for all $i$ and some threshold $c$ (the same applies to $\tilde{u}_i$, $\forall i$). Under the Bayesian setting, such a probability can be estimated based on the posterior sample for the spatial effects. This measure quantifies the variability of the random effect estimate around $c$, but it is also useful to assess unusual elevations in such a model component. For instance, we might be interested in computing $\mathbb{P}(u_i > 0)$, for all regions $i$ (the same for $\tilde{u}_i$, $\forall i$). This quantity can be used as a proxy for the risk level to which a group of patients (defined by their geographical location) is subjected, compared to the general population. Figures SF7, SF8, and SF9 (Supporting Information) show the computed relative exceedance probabilities for both time-level and hazard-level random effects, such that the threshold $c = 0$, for models RS-SGH LN ICAR, RS-SGH LN IID, and  RS-SGH LN BYM2, respectively.

\change{Lastly, we can also analyze the estimates for $\sigma_{u} = 1 / \sqrt{\tau_{u}}$ (and $\sigma_{\tilde{u}} = 1 / \sqrt{\tau_{\tilde{u}}}$) in all cases. Figures SF10, SF11, and SF12 (Supporting Information) show the estimated posterior densities for such parameter when fitting the RS-SGH LN model with ICAR, IID, and BYM2 random effects, respectively. From these figures (and TableST8), we can see that the two spatial structures were simultaneously and successfully estimated. As discussed in Section \ref{ssec:likelihood}, if (i) we have enough uncensored patients in each region and (ii) the baseline hazard does not belong to the Weibull family, then we may estimate well the random effects.}

\section{Case study} \label{sec:case_study}

In this section, we will analyze a data set that contains survival information about male and female patients diagnosed with colon cancer between 2015 and 2016 in England. Appendix \ref{appendix:data-applications-details} presents a complete description of the data set. More specifically, we analyze the survival of colon cancer patients in England with spatial structure defined in two different manners: (i) based on the administrative boundaries given by the Government Office Regions (as per Figure \ref{fig:map_england}) and (ii) based on the health boundaries determined by the Cancer Alliances \citep{canceralliances}. The main goal is assessing the impact of different geographies when accounting for the possible spatial correlation in the data.

For all scenarios, we know subject-specific prognostic factors, which include age at diagnosis, sex, deprivation level, and cancer stage. The population hazard term $h_{\text{P}}(\text{age}_{ij} + t; \mathbf{z}_{ij})$ was determined based on the life tables for England defined for the corresponding calendar year, and stratified by age, sex, deprivation level (according to the computed quintiles of such a score), and region of residence \citep{rachet2015multivariable, inequalitylifetables}. Also, for all models separately fitted for male and female individuals, we always set the time-level linear predictor to $\text{age}_{ij} \alpha + \tilde{u}_{i}$, and the hazard-level linear predictor to $\text{age}_{ij}\beta_1 + \sum_{k = 2}^K \mathds{1}_{\text{stage}_{ij}(k)} \beta_k + \text{deprivation}_{ij} \beta_{(K + 1)} + u_i$, where $\mathds{1}_{\text{stage}_{ij}(k)}$, for $2 \leq k \leq K$, is an indicator function for individuals who belong to the $k$-th cancer tumour stage, and, as in Section \ref{sec:simulations}, the $\tilde{u}_i$ and $u_i$ components (if any) are defined as one of the (spatially dependent) random structures introduced in Section \ref{ssec:spatial-model}. Finally, the variables ``age'' and ``deprivation'' were standardized for numerical stability.

\change{Given the setting we just described,} we will fit Model \eqref{eq:exc-haz} for $10,936$ males and $9,586$ females  with a diagnosis of colon cancer in 2016 in England. The linear predictor terms will be defined as mentioned above, such that we have $K = 4$ levels for the cancer stage (``1'' being \textit{early stage} and ``4'' \textit{late stage}). Also, regarding the baseline hazard distributions and the random effects, for each scenario we fit the following models: RS-SGH LL ICAR, RS-SGH LL BYM2, RS-SGH LN ICAR, RS-SGH LN BYM2, RS-SGH PGW ICAR, and RS-SGH PGW BYM2, as per the notation introduced in Table \ref{tab:all-models}. \change{We selected these models for two reasons. First, due to the computational cost associated with fitting them to such a large data set (Table ST10, Supporting Information, shows the fitting times for all cases), we decided to limit our investigation to models with clinical motivation (in particular, models that present some underlying spatial structure). Second, such models are flexible enough to cover many different hazard shapes and possible spatially dependent random effects.} For the MCMC-based code, we set the number of chains, the number of iterations and the burn-in size as 4, $10000$, and $8000$, respectively (the posterior chains were well mixed in all cases, except for the RS-SGH LL BYM2 model with male patients spatially distributed over the Government Office Regions---see Table ST9, Supporting Information). Next, the best model is selected according to the $\widehat{\text{elpd}}_{\text{PSIS-LOO}}$ criterion, as in Section \ref{ssec:modelselection}. Lastly, the spatial structure is defined according to two geographies: (i) the 9 Government Office Regions (GOR), as in Figure \ref{fig:map_england}, and (ii) the 19 Cancer Alliances Regions delimited during the calendar year of 2016. 

\begin{figure}[!ht]
	\centering
	\includegraphics[width = 0.75\textwidth]{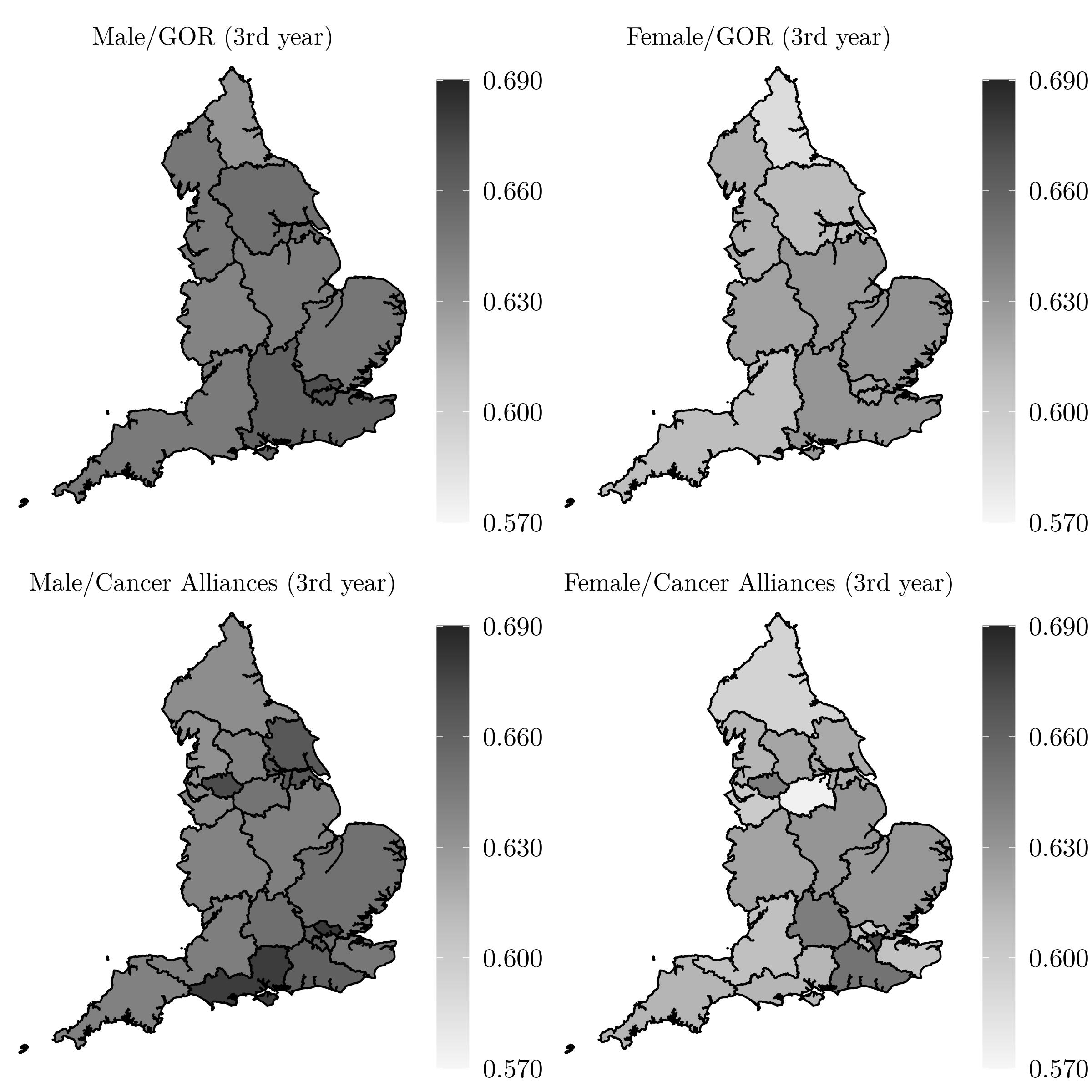}
    \caption{Net survival point estimate for $t = 3$ based on the {(i: top-left panel)} ``Government Office Regions'' spatial structure with fitted model RS-SGH LN BYM2 for male patients, {(ii: top-right panel)} ``Government Office Regions'' spatial structure with fitted model RS-SGH LN BYM2 for female patients, {(iii: bottom-left panel)} ``Cancer Alliances Regions'' spatial structure with fitted model RS-SGH LN BYM2 for male patients, and {(iv: bottom-right panel)} ``Cancer Alliances Regions'' spatial structure with fitted model RS-SGH LN BYM2 for female patients.}
	\label{fig:cs01-t3-mean}
\end{figure}

Considering these fitted models, Table ST9 (Supporting Information) shows the selected model (according to the $\widehat{\text{elpd}}_{\text{PSIS-LOO}}$ criterion) for each scenario. Thus, the following results are based on the highest-ranked modeling alternatives. Then, we compute the net survival for $t = 1$ and $3$ years (along with the estimated 95\% equal-tailed credible interval) for all regions. Here, it is important mentioning how the uncertainty for this quantity is being estimated; for each sampled vector of parameters $\boldsymbol{\xi}^{s}$, where $s$ is the index for the posterior sample, we determine the $2.5^{\text{th}}$ and $97.5^{\text{th}}$ percentiles of $S_{\text{N},i}(t \mid  \boldsymbol{\xi}^{s})$, for all $t$ and $i$, as per Equation \eqref{eq:netsur-strat}. Figure \ref{fig:cs01-t3-mean} shows the estimated net survival for the male and female groups, the two different geographies, and $t = 3$ years; also, Figures SF13 and SF14 (Supporting Information) report the associated uncertainty. Similarly, Figures SF15, SF16, and SF17 (Supporting Information) present the corresponding maps for $t = 1$ year.

From these figures, we can analyze (i) the rate of change for the net survival estimates, (ii) the difference between the net survival for male and female patients, and (iii) the impact of the chosen administrative boundaries when estimating the quantities of interest. Firstly, the net survival seems to decrease faster during the first years after diagnosis, such that the corresponding estimates after 1 and 3 years of diagnosis are, approximately, 0.75 and 0.63, respectively.  Secondly, regarding the differences between male and female patients, the female individuals are shown to be slightly more likely of dying than men---this difference can be seen in the maps for all follow-up time windows. Thirdly, analyzing the England territory based on a finer resolution (e.g., Cancer Alliances Regions) brings us more information if compared to the GOR-based results. In particular, ``South Yorkshire, Bassetlaw, North Derbyshire and Hardwick'' (as per Figure SF18, Supporting Information) shows a lower (for the female group) estimated net survival than the other regions---notice that, just by inspecting the GOR-based estimates, it would be difficult to identify these locations, and practical implications (e.g., allocation of resources for medically underserved areas) could not happen. 

\change{Furthermore, given that the random effects play a major role in the description of the results, Table ST11 (Supporting Information) shows the estimated values (with reported uncertainty) of $\sigma_{u} = 1 / \sqrt{\tau_u}$ (same for $\sigma_{\tilde{u}}$) and $\rho$ (same for $\tilde{\rho}$) for all highest-ranked BYM2 models. Additionally, Figures SF19--SF22 show the estimated posterior densities of $\sigma_{u}$ and $\sigma_{\tilde{u}}$ for the same models. From that table and figures, note that, first, the uncertainty associated with the estimation of $\sigma_u$ and $\sigma_{\tilde{u}}$ is not large. This indicates that we did not encounter any identifiability issues when recovering the underlying spatial structures. Second, the point estimates for $\rho$ and $\tilde{\rho}$ are around $0.55$ (with standard error around $0.35$) for all scenarios, which indicates that the proportion of the variance that comes from the structured random effects is similar to the contribution from the unstructured random effects. That is, the flexibility that comes with the BYM2 structure seems to be important to correctly characterize the variability that cannot be explained by the fixed effects, both in time- and hazard-level.}

We can also compute the net survival stratified by the categorical variables; in this section, we will consider the ``deprivation level'' and the ``cancer stage'' as the population strata. However, recall that the deprivation level is a continuously defined score, thus, if we want to aggregate the patients based on such information, we can compute its quintiles and classify the individuals according to the obtained intervals. In that way, the deprivation score will have 5 levels (``1'' being \textit{least deprived} and ``5'' \textit{most deprived}). As before, Figures SF23 and SF26 (Supporting Information) show the estimated net survival maps, for $t = 3$ years, stratified by the ``deprivation level'' and the ``cancer stage,'' respectively, such that we plotted and compared the deprivation levels ``1'' (least deprived level) and ``5'' (most deprived level), and the cancer stages ``1, 2, and 3'' (early stages) and ``4'' (late stage). Figures SF24, SF25, SF27, and SF28 (Supporting Information) report the associated uncertainty. Lastly, Figures SF29--SF34 (Supporting Information) present the corresponding maps for $t = 1$ year.

Firstly, based on the figures for the estimated net survival curves stratified by deprivation level, we can notice that, for all time points, not only the estimates vary over space (with less homogeneous spatial distribution when we consider the finer spatial resolution), but also the net survival for the different population strata decrease as the deprivation level gets larger; in particular, patients with a deprivation score of 5 have higher chances of dying than the least deprived group regardless of the time span, gender, and the Government Office Region (or Cancer Alliance)---which is likely to be associated with sub-optimal treatment strategies offered to this group. Secondly, by analyzing the figures that show the net survival estimates stratified by the cancer stage, we have similar conclusions, that is, the net survival decrease as the patients are diagnosed with later stages for the colon cancer. However, for this stratified analysis, we can notice a much larger difference in the chances of surviving between the groups with cancer stages ``1, 2, and 3'' (early stages) and ``4'' (late stage), regardless of the other factors. In fact, when plotting the corresponding net survival maps, we had to present the results in different scales for each level of severity (as stressed in all figures captions); otherwise, the spatial variability in each of these two groups would not be captured in the maps. This occurs since patients with \textit{stage 4}-cancer are less likely to be cured and, instead, only receive palliative care.

\section{Discussion} \label{sec:discussion}

In this work, we introduced the Relative Survival Spatial General Hazard (RS-SGH) class of models that generalizes, under the relative survival framework, other survival models. The proposed RS-SGH models account for spatial random effects both in the time-level and hazard-level components, such that these random structures can be modeled, among other approaches, according to the ICAR or BYM2 smoothing priors. The proposed class of models was implemented using \texttt{R} \citep{CRAN} and STAN \citep{carpenter:2017} and made available in a public repository, which allows for reproducibility of our research. Web Appendix 2 (Supporting Information) provides an example on how to use the scripts; in particular, Table ST1 lists all models that are currently possible to implement. Also, regarding model selection, we computed and used the $\widehat{\text{elpd}}_{\text{PSIS-LOO}}$ estimates (as per Section \ref{ssec:modelselection}), and tested its performance in Section \ref{ssec:simulation-model-selection}. This work also contains other minor contributions, such as (i) the prior distribution recommendations (as per Section \ref{ssec:priors}) for the model parameters and hyperparameters, (ii) some guidelines about the sample size, baseline hazard distribution misspecification, and censoring rate when fitting models of this kind (as per Section \ref{ssec:simulation-marginal-quantities}), and (iii) a simple extension of the ``exceedance probability'' idea to the computed (and interpreted) ``relative exceedance probabilities'' (as per Section \ref{ssec:simulation-spatial-effects}).

Aiming to validate the proposed model and inference tools, we conducted a simulation study that analyzed the effects of the sample size, censoring rate, and the baseline hazard distribution when estimating the model parameters and recovering the net survival curves. In this regard, the sample size and the censoring rate were shown to be the most important factors to control; for instance, in most cases, a minimum sample size of 500--1000 patients provided estimates with less variability for the net survival curves. Also, higher censoring rates (e.g., 50\%) with not large sample sizes (e.g., 200--500 patients) produced biased estimates for this same quantity. In fact, for 3-parameter distributions (e.g., Power Generalized Weibull), it might be difficult even to obtain well-mixed posterior chains when fitting the model. However, the misspecification of the baseline hazard distribution, provided that we have enough non-censored observations and a model that can capture the true hazard shape, had little impact in the estimation of marginal quantities.
As part of the simulation study, we also assessed the model selection performance and the ability to recover the true spatial effects. Also, based on these estimated random structures, we could compute the relative exceedance probabilities, which are functions of the spatial effects that can be used, depending on the set threshold, to compare specific locations to the general population with respect to their net survival. \change{As a note, our simulation study was conducted based on a spatial graph defined by 9 regions. Hence, it may be of interest to explore scenarios with a larger number of areas.}

We have also presented a case study aiming at answering genuine questions of interest in cancer epidemiology. In particular, we found that a finer spatial resolution brings us more important information about areas that present lower net survival than the overall country. Identifying these locations is crucial as, based on such knowledge, decision-makers can focus their resources on improving the lives of the vulnerable groups of the population. \change{Moreover, we have illustrated how to produce summaries for subgroups of the population of interest, such as those defined by ``deprivation level'' and ``cancer stage.'' For the former, we have found that most-deprived patients (deprivation level 5) exhibit lower chances of survival compared to the least deprived groups. For the latter, patients with late-stage cancer (stage 4) experience a significant reduction in their survival prospects; in fact, as we have briefly mentioned in Section \ref{sec:case_study}, in most cases, these patients only receive palliative care.}

The proposed methodology and results presented in this work can be extended in different directions. Firstly, we could also include a time-dependent component in Equation \eqref{eq:exc-haz} that explains the non-observed temporal variability associated to the year of diagnosis. Such an extension could be mainly useful for studies that take individuals diagnosed over a very large time window, as the treatment (thus, the chances of surviving) is likely to improve in the long term. \change{Secondly, in Section \ref{sec:case_study}, when modeling survival, it may be useful to simultaneously include spatial information not only about the patients' place of residence, but also about their local of treatment. As pointed out by \cite{quaresma:2022}, cancer incidence depends on where you live (as this is related to deprivation, and deprivation has a strong relationship with geographies), while survival also depends on where you are treated (as it depends on the quality of healthcare). Thus, future work might extend our model into this direction.} Thirdly, missing data is a prevalent problem in population studies. Thus, a possible extension of our work consists of developing multiple imputation strategies to account for missing data, while also accounting for spatial variability. Fourthly, less common smoothing priors for describing the possible spatial autocorrelations among regions could have been used; for instance, the directed acyclic graph auto-regressive (DAGAR) model \citep{datta:2019} is an alternative to the ICAR model that can also be used for modeling other data structures (e.g., images and networks). \change{However, while still using the ICAR formulation, the PC priors can be employed when specifying the precision parameters in the spatial random effects. As discussed by \cite{simpson:2017} (and references therein), a Gamma prior may not be the most suitable choice for this problem. Similarly, PC priors may also be used in the context of the BYM2 model \citep{riebler:2016}. Therefore, the implementation of such penalized complexity priors is a consideration for future work.} Finally, the idea of incorporating spatial (or spatio-temporal) random structures into the hazard model can also be implemented in other survival modeling frameworks, such as the competing risks models, cure models, and overall survival models. 

\section*{Declarations} 
\subsection*{Funding} Manuela Quaresma is funded by the Cancer Research UK Population Research Committee Funding Scheme: Cancer Research UK Population Research Committee-Programme Award C7923/A29018. The research work of Francisco J. Rodr\'iguez-Cort\'es has been partially supported by Universidad Nacional de Colombia, HERMES projects, Grant/Award Number: 56470.

\subsection*{Conflict of interest} The authors declare that they have no conflict of interest.

\subsection*{Data availability statement} Data may be obtained from a third party and are not publicly available. This study uses English cancer registration data managed by the National Cancer Registration and Analysis Service (NCRAS). The authors do not own these data and hence are not permitted to share them in the original form. The data are available from the Data Access Request Service (DARS) at NHS Digital (merged with NHS England from February 2023). 

\section*{References}
\addcontentsline{toc}{section}{References}
\bibliographystyle{cell}
\bibliography{references}

\clearpage    
\appendix

\section{Sub-models based on the RS-SGH approach} \label{appendix:submodels}

Table \ref{tab:models} shows eight possible sub-models that can be derived from the Relative Survival Spatial General Hazard (RS-SGH) model and that we believe are useful for researchers and practitioners working with survival data.

\begin{table}[!ht]
	\caption{Eight simpler models based on the Relative Survival Spatial General Hazard (RS-SGH) modeling approach. The ``Description'' column refers to the corresponding terms in Equation \eqref{eq:exc-haz}.}
    \centering
	\begin{tabular}{ l | c | l | c }
	Name & Description & Name & Description  \\ \hline 
	RS-SGH-I &  $\tilde{\mathbf{u}} = \mathbf{0}$& RS-GH & $\tilde{\mathbf{u}} = \mathbf{u} = \mathbf{0}$ \\
	RS-SGH-II &  $\tilde{\mathbf{u}} = \mathbf{u}$& RS-PH & $\tilde{\mathbf{u}} = \mathbf{u} = \mathbf{0}, \boldsymbol{\alpha} = \mathbf{0}$ \\
	RS-SPH &  $\tilde{\mathbf{u}} = \mathbf{0}, \boldsymbol{\alpha} = \mathbf{0}$ & RS-AFT & $\tilde{\mathbf{u}} = \mathbf{u} = \mathbf{0}, \boldsymbol{\alpha} = \boldsymbol{\beta}$ \\
	RS-SAFT &  $\tilde{\mathbf{u}} = \mathbf{u}, \boldsymbol{\alpha} = \boldsymbol{\beta}$ & RS-AH&  $\tilde{\mathbf{u}} = \mathbf{u} = \mathbf{0}, \boldsymbol{\beta} = \mathbf{0}$ \\
	\end{tabular}%
	\label{tab:models}
\end{table}

\section{Implemented models} \label{appendix:implemented-models}

Referring back to Model \eqref{eq:exc-haz} (and all variations from Table \ref{tab:models}) and assuming a parametric form for the baseline hazard function given by the models listed in Section \ref{ssec:hazard-model}, and considering all possible structures for the random effects defined in Section \ref{ssec:spatial-model}, we can enumerate at least 95 models to choose from. Table \ref{tab:all-models} lists all possible models.

\begin{table}[!ht]
	\caption{All implemented models. The ``Dist.'' column refers to the possible distributions for the baseline hazard function, ``Model'' refers to the implemented excess hazard models---as per Table \eqref{tab:models}, and ``R.E.'' refers to the spatially structured (and unstructured) random effects described in Section \ref{ssec:spatial-model}.}
	\resizebox{\textwidth}{!}{%
	    \centering
		\begin{tabular}{ c | c | l | l |  c |c | l | l | c | c | l | l | c | c | l | l | c | c | l | l }
		\#  & Dist.& Model      & R.E. & \#  & Dist. & Model      & R.E. & \#  & Dist. & Model      & R.E. & \#  & Dist. & Model      & R.E. & \#  & Dist. & Model      & R.E. \\ \hline
		01 & LN   & RS-SGH    & BYM2 & 20 & LL    & RS-SGH    & BYM2 & 39 & PGW   & RS-SGH    & BYM2 & 58 & GAM   & RS-SGH    & BYM2 & 77 & GG    & RS-SGH    & BYM2 \\
		02 & LN   & RS-SGH    & ICAR & 21 & LL    & RS-SGH    & ICAR & 40 & PGW   & RS-SGH    & ICAR & 59 & GAM   & RS-SGH    & ICAR & 78 & GG    & RS-SGH    & ICAR \\  
		03 & LN   & RS-SGH    & IID  & 22 & LL    & RS-SGH    & IID  & 41 & PGW   & RS-SGH    & IID  & 60 & GAM   & RS-SGH    & IID  & 79 & GG    & RS-SGH    & IID  \\  
		04 & LN   & RS-SGH-I  & BYM2 & 23 & LL    & RS-SGH-I  & BYM2 & 42 & PGW   & RS-SGH-I  & BYM2 & 61 & GAM   & RS-SGH-I  & BYM2 & 80 & GG    & RS-SGH-I  & BYM2 \\
		05 & LN   & RS-SGH-I  & ICAR & 24 & LL    & RS-SGH-I  & ICAR & 43 & PGW   & RS-SGH-I  & ICAR & 62 & GAM   &  RS-SGH-I & ICAR & 81 & GG    & RS-SGH-I  & ICAR \\
		06 & LN   & RS-SGH-I  & IID  & 25 & LL    & RS-SGH-I  & IID  & 44 & PGW   & RS-SGH-I  & IID  & 63 & GAM   & RS-SGH-I  & IID  & 82 & GG    & RS-SGH-I  & IID  \\
		07 & LN   & RS-SGH-II & BYM2 & 26 & LL    & RS-SGH-II & BYM2 & 45 & PGW   & RS-SGH-II & BYM2 & 64 & GAM   & RS-SGH-II & BYM2 & 83 & GG    & RS-SGH-II & BYM2 \\
		08 & LN   & RS-SGH-II & ICAR & 27 & LL    & RS-SGH-II & ICAR & 46 & PGW   & RS-SGH-II & ICAR & 65 & GAM   & RS-SGH-II & ICAR & 84 & GG    & RS-SGH-II & ICAR \\
		09 & LN   & RS-SGH-II & IID  & 28 & LL    & RS-SGH-II & IID  & 47 & PGW   & RS-SGH-II & IID  & 66 & GAM   & RS-SGH-II & IID  & 85 & GG    & RS-SGH-II & IID  \\
		10 & LN   & RS-SPH    & BYM2 & 29 & LL    & RS-SPH    & BYM2 & 48 & PGW   & RS-SPH    & BYM2 & 67 & GAM   & RS-SPH    & BYM2 & 86 & GG    & RS-SPH    & BYM2 \\  
		11 & LN   & RS-SPH    & ICAR & 30 & LL    & RS-SPH    & ICAR & 49 & PGW   & RS-SPH    & ICAR & 68 & GAM   & RS-SPH    & ICAR & 87 & GG    & RS-SPH    & ICAR \\
		12 & LN   & RS-SPG    & IID  & 31 & LL    & RS-SPH    & IID  & 50 & PGW   & RS-SPH    & IID  & 69 & GAM   & RS-SPG    & IID  & 88 & GG    & RS-SPG    & IID  \\
		13 & LN   & RS-SAFT   & BYM2 & 32 & LL    & RS-SAFT   & BYM2 & 51 & PGW   & RS-SAFT   & BYM2 & 70 & GAM   & RS-SAFT   & BYM2 & 89 & GG    & RS-SAFT   & BYM2 \\
		14 & LN   & RS-SAFT   & ICAR & 33 & LL    & RS-SAFT   & ICAR & 52 & PGW   & RS-SAFT   & ICAR & 71 & GAM   & RS-SAFT   & ICAR & 90 & GG    & RS-SAFT   & ICAR \\
		15 & LN   & RS-SAFT   & IID  & 34 & LL    & RS-SAFT   & IID  & 53 & PGW   & RS-SAFT   & IID  & 72 & GAM   & RS-SAFT   & IID  & 91 & GG    & RS-SAFT   & IID  \\
		16 & LN   & RS-GH     & ---  & 35 & LL    & RS-GH     & ---  & 54 & PGW   & RS-GH     & ---  & 73 & GAM   & RS-GH     & ---  & 92 & GG    & RS-GH     & ---  \\
		17 & LN   & RS-PH     & ---  & 36 & LL    & RS-PH     & ---  & 55 & PGW   & RS-PH     & ---  & 74 & GAM   & RS-PH     & ---  & 93 & GG    & RS-PH     & ---  \\ 
		18 & LN   & RS-AFT    & ---  & 37 & LL    & RS-AFT    & ---  & 56 & PGW   & RS-AFT    & ---  & 75 & GAM   & RS-AFT    & ---  & 94 & GG    & RS-AFT    & ---  \\
		19 & LN   & RS-AH     & ---  & 38 & LL    & RS-AH     & ---  & 57 & PGW   & RS-AH     & ---  & 76 & GAM   & RS-AH     & ---  & 95 & GG    & RS-AH     & ---  
		\end{tabular}%
	}
	\label{tab:all-models}
\end{table}

The code, available on \href{https://github.com/avramaral/relative_survival}{\url{https://github.com/avramaral/relative\_survival}}, implements all such models, and the fitting procedure, using \texttt{RStan} \citep{rstanpackage} in the background, can be performed as in Web Appendix 2 (Supporting Information). In that section, we provide the code snippet that can be used for fitting Model \eqref{eq:exc-haz} for an example based on observed leukemia-diagnosed patients \citep{henderson:2002}.

\section{Simulation details} \label{appendix:simulation-details}

\begin{figure}[!ht]
	\centering
	\includegraphics[width = 0.4\textwidth]{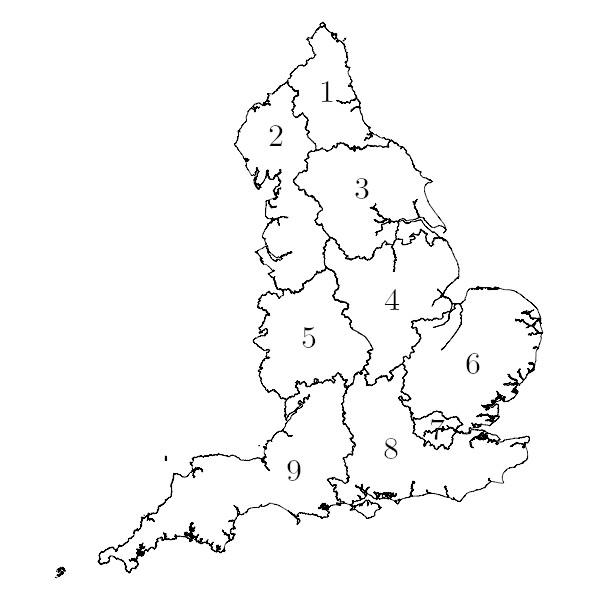}
    \caption{Map of England divided into the 1--9 Government Office Regions, namely, North East, North West, Yorkshire and The Humber, East Midlands, West Midlands, East of England, London, South East, and South West, respectively.}
	\label{fig:map_england}
\end{figure}

The covariates for the simulated data will be based on the lung cancer estimates in London, obtained (and implemented in the \texttt{SimLT} package) by \cite{simlt}. In particular, we will generate synthetic data for $n$ patients---$0.5n$ male and $0.5n$ female patients, such that we will have information about the ``date of diagnosis,'' ``deprivation level'' (1 to 5, where 1 is ``least deprived'' and 5 is ``most deprived''), ``region'' (9 regions of England, as per Figure \ref{fig:map_england}), and ``age.'' Based on it, and given the life tables for England (for the corresponding period), we can simulate the survival times $t_{ij}^{\text{P}}$ associated to the population hazard.

Next, we can simulate the survival times $t_{ij}^{\text{E}}$ associated to the excess hazard with parameters that we detail now. The excess hazard model was defined as follows
\begin{align*}
    h_{\text{E}}(t; \mathbf{x}_{ij} \mid \boldsymbol{\theta}, \alpha, \boldsymbol{\beta}, \tilde{u}_i, u_i) = h_0(t \exp\{\text{age}_{ij} \alpha + \tilde{u}_i\} \mid \boldsymbol{\theta}) \exp\left\{\text{age}_{ij} \beta_1 + \sum_{k = 2}^{5} \mathds{1}_{\text{dep}_{ij}(k)} \beta_k + \text{sex}_{ij} \beta_6 + u_i\right\},
\end{align*}
where $\boldsymbol{\beta} = (\beta_1, \beta_2, \beta_3, \beta_4, \beta_5, \beta_6)^{\top}$ and $\textbf{x}_{ij} = (\text{age}_{ij}, \mathds{1}_{\text{dep}_{ij}(2)}, \mathds{1}_{\text{dep}_{ij}(3)}, \mathds{1}_{\text{dep}_{ij}(4)}, \mathds{1}_{\text{dep}_{ij}(5)}, \text{sex}_{ij})^{\top}$, such that $\mathds{1}_{\text{dep}_{ij}(k)}$, for $2 \leq k \leq 5$, is an indicator function for individuals who belong to the $k$-th deprivation level group (notice that ``deprivation level 1'' is our reference class). For LN baseline hazard distribution, we set the parameters, according to the parameterization in Web Appendix 1 (Supporting Information), as $\mu = 0.65$ and $\sigma = 1.15$; and for the PGW, we set them as $\eta = 0.5$, $\nu = 3.75$, and $\kappa = 8$. The true coefficients were $\alpha = 1.0$ and $\boldsymbol{\beta} = (1.0, -1.0, -1.0, -1.0, -1.0, 2.0)^{\top}$ in all cases. For the spatial effects, we set both $\tilde{\mathbf{u}}$ and $\mathbf{u}$ as following the ICAR model with $\tau_{\tilde{u}} = \tau_{u} = 10$. Finally, for the excess hazard simulated survival times, we apply two sources of censoring: 1) 1.5-year (50\% censoring rate) and 4-year (25\% censoring rate) administrative censoring for all individuals (that corresponds to the end of the study), and 2) a random censoring given by an $\text{Exponential}(\text{rate} = 0.01)$ model (that represents the individuals who, for any reason, dropped the study). The final survival times were set as $t_{ij} = \min(t_{ij}^{\text{P}}, t_{ij}^{\text{E}})$, $\forall i, j$, with the corresponding censoring indicators.

\section{Data description for Applications} \label{appendix:data-applications-details}

For Section \ref{sec:case_study}, we obtained information on all adult (aged 15--99 years) colon cancer patients (International Classification of Diseases for Oncology, third edition, ICD-O-3 codes 18.0--18.9) diagnosed in England between 2015 and 2016, such that we extracted the data from the National Cancer Registration and Analysis Service (NCRAS) data base linked to Hospital Episode Statistics (HES), including basic information on patient, tumour characteristics, and area of residence.
All patients were followed up to update their vital status until 31 December 2018. The data variables available for analysis were sex, age at diagnosis, follow-up time (measured in years from diagnosis), vital status indicator (dead or censored as alive at the end of follow-up), 
Government Office Region (GOR) of residence at diagnosis, Cancer Alliance of residence at diagnosis, deprivation score (based on the Income Domain scores of the 2011 Indices of Multiple Deprivation, IMD), deprivation category (defined according to the quintiles of the IMD Income Domain scores distribution, such that ``1'' is the \textit{least deprived} group and ``2, 3, 4, 5'' are the \textit{most deprived} groups), and colon cancer stage at diagnosis (``1'' being \textit{localised cancer stage}, and ``2, 3, 4'' corresponding to the \textit{metastatic cancer stage}). 

\end{document}


\pagenumbering{gobble}
	
	\begin{center}
		{\Large\textbf{Supporting Information for ``Extended Excess Hazard Model for Spatially Dependent Survival Data''}} \\
		\vspace{24pt}
		{\large Andr\'e Victor Ribeiro Amaral ${}^{\dagger, *}$, Francisco Javier Rubio ${}^{\ddagger}$, Manuela Quaresma ${}^{\mathsection}$, Francisco J. Rodr\'iguez-Cort\'es ${}^{\mathparagraph}$ and Paula Moraga ${}^{\dagger}$}\\
	    \vspace{24pt}
	    ${}^{\dagger}$\hspace{2pt}CEMSE Division, King Abdullah University of Science and Technology. Thuwal, Saudi Arabia. \\
	    ${}^{\ddagger}$\hspace{2pt}Department of Statistical Science, University College London. London, UK. \\
	    ${}^{\mathsection}$\hspace{2pt}Department of Non-Communicable Disease Epidemiology, London School of Hygiene \& Tropical Medicine. London, UK. \\
	    ${}^{\mathparagraph}$\hspace{2pt}Escuela de Estad\'istica, Universidad Nacional de Colombia.
Medell\'in, Colombia. \\
		${}^{*}$ Corresponding author. E-mail: \texttt{\href{mailto:andre.ribeiroamaral@kaust.edu.sa}{andre.ribeiroamaral@kaust.edu.sa}}
		\vspace{18pt}
	\end{center}

\newpage
\pagenumbering{arabic}
\setcounter{footnote}{0} 

\section{Baseline distributions} \label{S-sec:baseline-dist}

Below, we detail the chosen distributions for the baseline function $h_0(\cdot)$. In particular, we define the Log-normal (LN), Log-logistic (LL), Power Generalized Weibull (PGW), Gamma (GAM), and Generalized Gamma (GG) distributions.
For each distribution, the probability density function $f(\cdot)$, hazard function $h(\cdot)$, cumulative hazard function $H(\cdot)$, and survival function $S(\cdot)$ are specified as in the following sections.
    
\subsection*{Log-normal distribution}
\begin{align*}
    f(t\mid\boldsymbol{\theta}) &= \frac{1}{t \sigma \sqrt{2\pi}}\exp\left\{-\frac{(\log(t) - \mu) ^ 2}{2 \sigma^2}\right\}, \text{ for } t > 0, \\
    h(t\mid\boldsymbol{\theta}) &= \frac{\left(\frac{1}{t\sigma}\right)\phi\left(\frac{\log(t)}{\sigma}\right)}{\Phi\left(-\frac{\log{x}}{\sigma}\right)}, \\
    H(t\mid\boldsymbol{\theta}) &= -\log\left(1 - \Phi\left(\frac{\log(t)}{\sigma}\right)\right), \text{ and } \\
    S(t\mid\boldsymbol{\theta}) &= 1 - \Phi\left(\frac{\log(t)}{\sigma}\right),
\end{align*}
where $\boldsymbol{\theta} = (\mu, \sigma^2)$, such that $\mu \in \mathbb{R}$ and $\sigma^2 > 0$, $\phi(\cdot)$ is the probability density function of the standard Normal distribution, and $\Phi(\cdot)$ is the cumulative distribution function of the standard Normal distribution.

\subsection*{Log-logistic distribution}
\begin{align*}
    f(t\mid\boldsymbol{\theta}) &= \frac{g\left(\frac{\log(t) - \mu}{\sigma}\right)}{t\sigma}, \text{ for } t > 0, \\
    h(t\mid\boldsymbol{\theta}) &= \frac{g\left(\frac{\log(t) - \mu}{\sigma}\right)}{t\sigma \left(1 - G\left(\frac{\log(t) - \mu}{\sigma}\right)\right)}, \\
    H(t\mid\boldsymbol{\theta}) &= -\log\left(1 - G\left(\frac{\log(t) - \mu}{\sigma}\right)\right), \text{ and } \\
    S(t\mid\boldsymbol{\theta}) &= 1 - G\left(\frac{\log(t) - \mu}{\sigma}\right),
\end{align*}
where $\boldsymbol{\theta} = (\mu, \sigma)$, such that $\mu \in \mathbb{R}$ and $\sigma > 0$, $g(t) = \exp\{-t\}(1 + \exp\{-t\})^{-2}$, and $G(t) = (1 + \exp\{-t\})^{-1}$.

\subsection*{Power Generalized Weibull distribution}
\begin{align*}
    f(t\mid\boldsymbol{\theta}) &= \frac{\nu}{\kappa \eta^{\nu}} t^{\nu - 1} \left(1 + \left(\frac{t}{\eta}\right)^{\nu}\right)^{\left(\frac{1}{\kappa} - 1\right)} \exp\left\{1 - \left(1 + \left(\frac{t}{\eta}\right)^{\nu}\right)^{\frac{1}{\kappa}}\right\}, \text{ for } t > 0, \\
    h(t\mid\boldsymbol{\theta}) &= \frac{\nu}{\kappa \eta^{\nu}} t^{\nu - 1 } \left(1 + \left(\frac{t}{\eta}\right)^{\nu}\right)^{\left(\frac{1}{\kappa} - 1\right)}, \\
    H(t\mid\boldsymbol{\theta}) &= - 1 + \left(1 + \left(\frac{t}{\eta}\right)^{\nu}\right)^{\frac{1}{\kappa}}, \text{ and } \\
    S(t\mid\boldsymbol{\theta}) &= \exp\left\{1 - \left(1 + \left(\frac{t}{\eta}\right)^{\nu}\right)^{\frac{1}{\kappa}}\right\},
\end{align*}
where $\boldsymbol{\theta} = (\eta, \nu, \kappa)$, such that $\eta > 0$ is a scale parameter and  $\nu, \kappa > 0$ are shape parameters.

\subsection*{Gamma distribution}
\begin{align*}
    f(t\mid\boldsymbol{\theta}) &= \frac{1}{\Gamma(\nu)\eta^{\nu}}t^{\nu -1}\exp\left\{-\frac{t}{\eta}\right\}, \text{ for } t > 0, \\
    h(t\mid\boldsymbol{\theta}) &= \frac{t^{\nu -1} \exp\left\{-\frac{t}{\eta}\right\}}{\eta^{\nu}\left[\Gamma\left(\nu\right) - \gamma\left(\nu, \frac{t}{\eta}\right)\right]} \\
    H(t\mid\boldsymbol{\theta}) &= -\log\left(1 - \frac{\gamma\left(\nu, \frac{t}{\eta}\right)}{\Gamma(\nu)}\right), \text{ and } \\
    S(t\mid\boldsymbol{\theta}) &= 1 - \frac{\gamma\left(\nu, \frac{t}{\eta}\right)}{\Gamma(\nu)},
\end{align*}
where $\boldsymbol{\theta} = (\eta, \nu)$, such that $\eta > 0$ is a scale parameter and  $\nu > 0$ is a shape parameter, $\Gamma(\cdot)$ is gamma function, and $\gamma(\cdot, \cdot)$ is the lower incomplete gamma function.

\subsection*{Generalized Gamma distribution}
\begin{align*}
    f(t\mid\boldsymbol{\theta}) &= \frac{\kappa}{\Gamma\left(\frac{\nu}{\kappa}\right)\eta^{\nu}}t^{\nu - 1} \exp\left\{-\left(\frac{t}{\eta}\right)^{\kappa}\right\}, \text{ for } t > 0, \\
    h(t\mid\boldsymbol{\theta}) &= \frac{t^{\nu -1} \exp\left\{-\left(\frac{t}{\eta}\right)^{\kappa}\right\}}{\eta^{\nu}\left[\Gamma\left(\frac{\nu}{\kappa}\right) - \gamma\left(\frac{\nu}{\kappa}, \left(\frac{t}{\eta}\right)^{\kappa}\right)\right]}, \\
    H(t\mid\boldsymbol{\theta}) &= - \log \left(1 - \frac{\gamma\left(\frac{\nu}{\kappa}, \left(\frac{t}{\eta}\right)^{\kappa}\right)}{\Gamma\left(\frac{\nu}{\kappa}\right)}\right), \text{ and } \\
    S(t\mid\boldsymbol{\theta}) &= 1 - \frac{\gamma\left(\frac{\nu}{\kappa}, \left(\frac{t}{\eta}\right)^{\kappa}\right)}{\Gamma\left(\frac{\nu}{\kappa}\right)},
\end{align*}
where $\boldsymbol{\theta} = (\eta, \nu, \kappa)$, such that $\eta > 0$ is a scale parameter and  $\nu, \kappa > 0$ are shape parameters, $\Gamma(\cdot)$ is gamma function, and $\gamma(\cdot, \cdot)$ is the lower incomplete gamma function. 

From a practical point of view, it might be useful to use that $F(t\mid \eta, \nu, \kappa) = G(t^{\kappa}\mid \eta^{\kappa}, (\nu/\kappa))$, such that $F(\cdot)$ is the cumulative distribution function (CDF) of the Generalized Gamma distribution (as defined above), and $G(t\mid \eta, \nu)$ is the CDF of the Gamma distribution with scale and shape parameters given by $\eta$ and $\nu$, respectively.

\clearpage

\section{Code snippet} \label{S-sec:code-snippet}

As an example, we will fit the RS-SGH model for leukemia-diagnosed patients \citep{henderson:2002}, such that the excess hazard component will be given by the following expression 
\begin{align*}
    h_{\text{E}}(t; \mathbf{x}_{ij} \mid \boldsymbol{\theta}, \alpha, \boldsymbol{\beta}, \tilde{u}_i, u_i) = h_0(t \exp\{\text{age}_{ij} \alpha + \tilde{u}_i\} \mid \boldsymbol{\theta}) \exp\left\{\text{age}_{ij}\beta_1 + \text{wbc}_{ij}\beta_2 + \text{sex}_{ij} \beta_3 + \text{dep}_{ij} \beta_4 + u_i\right\},
\end{align*}
where $\boldsymbol{\theta}$ collects the corresponding distribution parameters, $\boldsymbol{\beta} = (\beta_1, \beta_2, \beta_3, \beta_4)^{\top}$ and $\textbf{x}_{ij} = (\text{age}_{ij}$, $\text{wbc}_{ij}$, $\text{sex}_{ij}$, $\text{dep}_{ij})^{\top}$, such that ``$\text{wbc}$'' stands for ``white blood count,'' and ``dep'' corresponds to the Townsend Score (a index of social deprivation, such that higher values indicates less affluent areas). The baseline hazard $h_0(\cdot | \boldsymbol{\theta})$ will be specified according to a log-normal distribution, and the random effects $\tilde{\mathbf{u}}$ and $\mathbf{u}$ will follow the ICAR model. 

Based on the code from \href{https://github.com/avramaral/relative\_survival}{this repository} (\href{https://github.com/avramaral/relative_survival}{\url{https://github.com/avramaral/relative\_survival}}), we can fit such a model as follows:

\begin{lstlisting}
source("header.R") # load libraries and needed functions

data <- readRDS(file = "DATA/leuk.rds") # load the "leukemia" data

# Optional
data$age <- scale(data$age)
data$wbc <- scale(data$wbc)
data$dep <- scale(data$dep)

map <- readRDS(file = "DATA/nwengland_map.rds") # load the England map
adj_info <- adj_list(map = map) # create an object with information about the neighborhood structure

model <- "LN_ABST" 
dist <- gsub(pattern = "_", replacement = "", x = substring(text = model, first = c(1, 4), last = c(3, 7))[1]) # extract the distribution code from "model"

d <- data_stan(data = data, model = model, cov.tilde = c("age"), cov = c("age", "wbc", "sex", "dep"), nonlinear = c(), adj_info = adj_info) # create the data object
m <- compile_model(model = model) # compile the Stan model
r <- fit_stan(mod = m, data = d) # fit the model
\end{lstlisting}
From the above code, notice that the variable \texttt{model} specifies the fitted model. In particular, we can set it as \texttt{\textcolor{red}{XXX}\textcolor{cyan}{YY}\textcolor{violet}{ZZ}}, where \texttt{\textcolor{red}{XXX}} specifies the model for the baseline term (the options are \texttt{LN\_}, \texttt{LL\_}, \texttt{PGW}, \texttt{GAM}, and \texttt{GG\_}, such that they correspond to the LN, LL, PGW, Gamma, and GG distributions, respectively), \texttt{\textcolor{cyan}{YY}} specifies the structure for the fixed coefficients (these two characters refer to the vector of coefficients in the time-level and hazard-level components, respectively, such that the allowed combinations are listed in the \texttt{models.R} file), and \texttt{\textcolor{violet}{ZZ}} specifies the possibly different random effects structures for $\tilde{\mathbf{u}}$ and $\mathbf{u}$, respectively (the letters \texttt{C} and \texttt{D} refer to the IID model, \texttt{S} and \texttt{T} refer to the ICAR model, and \texttt{Y} and \texttt{Z} refer to the BYM2 model. If one does not want to include the random effects structure in either time- or hazard-level components, they can set the corresponding character as \texttt{X}). Table \ref{tab:S-all-models-code} lists all implemented models (along with the \texttt{model} codes).

\begin{table}[!ht]
	\caption{All implemented models with the corresponding \texttt{model} code. For each of the RS-SGH, RS-SGH-I, RS-SGH-II, RS-SPH, RS-SAFT, RS-GH, RS-PH, RS-AFT, and RS-AH models, we specify the baseline hazard distribution, and the random effects structure.}
	\resizebox{\textwidth}{!}{%
	    \centering
		\begin{tabular}{ c | l | c | c | l | c | c | l | c | c | l | c | c | l | c } 
		\# & Model             & Code             & \# & Model             & Code             & \# & Model              & Code             & \# & Model                & Code             & \# & Model             & Code             \\ \hline
        01 & RS-SGH    LN BYM2 & \texttt{LN\_ABYZ} & 20 & RS-SGH    LL BYM2 & \texttt{LL\_ABYZ} & 39 & RS-SGH    PGW BYM2 & \texttt{PGWABYZ} & 58 & RS-SGH    Gamma BYM2 & \texttt{GAMABYZ} & 77 & RS-SGH    GG BYM2 & \texttt{GG\_ABYZ} \\
        02 & RS-SGH    LN ICAR & \texttt{LN\_ABST} & 21 & RS-SGH    LL ICAR & \texttt{LL\_ABST} & 40 & RS-SGH    PGW ICAR & \texttt{PGWABST} & 59 & RS-SGH    Gamma ICAR & \texttt{GAMABST} & 78 & RS-SGH    GG ICAR & \texttt{GG\_ABST} \\
        03 & RS-SGH    LN IID  & \texttt{LN\_ABCD} & 22 & RS-SGH    LL IID  & \texttt{LL\_ABCD} & 41 & RS-SGH    PGW IID  & \texttt{PGWABCD} & 60 & RS-SGH    Gamma IID  & \texttt{GAMABCD} & 79 & RS-SGH    GG IID  & \texttt{GG\_ABCD} \\
        04 & RS-SGH-I  LN BYM2 & \texttt{LN\_ABXZ} & 23 & RS-SGH-I  LL BYM2 & \texttt{LL\_ABXZ} & 42 & RS-SGH-I  PGW BYM2 & \texttt{PGWABXZ} & 61 & RS-SGH-I  Gamma BYM2 & \texttt{GAMABXZ} & 80 & RS-SGH-I  GG BYM2 & \texttt{GG\_ABXZ} \\
        05 & RS-SGH-I  LN ICAR & \texttt{LN\_ABXT} & 24 & RS-SGH-I  LL ICAR & \texttt{LL\_ABXT} & 43 & RS-SGH-I  PGW ICAR & \texttt{PGWABXT} & 62 & RS-SGH-I  Gamma ICAR & \texttt{GAMABXT} & 81 & RS-SGH-I  GG ICAR & \texttt{GG\_ABXT} \\
        06 & RS-SGH-I  LN IID  & \texttt{LN\_ABXD} & 25 & RS-SGH-I  LL IID  & \texttt{LL\_ABXD} & 44 & RS-SGH-I  PGW IID  & \texttt{PGWABXD} & 63 & RS-SGH-I  Gamma IID  & \texttt{GAMABXD} & 82 & RS-SGH-I  GG IID  & \texttt{GG\_ABXD} \\
        07 & RS-SGH-II LN BYM2 & \texttt{LN\_ABYY} & 26 & RS-SGH-II LL BYM2 & \texttt{LL\_ABYY} & 45 & RS-SGH-II PGW BYM2 & \texttt{PGWABYY} & 64 & RS-SGH-II Gamma BYM2 & \texttt{GAMABYY} & 83 & RS-SGH-II GG BYM2 & \texttt{GG\_ABYY} \\
        08 & RS-SGH-II LN ICAR & \texttt{LN\_ABSS} & 27 & RS-SGH-II LL ICAR & \texttt{LL\_ABSS} & 46 & RS-SGH-II PGW ICAR & \texttt{PGWABSS} & 65 & RS-SGH-II Gamma ICAR & \texttt{GAMABSS} & 84 & RS-SGH-II GG ICAR & \texttt{GG\_ABSS} \\
        09 & RS-SGH-II LN IID  & \texttt{LN\_ABCC} & 28 & RS-SGH-II LL IID  & \texttt{LL\_ABCC} & 47 & RS-SGH-II PGW IID  & \texttt{PGWABCC} & 66 & RS-SGH-II Gamma IID  & \texttt{GAMABCC} & 85 & RS-SGH-II GG IID  & \texttt{GG\_ABCC} \\
        10 & RS-SPH    LN BYM2 & \texttt{LN\_XBXZ} & 29 & RS-SPH    LL BYM2 & \texttt{LL\_XBXZ} & 48 & RS-SPH    PGW BYM2 & \texttt{PGWXBXZ} & 67 & RS-SPH    Gamma BYM2 & \texttt{GAMXBXZ} & 86 & RS-SPH    GG BYM2 & \texttt{GG\_XBXZ} \\
        11 & RS-SPH    LN ICAR & \texttt{LN\_XBXT} & 30 & RS-SPH    LL ICAR & \texttt{LL\_XBXT} & 49 & RS-SPH    PGW ICAR & \texttt{PGWXBXT} & 68 & RS-SPH    Gamma ICAR & \texttt{GAMXBXT} & 87 & RS-SPH    GG ICAR & \texttt{GG\_XBXT} \\
        12 & RS-SPH    LN IID  & \texttt{LN\_XBXD} & 31 & RS-SPH    LL IID  & \texttt{LL\_XBXD} & 50 & RS-SPH    PGW IID  & \texttt{PGWXBXD} & 69 & RS-SPH    Gamma IID  & \texttt{GAMXBXD} & 88 & RS-SPH    GG IID  & \texttt{GG\_XBXD} \\
        13 & RS-SAFT   LN BYM2 & \texttt{LN\_AAYY} & 32 & RS-SAFT   LL BYM2 & \texttt{LL\_AAYY} & 51 & RS-SAFT   PGW BYM2 & \texttt{PGWAAYY} & 70 & RS-SAFT   Gamma BYM2 & \texttt{GAMAAYY} & 89 & RS-SAFT   GG BYM2 & \texttt{GG\_AAYY} \\
        14 & RS-SAFT   LN ICAR & \texttt{LN\_AASS} & 33 & RS-SAFT   LL ICAR & \texttt{LL\_AASS} & 52 & RS-SAFT   PGW ICAR & \texttt{PGWAASS} & 71 & RS-SAFT   Gamma ICAR & \texttt{GAMAASS} & 90 & RS-SAFT   GG ICAR & \texttt{GG\_AASS} \\
        15 & RS-SAFT   LN IID  & \texttt{LN\_AACC} & 34 & RS-SAFT   LL IID  & \texttt{LL\_AACC} & 53 & RS-SAFT   PGW IID  & \texttt{PGWAACC} & 72 & RS-SAFT   Gamma IID  & \texttt{GAMAACC} & 91 & RS-SAFT   GG IID  & \texttt{GG\_AACC} \\
        16 & RS-GH     LN ---  & \texttt{LN\_ABXX} & 35 & RS-GH     LL ---  & \texttt{LL\_ABXX} & 54 & RS-GH     PGW ---  & \texttt{PGWABXX} & 73 & RS-GH     Gamma ---  & \texttt{GAMABXX} & 92 & RS-GH     GG ---  & \texttt{GG\_ABXX} \\
        17 & RS-PH     LN ---  & \texttt{LN\_XBXX} & 36 & RS-PH     LL ---  & \texttt{LL\_XBXX} & 55 & RS-PH     PGW ---  & \texttt{PGWXBXX} & 74 & RS-PH     Gamma ---  & \texttt{GAMXBXX} & 93 & RS-PH     GG ---  & \texttt{GG\_XBXX} \\
        18 & RS-AFT    LN ---  & \texttt{LN\_AAXX} & 37 & RS-AFT    LL ---  & \texttt{LL\_AAXX} & 56 & RS-AFT    PGW ---  & \texttt{PGWAAXX} & 75 & RS-AFT    Gamma ---  & \texttt{GAMAAXX} & 94 & RS-AFT    GG ---  & \texttt{GG\_AAXX} \\
        19 & RS-AH     LN ---  & \texttt{LN\_AXXX} & 38 & RS-AH     LL ---  & \texttt{LL\_AXXX} & 57 & RS-AH     PGW ---  & \texttt{PGWAXXX} & 76 & RS-AH     Gamma ---  & \texttt{GAMAXXX} & 95 & RS-AH     GG ---  & \texttt{GG\_AXXX} \\
		\end{tabular}%
	}
	\label{tab:S-all-models-code}
\end{table}

\clearpage

\section{RS-SGH data simulation} \label{S-sec:RS-SGH_data_simulation}

In this section, we will see how to simulate the survival data from the Relative Survival Spatial General Hazard (RS-SGH) model. To do so, first, we have to simulate the survival times $t_{ij}^{\text{P}}$ associated to the population hazard, and second, we have to simulate the survival times $t_{ij}^{\text{E}}$ that corresponds to the excess hazard model. Then, at the end, all we have to do is setting $t_{ij} = \min\{t_{ij}^{\text{P}}, t_{ij}^{\text{E}}\}$. Also, throughout this section, we will assume known covariates for all individuals and life table for the corresponding region and time-window. To simulate (colon and lung) cancer-related covariates in England, one can refer, for example, to \cite{simlt}.

To simulate $t_{ij}^{\text{P}}$, notice that the problem boils down to simulating from a piecewise constant hazard model, as this is often the information we have available from the life tables. Finally, recall that this is equivalent to simulate from a piecewise exponential (piecewise constant rate) distribution. The \texttt{rsim.pwexp()} function from the \texttt{SimLT} package \citep{simlt} implements such a procedure.

Secondly, we will simulate the survival times associated to the excess hazard model. To do so, we will rely on the Probability Integral Transform (PIT). Here, the idea is simulating $z \sim \text{Uniform}(0, 1)$, and apply the PIT for the corresponding survival model. This means solving Equation \eqref{eq:exc-sim} for $t^{\text{E}}_{ij}$:
\begin{align} \label{eq:exc-sim}
    S_{\text{N}}(t_{ij}^{\text{E}}; \mathbf{x}_{ij} \mid \boldsymbol{\xi} \tilde{u}_i, u_i) = \exp\left\{-H_{\text{E}}(t_{ij}; \mathbf{x}_{ij} \mid \boldsymbol{\xi}, \tilde{u}_i, u_i)\right\} = (1 - z),
\end{align}
where $\boldsymbol{\xi} = (\boldsymbol{\theta}^\top,\boldsymbol{\alpha}^\top,\boldsymbol{\beta}^\top,\boldsymbol{\gamma}^\top)^\top$, and $\boldsymbol{\theta}$ collects the corresponding distribution parameters. To do so, we can proceed as follows,
\begin{align*}
    \exp\left\{-H_{\text{E}}(t_{ij}^{\text{E}}; \mathbf{x}_{ij} \mid \boldsymbol{\xi}, \tilde{u}_i, u_i)\right\} & = (1 - z) \\
    \implies \exp\left\{-H_0(t_{ij}^{\text{E}} \exp\{\tilde{\mathbf{x}}_{ij}^{\top} \boldsymbol{\alpha} + \tilde{u}_i\} \mid \boldsymbol{\theta}) \exp\{\mathbf{x}_{ij}^{\top} \boldsymbol{\beta} - \tilde{\mathbf{x}}_{ij}^{\top} \boldsymbol{\alpha} + u_i - \tilde{u}_i\}\right\} & = (1 - z) \\
    \implies S_0(t_{ij}^{\text{E}} \exp\{\tilde{\mathbf{x}}_{ij}^{\top} \boldsymbol{\alpha} + \tilde{u}_i\}\mid \boldsymbol{\theta})^{\exp\{\mathbf{x}_{ij}^{\top} \boldsymbol{\beta} - \tilde{\mathbf{x}}_{ij}^{\top} \boldsymbol{\alpha} + u_i - \tilde{u}_i\}} & = (1 - z).
\end{align*}
This implies that
\begin{align*}
    \implies 1 - F_0(t_{ij}^{\text{E}} \exp\{\tilde{\mathbf{x}}_{ij}^{\top} \boldsymbol{\alpha} + \tilde{u}_i\}\mid \boldsymbol{\theta}) & = \exp\left\{\frac{\log(1 - z)}{\exp\{\mathbf{x}_{ij}^{\top} \boldsymbol{\beta} - \tilde{\mathbf{x}}_{ij}^{\top} \boldsymbol{\alpha} + u_i - \tilde{u}_i\}}\right\} \\
    \implies 1 - \exp\left\{\frac{\log(1 - z)}{\exp\{\mathbf{x}_{ij}^{\top} \boldsymbol{\beta} - \tilde{\mathbf{x}}_{ij}^{\top} \boldsymbol{\alpha} + u_i - \tilde{u}_i\}}\right\} & = F_0(t_{ij}^{\text{E}} \exp\{\tilde{\mathbf{x}}_{ij}^{\top} \boldsymbol{\alpha} + \tilde{u}_i\}\mid \boldsymbol{\theta}) \\
    \implies F_0^{-1}\left[1 - \exp\left\{\frac{\log(1 - z)}{\exp\{\mathbf{x}_{ij}^{\top} \boldsymbol{\beta} - \tilde{\mathbf{x}}_{ij}^{\top} \boldsymbol{\alpha} + u_i - \tilde{u}_i\}}\right\} \Big| \boldsymbol{\theta}\right] & = t_{ij}^{\text{E}} \exp\{\tilde{\mathbf{x}}_{ij}^{\top} \boldsymbol{\alpha} + \tilde{u}_i\},
\end{align*}
such that $F_0^{-1}(\cdot; \boldsymbol{\theta})$ is the quantile function for the baseline hazard model. Lastly,
\begin{align*}
    t_{ij}^{\text{E}} = \frac{F_0^{-1}\left[1 - \exp\left\{\log(1 - z)  \exp\{\tilde{\mathbf{x}}_{ij}^{\top} \boldsymbol{\alpha} - \mathbf{x}_{ij}^{\top} \boldsymbol{\beta} + \tilde{u}_i -  u_i\}\right\} \mid \boldsymbol{\theta}\right]}{t_{ij} \exp\{\tilde{\mathbf{x}}_{ij}^{\top} \boldsymbol{\alpha} + \tilde{u}_i\}}.
\end{align*}

As mentioned before, once we have simulated $t_{ij}^{\text{P}}$ and $t_{ij}^{\text{E}}$, we simply set $t_{ij} = \min(t_{ij}^{\text{P}}, t_{ij}^{\text{E}})$, $\forall i, j$.

\clearpage

\section{Analysis of marginal quantities (Simulation)} \label{S-sec:supp-simulation-marginal}

\begin{figure}[!ht]
	\centering
	\includegraphics[width = 1\textwidth]{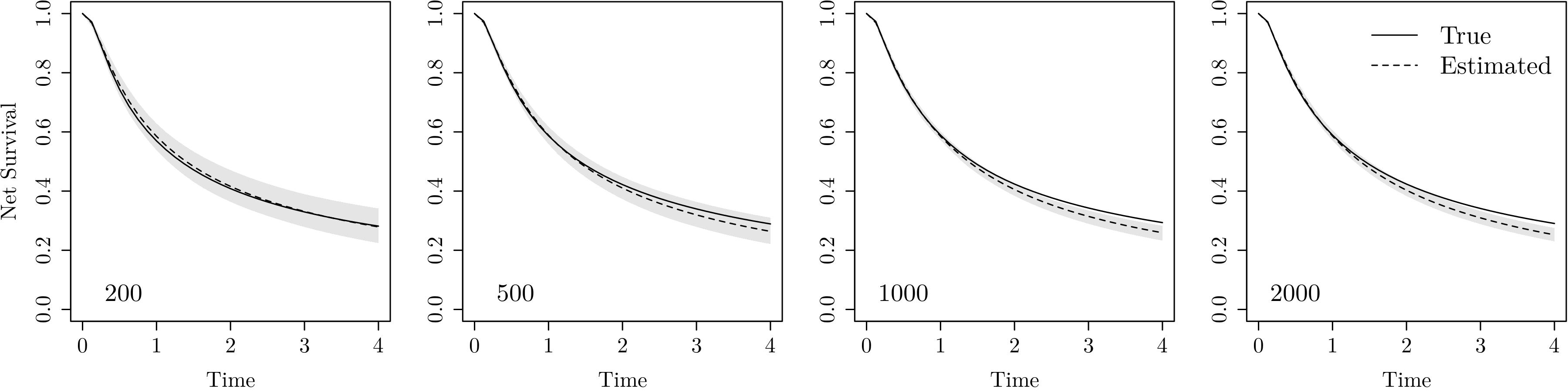}
    \caption{True and estimated (along with a 95\% equal-tailed credible interval) net survival curves based on the fitted RS-SGH LN ICAR model. The data were generated from the RS-SGH PGW ICAR model with 50\% censoring rate and sample size set to $200$, $500$, $1000$, and $2000$ patients. Such estimates were obtained by averaging over the $100$ simulated data sets and all regions for each scenario (the corresponding uncertainty was computed based on the percentiles for the curves that average the regions' net survival).}
	\label{fig:S-third-block-netSur}
\end{figure}

\begin{figure}[!ht]
	\centering
	\includegraphics[width = 0.5\textwidth]{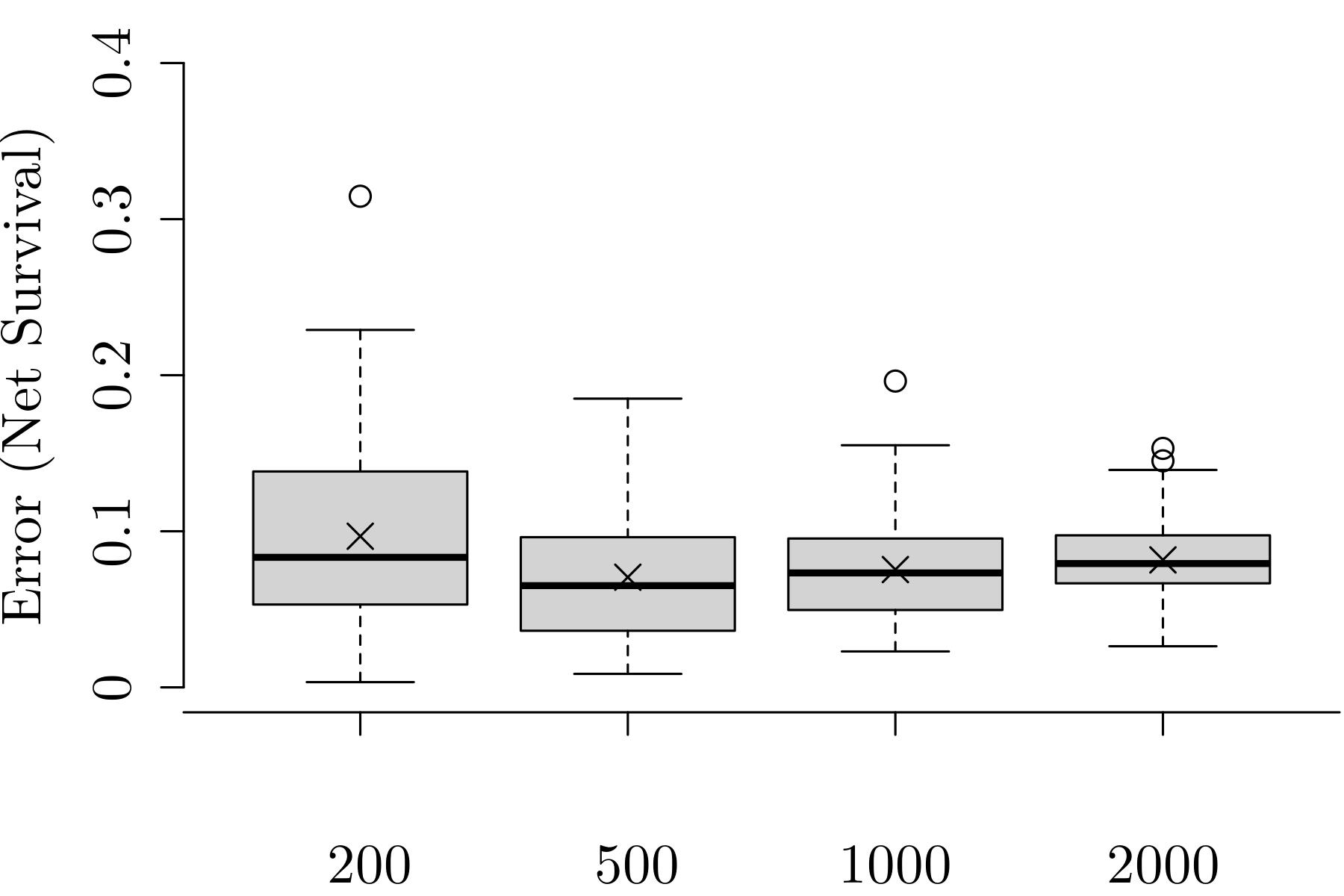}
    \caption{Error in estimating the true net survival function based on the fitted RS-SGH LN ICAR model. The data were generated from the RS-SGH PGW ICAR model with 50\% censoring rate and sample size set to $200$, $500$, $1000$, and $2000$ patients. The computed errors aggregate the $100$ simulated data sets and all regions for each scenario. The crosses ($\times$) correspond to the boxplot values mean.}
	\label{fig:S-third-block-netSur-boxplot}
\end{figure}

\begin{figure}[!ht]
	\centering
	\includegraphics[width = 1\textwidth]{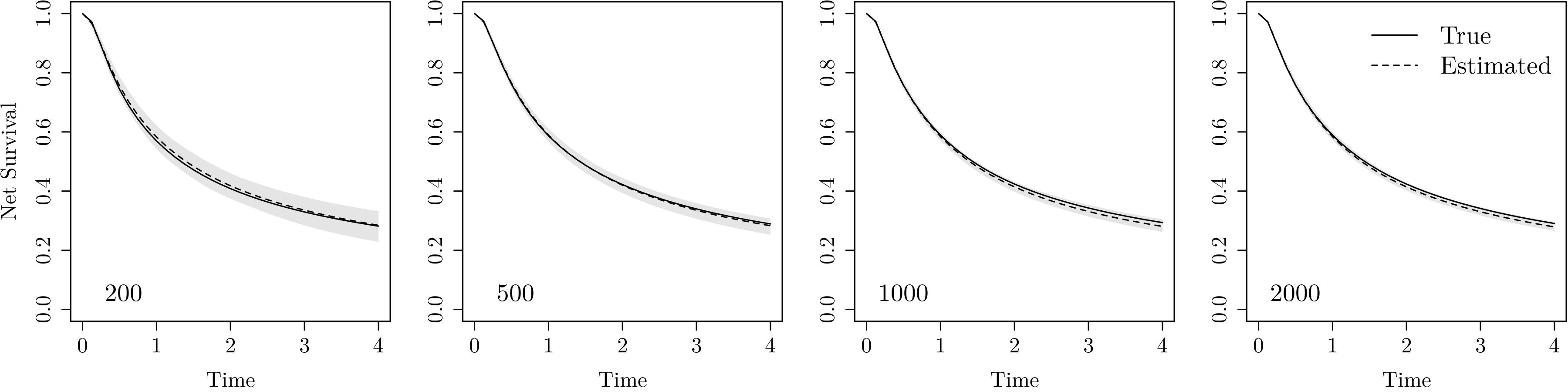}
    \caption{True and estimated (along with a 95\% equal-tailed credible interval) net survival curves based on the fitted RS-SGH PGW ICAR model. The data were generated from the same model with 25\% censoring rate and sample size set to $200$, $500$, $1000$, and $2000$ patients. Such estimates were obtained by averaging over the $100$ simulated data sets and all regions for each scenario (the corresponding uncertainty was computed based on the percentiles for the curves that average the regions' net survival).}
	\label{fig:S-fourth-block-netSur}
\end{figure}

\begin{figure}[!ht]
	\centering
	\includegraphics[width = 0.5\textwidth]{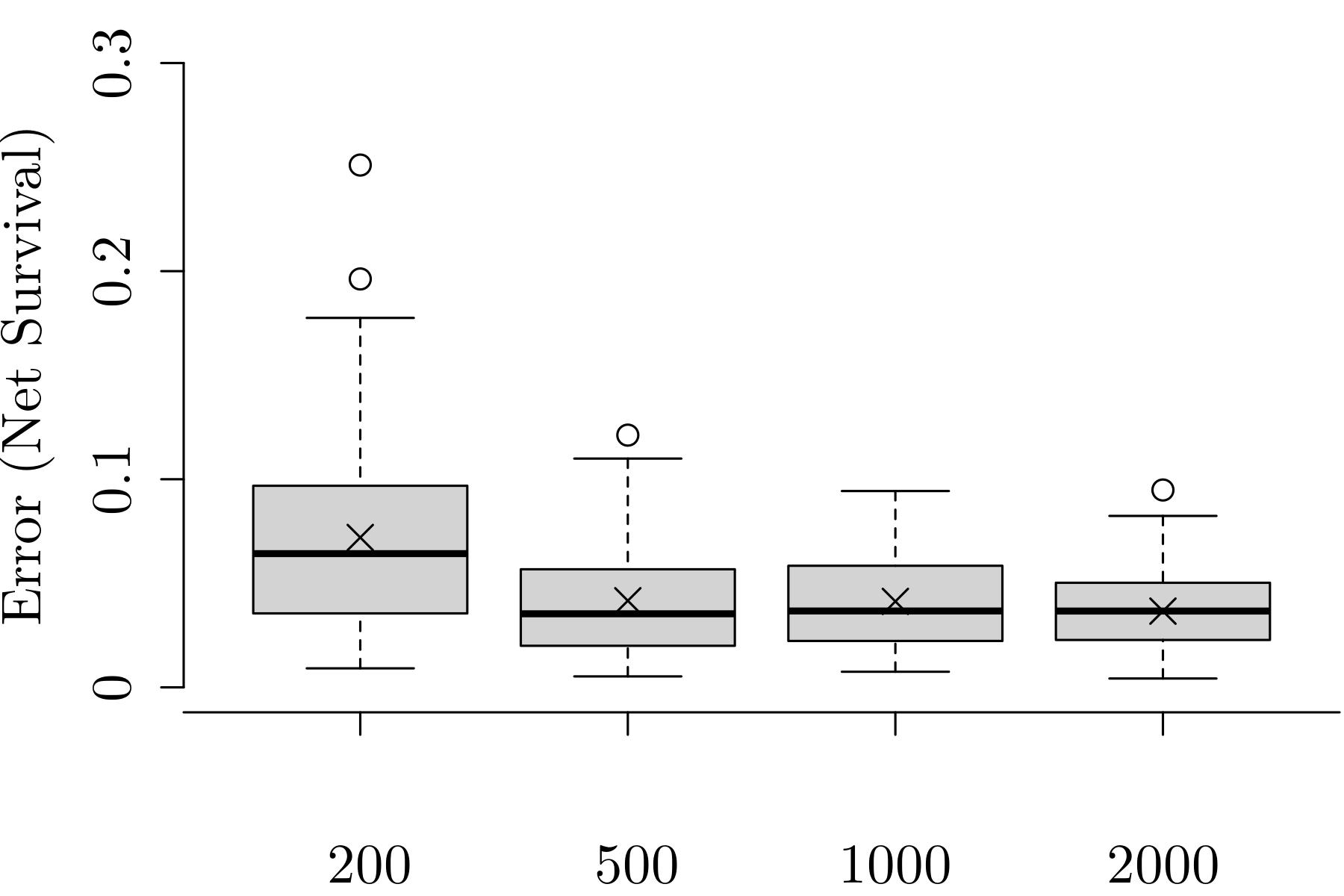}
    \caption{Error in estimating the true net survival function based on the fitted RS-SGH PGW ICAR model. The data were generated from the same model with 25\% censoring rate and sample size set to $200$, $500$, $1000$, and $2000$ patients. The computed errors aggregate the $100$ simulated data sets and all regions for each scenario. The crosses ($\times$) correspond to the boxplot values mean.}
	\label{fig:S-fourth-block-netSur-boxplot}
\end{figure}

This section presents complementary results for Section \textcolor{black}{4.1}. Table \ref{tab:S-simulated_scenarios_time} shows the computational cost for fitting the models. Figures \ref{fig:S-third-block-netSur} and \ref{fig:S-fourth-block-netSur} show the estimated and true net survival curves for different scenarios, and \ref{fig:S-third-block-netSur-boxplot} and \ref{fig:S-fourth-block-netSur-boxplot} plot the corresponding errors based on these simulated data sets.

\begin{table}[!ht]
	\caption{Fitting time for all simulated scenarios for Section  \textcolor{black}{4.1}. ``Fitting time (sec.)'' presents the \textbf{average} time (in seconds) to fit the corresponding model based on the 100 simulated data sets for each scenario. The models were fitted on a Intel--Xeon Gold 6230R CPU at 2.10 Ghz.}
    \resizebox{\textwidth}{!}{%
    \centering
	\begin{tabular}{ c | c | c | c | l | c | c | c | c | c | l | c }
	\# & Data Generating model & Cens. rate & Sample size & Fitted model & Fitting time (sec.) & \# & Data Generating model & Cens. rate & Sample size & Fitted model & Fitting time (sec.) \\ \hline 
	01 & RS-SGH LN ICAR & 25\% & \phantom{1}200 & RS-SGH LN  ICAR & \phantom{1}816.46 & 13 & RS-SGH PGW ICAR & 25\% & \phantom{1}200 & RS-SGH LN  ICAR & \phantom{1}639.69 \\
	02 & RS-SGH LN ICAR & 25\% & \phantom{1}500 & RS-SGH LN  ICAR &           1269.51 & 14 & RS-SGH PGW ICAR & 25\% & \phantom{1}500 & RS-SGH LN  ICAR & \phantom{1}988.41 \\
	03 & RS-SGH LN ICAR & 25\% &           1000 & RS-SGH LN  ICAR &           1873.57 & 15 & RS-SGH PGW ICAR & 25\% &           1000 & RS-SGH LN  ICAR &           1600.08 \\
	04 & RS-SGH LN ICAR & 25\% &           2000 & RS-SGH LN  ICAR &           2612.08 & 16 & RS-SGH PGW ICAR & 25\% &           2000 & RS-SGH LN  ICAR &           2198.06 \\ \hline
	05 & RS-SGH LN ICAR & 25\% & \phantom{1}200 & RS-SGH PGW ICAR &           1139.34 & 17 & RS-SGH PGW ICAR & 25\% & \phantom{1}200 & RS-SGH PGW ICAR & \phantom{1}848.25 \\
	06 & RS-SGH LN ICAR & 25\% & \phantom{1}500 & RS-SGH PGW ICAR &           1849.68 & 18 & RS-SGH PGW ICAR & 25\% & \phantom{1}500 & RS-SGH PGW ICAR &           1271.65 \\
	07 & RS-SGH LN ICAR & 25\% &           1000 & RS-SGH PGW ICAR &           2618.80 & 19 & RS-SGH PGW ICAR & 25\% &           1000 & RS-SGH PGW ICAR &           1989.32 \\
	08 & RS-SGH LN ICAR & 25\% &           2000 & RS-SGH PGW ICAR &           3535.60 & 20 & RS-SGH PGW ICAR & 25\% &           2000 & RS-SGH PGW ICAR &           2374.49 \\ \hline
	09 & RS-SGH LN ICAR & 50\% & \phantom{1}200 & RS-SGH LN  ICAR & \phantom{1}908.27 & 21 & RS-SGH PGW ICAR & 50\% & \phantom{1}200 & RS-SGH LN  ICAR & \phantom{1}674.08 \\
	10 & RS-SGH LN ICAR & 50\% & \phantom{1}500 & RS-SGH LN  ICAR &           1391.55 & 22 & RS-SGH PGW ICAR & 50\% & \phantom{1}500 & RS-SGH LN  ICAR &           1039.07 \\
	11 & RS-SGH LN ICAR & 50\% &           1000 & RS-SGH LN  ICAR &           2016.23 & 23 & RS-SGH PGW ICAR & 50\% &           1000 & RS-SGH LN  ICAR &           1654.65 \\
	12 & RS-SGH LN ICAR & 50\% &           2000 & RS-SGH LN  ICAR &           2730.82 & 24 & RS-SGH PGW ICAR & 50\% & \phantom{1}500 & RS-SGH LN  ICAR &           2314.54 
	\end{tabular}%
    }
	\label{tab:S-simulated_scenarios_time}
\end{table}

\clearpage

\section{Model selection (Simulation)} \label{S-sec:supp-simulation-model-selection}

This section presents complementary material for Section \textcolor{black}{4.2}. Table \ref{tab:S-scenarios-model-selection} lists all scenarios for which we will simulate data and fit the corresponding model. Then, based on such results, we will compare the equivalent scenarios with respect to the estimated $\widehat{\text{elpd}}_{\text{PSIS-LOO}}$. Also, tables \ref{tab:S-winning-model-selection-200}, \ref{tab:S-winning-model-selection-500}, and \ref{tab:S-winning-model-selection-1000} report the ``best-model proportions'' for all scenarios with sample size set to $200$, $500$, and $1000$ patients, respectively.

\begin{table}[!ht]
	\caption{All simulated scenarios for Section \textcolor{black}{4.2}. ``Data Generating model'' refers to the model assumed for the data generating procedure, and ``Fitted model'' defines the baseline hazard distribution and the random effects structure. For all scenarios, the censoring rate is assumed to be 25\%.}
    \resizebox{\textwidth}{!}{%
    \centering
	\begin{tabular}{ c | c | c | l | c | c | c | l }
	\# & Data Generating model & Sample size & Fitted model & \# & Data Generating model & Sample size & Fitted model \\ \hline 
	01 & RS-SGH LN ICAR & \phantom{1}200 & RS-SGH LN  ---  & 33 & RS-SGH PGW ICAR & \phantom{1}200 & RS-SGH LN  ---  \\
	02 & RS-SGH LN ICAR & \phantom{1}500 & RS-SGH LN  ---  & 34 & RS-SGH PGW ICAR & \phantom{1}500 & RS-SGH LN  ---  \\
	03 & RS-SGH LN ICAR &           1000 & RS-SGH LN  ---  & 35 & RS-SGH PGW ICAR &           1000 & RS-SGH LN  ---  \\
	04 & RS-SGH LN ICAR &           2000 & RS-SGH LN  ---  & 36 & RS-SGH PGW ICAR &           2000 & RS-SGH LN  ---  \\
	05 & RS-SGH LN ICAR & \phantom{1}200 & RS-SGH LN IID   & 37 & RS-SGH PGW ICAR & \phantom{1}200 & RS-SGH LN IID   \\
	06 & RS-SGH LN ICAR & \phantom{1}500 & RS-SGH LN IID   & 38 & RS-SGH PGW ICAR & \phantom{1}500 & RS-SGH LN IID   \\
	07 & RS-SGH LN ICAR &           1000 & RS-SGH LN IID   & 39 & RS-SGH PGW ICAR &           1000 & RS-SGH LN IID   \\
	08 & RS-SGH LN ICAR &           2000 & RS-SGH LN IID   & 40 & RS-SGH PGW ICAR &           2000 & RS-SGH LN IID   \\
	09 & RS-SGH LN ICAR & \phantom{1}200 & RS-SGH LN ICAR  & 41 & RS-SGH PGW ICAR & \phantom{1}200 & RS-SGH LN ICAR  \\
	10 & RS-SGH LN ICAR & \phantom{1}500 & RS-SGH LN ICAR  & 42 & RS-SGH PGW ICAR & \phantom{1}500 & RS-SGH LN ICAR  \\
	11 & RS-SGH LN ICAR &           1000 & RS-SGH LN ICAR  & 43 & RS-SGH PGW ICAR &           1000 & RS-SGH LN ICAR  \\
	12 & RS-SGH LN ICAR &           2000 & RS-SGH LN ICAR  & 44 & RS-SGH PGW ICAR & \phantom{1}500 & RS-SGH LN ICAR  \\
 	13 & RS-SGH LN ICAR & \phantom{1}200 & RS-SGH LN BYM2  & 45 & RS-SGH PGW ICAR & \phantom{1}200 & RS-SGH LN BYM2  \\
	14 & RS-SGH LN ICAR & \phantom{1}500 & RS-SGH LN BYM2  & 46 & RS-SGH PGW ICAR & \phantom{1}500 & RS-SGH LN BYM2  \\
	15 & RS-SGH LN ICAR &           1000 & RS-SGH LN BYM2  & 47 & RS-SGH PGW ICAR &           1000 & RS-SGH LN BYM2  \\
	16 & RS-SGH LN ICAR &           2000 & RS-SGH LN BYM2  & 48 & RS-SGH PGW ICAR &           2000 & RS-SGH LN BYM2  \\ \hline
	17 & RS-SGH LN ICAR & \phantom{1}200 & RS-SGH PGW ---  & 49 & RS-SGH PGW ICAR & \phantom{1}200 & RS-SGH PGW ---  \\
	18 & RS-SGH LN ICAR & \phantom{1}500 & RS-SGH PGW ---  & 50 & RS-SGH PGW ICAR & \phantom{1}500 & RS-SGH PGW ---  \\
	19 & RS-SGH LN ICAR &           1000 & RS-SGH PGW ---  & 51 & RS-SGH PGW ICAR &           1000 & RS-SGH PGW ---  \\
	20 & RS-SGH LN ICAR &           2000 & RS-SGH PGW ---  & 52 & RS-SGH PGW ICAR &           2000 & RS-SGH PGW ---  \\
	21 & RS-SGH LN ICAR & \phantom{1}200 & RS-SGH PGW IID  & 53 & RS-SGH PGW ICAR & \phantom{1}200 & RS-SGH PGW IID  \\
	22 & RS-SGH LN ICAR & \phantom{1}500 & RS-SGH PGW IID  & 54 & RS-SGH PGW ICAR & \phantom{1}500 & RS-SGH PGW IID  \\
	23 & RS-SGH LN ICAR &           1000 & RS-SGH PGW IID  & 55 & RS-SGH PGW ICAR &           1000 & RS-SGH PGW IID  \\
	24 & RS-SGH LN ICAR &           2000 & RS-SGH PGW IID  & 56 & RS-SGH PGW ICAR & \phantom{1}500 & RS-SGH PGW IID  \\
 	25 & RS-SGH LN ICAR & \phantom{1}200 & RS-SGH PGW ICAR & 57 & RS-SGH PGW ICAR & \phantom{1}200 & RS-SGH PGW ICAR \\
	26 & RS-SGH LN ICAR & \phantom{1}500 & RS-SGH PGW ICAR & 58 & RS-SGH PGW ICAR & \phantom{1}500 & RS-SGH PGW ICAR \\
	27 & RS-SGH LN ICAR &           1000 & RS-SGH PGW ICAR & 59 & RS-SGH PGW ICAR &           1000 & RS-SGH PGW ICAR \\
	28 & RS-SGH LN ICAR &           2000 & RS-SGH PGW ICAR & 60 & RS-SGH PGW ICAR &           2000 & RS-SGH PGW ICAR \\
	29 & RS-SGH LN ICAR & \phantom{1}200 & RS-SGH PGW BYM2 & 61 & RS-SGH PGW ICAR & \phantom{1}200 & RS-SGH PGW BYM2 \\
	30 & RS-SGH LN ICAR & \phantom{1}500 & RS-SGH PGW BYM2 & 62 & RS-SGH PGW ICAR & \phantom{1}500 & RS-SGH PGW BYM2 \\
	31 & RS-SGH LN ICAR &           1000 & RS-SGH PGW BYM2 & 63 & RS-SGH PGW ICAR &           1000 & RS-SGH PGW BYM2 \\
	32 & RS-SGH LN ICAR &           2000 & RS-SGH PGW BYM2 & 64 & RS-SGH PGW ICAR &           2000 & RS-SGH PGW BYM2 
	\end{tabular}%
    }
	\label{tab:S-scenarios-model-selection}
\end{table}

\begin{table}[!ht]
	\caption{``Best-model proportions'' for model selection based on the estimated $\widehat{\text{elpd}}_{\text{PSIS-LOO}}$. In all scenarios, we assumed a 25\% censoring rate and set the sample size to $200$ patients.}
    \resizebox{\textwidth}{!}{%
    \centering
	\begin{tabular}{ c | c | l | c | c | c | l | c }
	\# & Data Generating model & Fitted model & Best-model proportions & \# & Data Generating model & Fitted model & Best-model proportions \\ \hline 
	01 & RS-SGH LN ICAR & RS-SGH LN  ---  &            30\% & 09 & RS-SGH PGW ICAR & RS-SGH LN  ---  &            26\% \\
	02 & RS-SGH LN ICAR & RS-SGH LN  IID  &  \phantom{0}7\% & 10 & RS-SGH PGW ICAR & RS-SGH LN  IID  &  \phantom{0}8\% \\
	03 & RS-SGH LN ICAR & RS-SGH LN  ICAR &            22\% & 11 & RS-SGH PGW ICAR & RS-SGH LN  ICAR &            24\% \\
	04 & RS-SGH LN ICAR & RS-SGH LN  BYM2 &            41\% & 12 & RS-SGH PGW ICAR & RS-SGH LN  BYM2 &            42\% \\ \hline
	05 & RS-SGH LN ICAR & RS-SGH PGW ---  &            32\% & 13 & RS-SGH PGW ICAR & RS-SGH PGW ---  &            30\% \\
	06 & RS-SGH LN ICAR & RS-SGH PGW IID  &            22\% & 14 & RS-SGH PGW ICAR & RS-SGH PGW IID  &            21\% \\
	07 & RS-SGH LN ICAR & RS-SGH PGW ICAR &            24\% & 15 & RS-SGH PGW ICAR & RS-SGH PGW ICAR &            25\% \\
	08 & RS-SGH LN ICAR & RS-SGH PGW BYM2 &            22\% & 16 & RS-SGH PGW ICAR & RS-SGH PGW BYM2 &            24\% 
	\end{tabular}%
    }
	\label{tab:S-winning-model-selection-200}
\end{table}

\begin{table}[!ht]
	\caption{``Best-model proportions'' for model selection based on the estimated $\widehat{\text{elpd}}_{\text{PSIS-LOO}}$. In all scenarios, we assumed a 25\% censoring rate and set the sample size to $500$ patients.}
    \resizebox{\textwidth}{!}{%
    \centering
	\begin{tabular}{ c | c | l | c | c | c | l | c }
	\# & Data Generating model & Fitted model & Best-model proportions & \# & Data Generating model & Fitted model & Best-model proportions \\ \hline 
	01 & RS-SGH LN ICAR & RS-SGH LN  ---  &            18\% & 09 & RS-SGH PGW ICAR & RS-SGH LN  ---  &            10\% \\
	02 & RS-SGH LN ICAR & RS-SGH LN  IID  &  \phantom{0}3\% & 10 & RS-SGH PGW ICAR & RS-SGH LN  IID  &            20\% \\
	03 & RS-SGH LN ICAR & RS-SGH LN  ICAR &            32\% & 11 & RS-SGH PGW ICAR & RS-SGH LN  ICAR &            23\% \\
	04 & RS-SGH LN ICAR & RS-SGH LN  BYM2 &            47\% & 12 & RS-SGH PGW ICAR & RS-SGH LN  BYM2 &            47\% \\ \hline
	05 & RS-SGH LN ICAR & RS-SGH PGW ---  &            24\% & 13 & RS-SGH PGW ICAR & RS-SGH PGW ---  &            16\% \\
	06 & RS-SGH LN ICAR & RS-SGH PGW IID  &            23\% & 14 & RS-SGH PGW ICAR & RS-SGH PGW IID  &            22\% \\
	07 & RS-SGH LN ICAR & RS-SGH PGW ICAR &            29\% & 15 & RS-SGH PGW ICAR & RS-SGH PGW ICAR &            30\% \\
	08 & RS-SGH LN ICAR & RS-SGH PGW BYM2 &            24\% & 16 & RS-SGH PGW ICAR & RS-SGH PGW BYM2 &            32\% 
	\end{tabular}%
    }
	\label{tab:S-winning-model-selection-500}
\end{table}

\begin{table}[!ht]
	\caption{``Best-model proportions'' for model selection based on the estimated $\widehat{\text{elpd}}_{\text{PSIS-LOO}}$. In all scenarios, we assumed a 25\% censoring rate and set the sample size to $1{,}000$ patients.}
    \resizebox{\textwidth}{!}{%
    \centering
	\begin{tabular}{ c | c | l | c | c | c | l | c }
	\# & Data Generating model & Fitted model & Best-model proportions & \# & Data Generating model & Fitted model & Best-model proportions \\ \hline 
	01 & RS-SGH LN ICAR & RS-SGH LN  ---  &  \phantom{0}4\% & 09 & RS-SGH PGW ICAR & RS-SGH LN  ---  &  \phantom{0}1\% \\
	02 & RS-SGH LN ICAR & RS-SGH LN  IID  &  \phantom{0}8\% & 10 & RS-SGH PGW ICAR & RS-SGH LN  IID  &            35\% \\
	03 & RS-SGH LN ICAR & RS-SGH LN  ICAR &            47\% & 11 & RS-SGH PGW ICAR & RS-SGH LN  ICAR &            18\% \\
	04 & RS-SGH LN ICAR & RS-SGH LN  BYM2 &            41\% & 12 & RS-SGH PGW ICAR & RS-SGH LN  BYM2 &            46\% \\ \hline
	05 & RS-SGH LN ICAR & RS-SGH PGW ---  &            14\% & 13 & RS-SGH PGW ICAR & RS-SGH PGW ---  &  \phantom{0}3\% \\
	06 & RS-SGH LN ICAR & RS-SGH PGW IID  &  \phantom{0}5\% & 14 & RS-SGH PGW ICAR & RS-SGH PGW IID  &            18\% \\
	07 & RS-SGH LN ICAR & RS-SGH PGW ICAR &            36\% & 15 & RS-SGH PGW ICAR & RS-SGH PGW ICAR &            49\% \\
	08 & RS-SGH LN ICAR & RS-SGH PGW BYM2 &            45\% & 16 & RS-SGH PGW ICAR & RS-SGH PGW BYM2 &            30\% 
	\end{tabular}%
    }
	\label{tab:S-winning-model-selection-1000}
\end{table}

\clearpage

\section{Analysis of the spatial effects (Simulation)} \label{S-sec:supp-simulation-spatial-effects}

\begin{figure}[!ht]
	\centering
	\includegraphics[width = 0.825\textwidth]{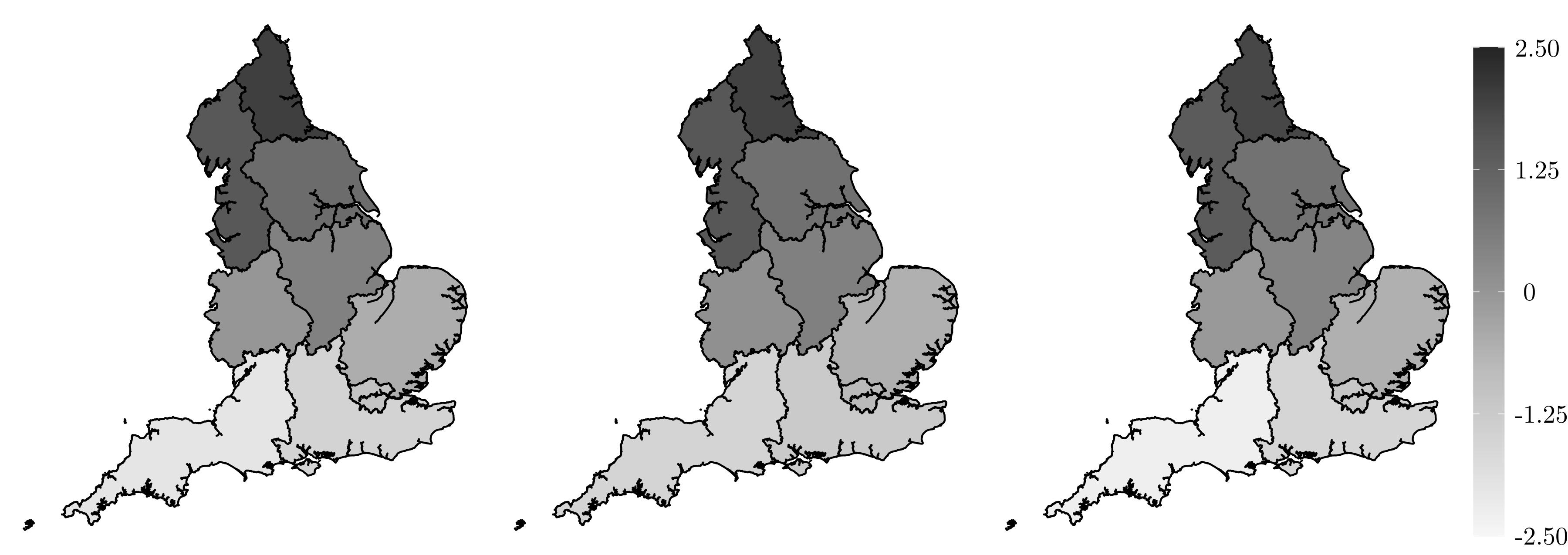}
    \caption{Spatial effects for the RS-SGH LN IID model presented in Section \textcolor{black}{4.3}. {Left panel:} True spatial effects $\tilde{\mathbf{u}} = \mathbf{u} = (2.0$, $1.5$, $1.0$, $0.5$, $0$, $-0.5$, $-1.0$, $-1.5$, $-2.0)^{\top}$. {Middle panel:} Estimated time-level spatial effects $\tilde{\mathbf{u}} = (1.94$, $1.54$, $0.90$, $0.52$, $0.15$, $-0.56$, $-0.78$, $-1.23$, $-1.51)^{\top}$. {Right panel:} Estimated hazard-level spatial effects $\mathbf{u} = (1.86$, $1.44$, $0.86$, $0.43$, $-0.07$, $-0.56$, $-1.12$, $-1.54$, $-2.27)^{\top}$.}
	\label{fig:S-re_LN_ABCD}
\end{figure}

\begin{figure}[!ht]
	\centering
	\includegraphics[width = 0.825\textwidth]{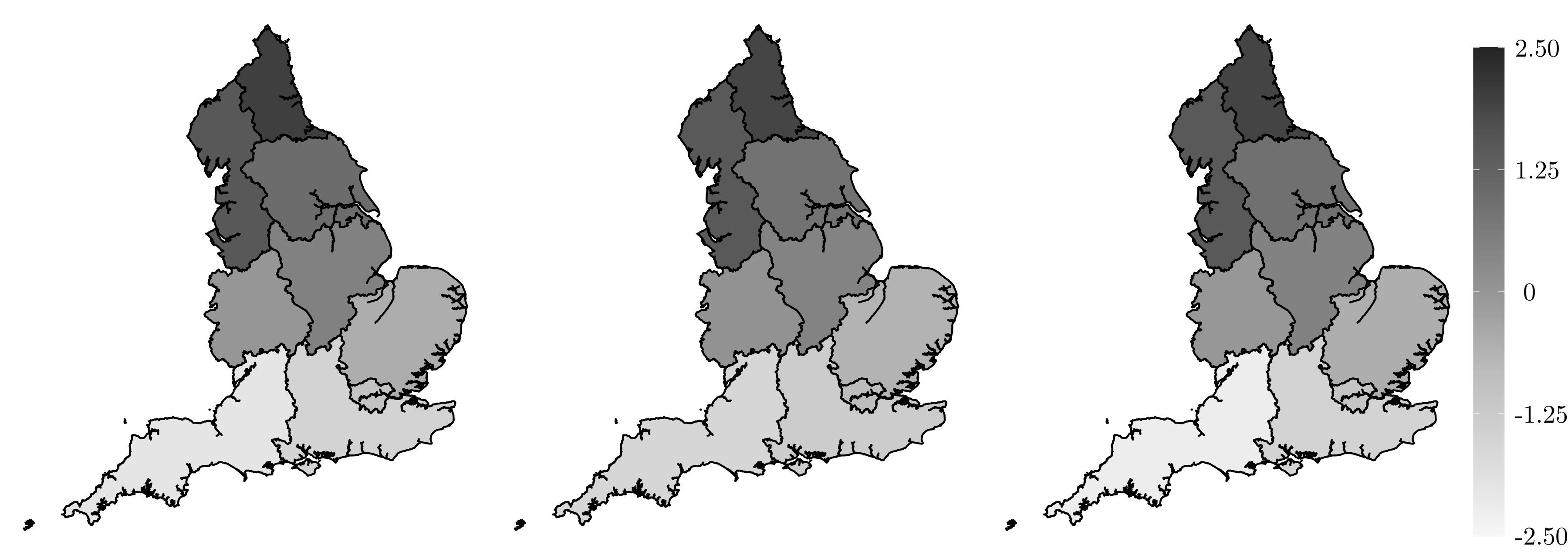}
    \caption{Spatial effects for the RS-SGH LN BYM2 model presented in Section \textcolor{black}{4.3}. {Left panel:} True spatial effects $\tilde{\mathbf{u}} = \mathbf{u} = (2.0$, $1.5$, $1.0$, $0.5$, $0$, $-0.5$, $-1.0$, $-1.5$, $-2.0)^{\top}$. {Middle panel:} Estimated time-level spatial effects $\tilde{\mathbf{u}} = (1.90$, $1.49$, $0.85$, $0.46$, $0.09$, $-0.63$, $-0.85$, $-1.29$, $-1.59)^{\top}$. {Right panel:} Estimated hazard-level spatial effects $\mathbf{u} = (1.91$, $1.49$, $0.91$, $0.48$, $-0.01$, $-0.51$, $-1.07$, $-1.49$, $-2.21)^{\top}$.}
	\label{fig:S-re_LN_ABYZ}
\end{figure}

\begin{figure}[!ht]
	\centering
	\includegraphics[width = 0.55\textwidth]{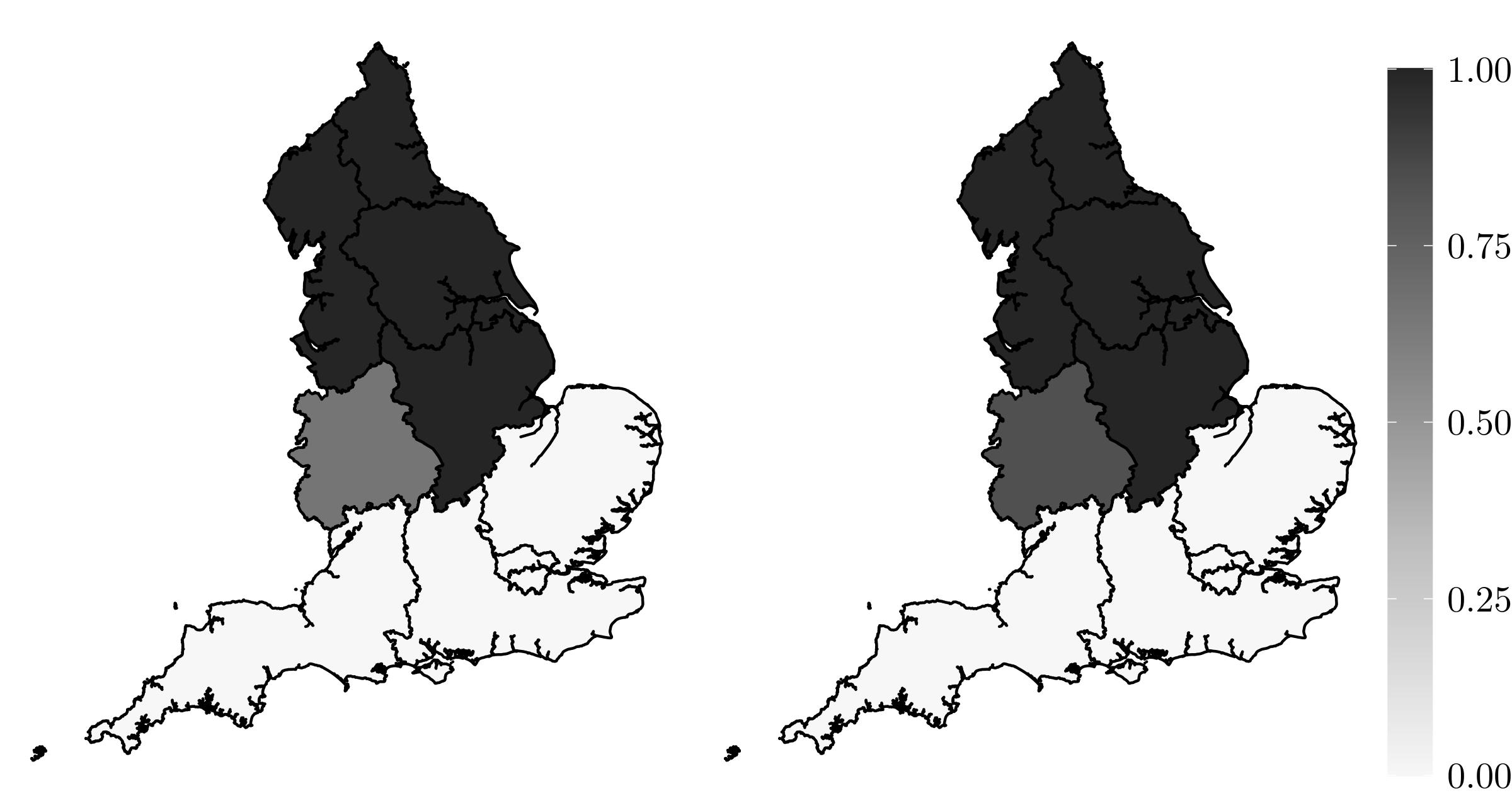}
    \caption{Relative exceedance probability for the RS-SGH LN ICAR model presented in Section \textcolor{black}{4.3}. {Left panel:} Estimated time-level relative exceedance probabilities $f(\tilde{\mathbf{u}}) = (1$, $1$, $1$, $1$, $0.654750$, $0$, $0$, $0$, $0)^{\top}$, such that $f(\tilde{u}_i) = \mathbb{P}(\tilde{u}_i > 0)$, $\forall i$. {Right panel:} Estimated hazard-level relative exceedance probabilities $f(\mathbf{u}) = (1$, $1$, $1$, $1$, $0.841625$, $0$, $0$, $0$, $0)^{\top}$, such that $f(u_i) = \mathbb{P}(u_i > 0)$, $\forall i$.}
	\label{fig:S-exc_prob_LN_ABST}
\end{figure}

\begin{figure}[!ht]
	\centering
	\includegraphics[width = 0.55\textwidth]{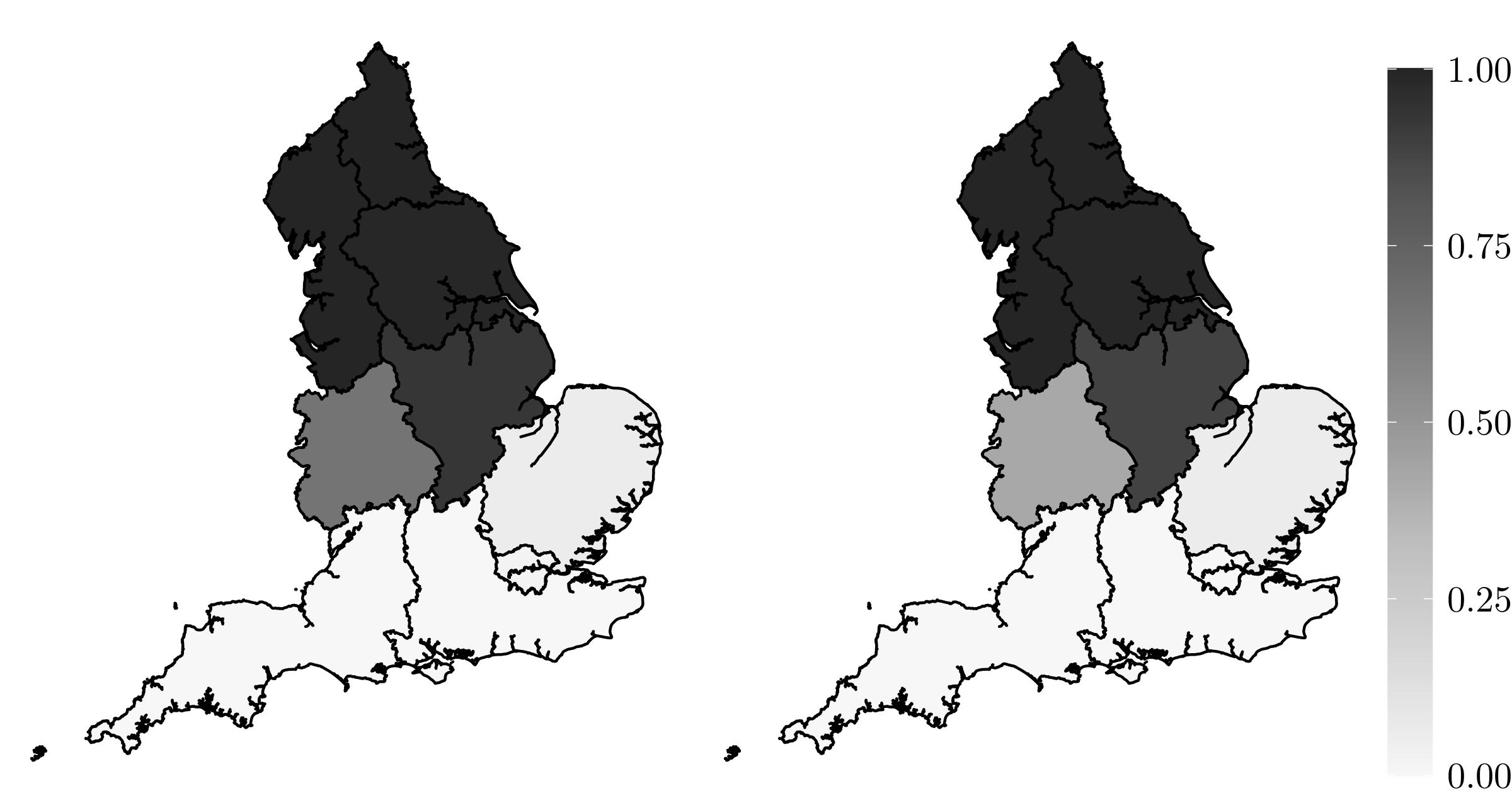}
    \caption{Relative exceedance probability for the RS-SGH LN IID model presented in Section \textcolor{black}{4.3}. {Left panel:} Estimated time-level relative exceedance probabilities $f(\tilde{\mathbf{u}}) = (1$, $1$, $0.992375$, $0.932875$, $0.663250$, $0.061500$, $0.023125$, $0.001375$, $0.000375)^{\top}$, such that $f(\tilde{u}_i) = \mathbb{P}(\tilde{u}_i > 0)$, $\forall i$. {Right panel:} Estimated hazard-level relative exceedance probabilities $f(\mathbf{u}) = (1$, $1$, $0.990875$, $0.889750$, $0.422500$, $0.060500$, $0.001625$, $0$, $0)^{\top}$, such that $f(u_i) = \mathbb{P}(u_i > 0)$, $\forall i$.}
	\label{fig:S-exc_prob_LN_ABCD}
\end{figure}

\begin{figure}[!ht]
	\centering
	\includegraphics[width = 0.55\textwidth]{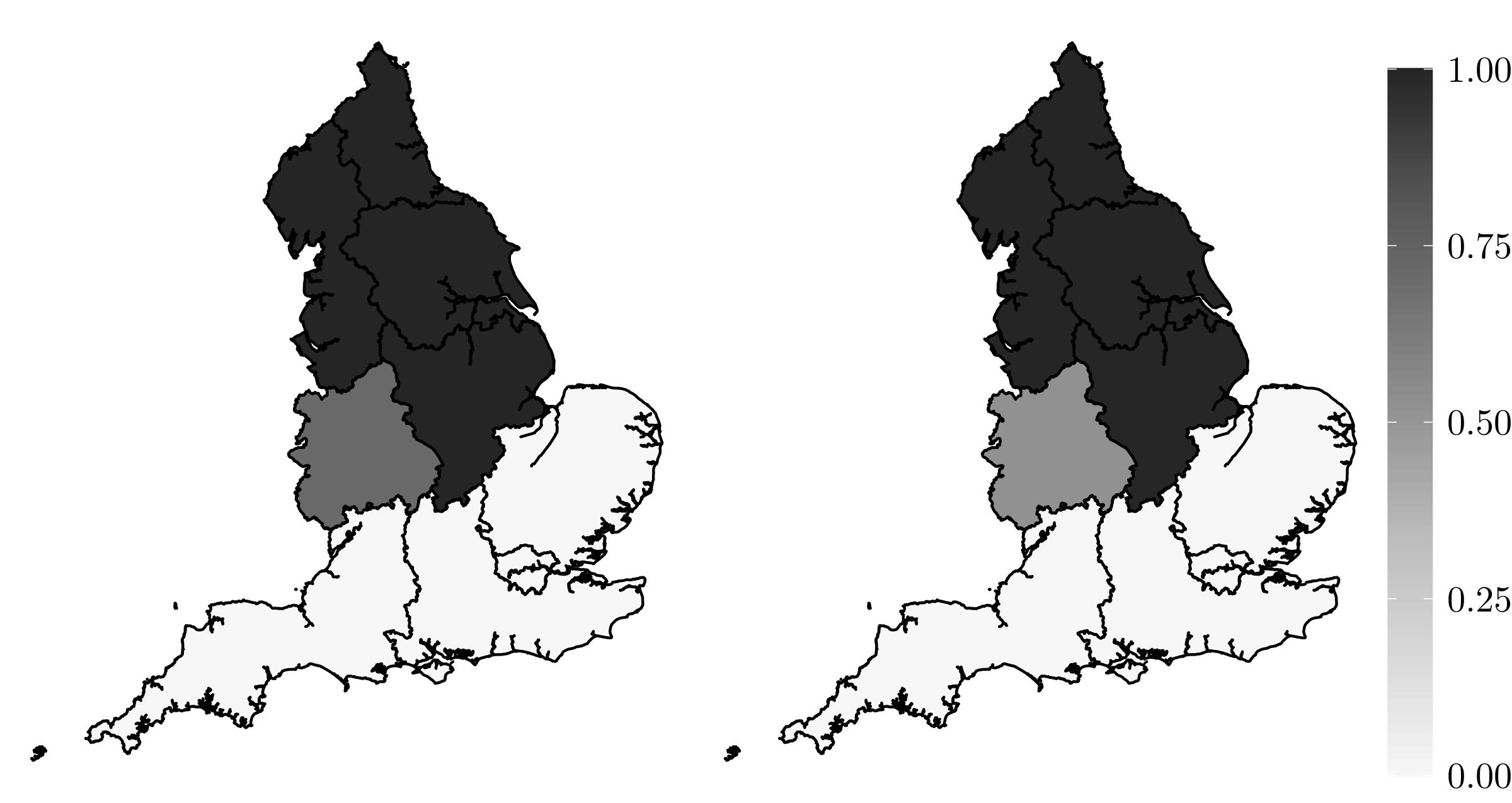}
    \caption{Relative exceedance probability for the RS-SGH LN BYM2 model presented in Section \textcolor{black}{4.3}. {Left panel:} Estimated time-level relative exceedance probabilities $f(\tilde{\mathbf{u}}) = (1$, $1$, $1$, $0.998875$, $0.715125$, $0.001125$, $0.000125$, $0$, $0)^{\top}$, such that $f(\tilde{u}_i) = \mathbb{P}(\tilde{u}_i > 0)$, $\forall i$. {Right panel:} Estimated hazard-level relative exceedance probabilities $f(\mathbf{u}) = (1$, $1$, $1$, $0.997375$, $0.523250$, $0.000125$, $0$, $0$, $0)^{\top}$, such that $f(u_i) = \mathbb{P}(u_i > 0)$, $\forall i$.}
	\label{fig:S-exc_prob_LN_ABYZ}
\end{figure}

\begin{figure}[!ht]
	\centering
	\includegraphics[width = 1\textwidth]{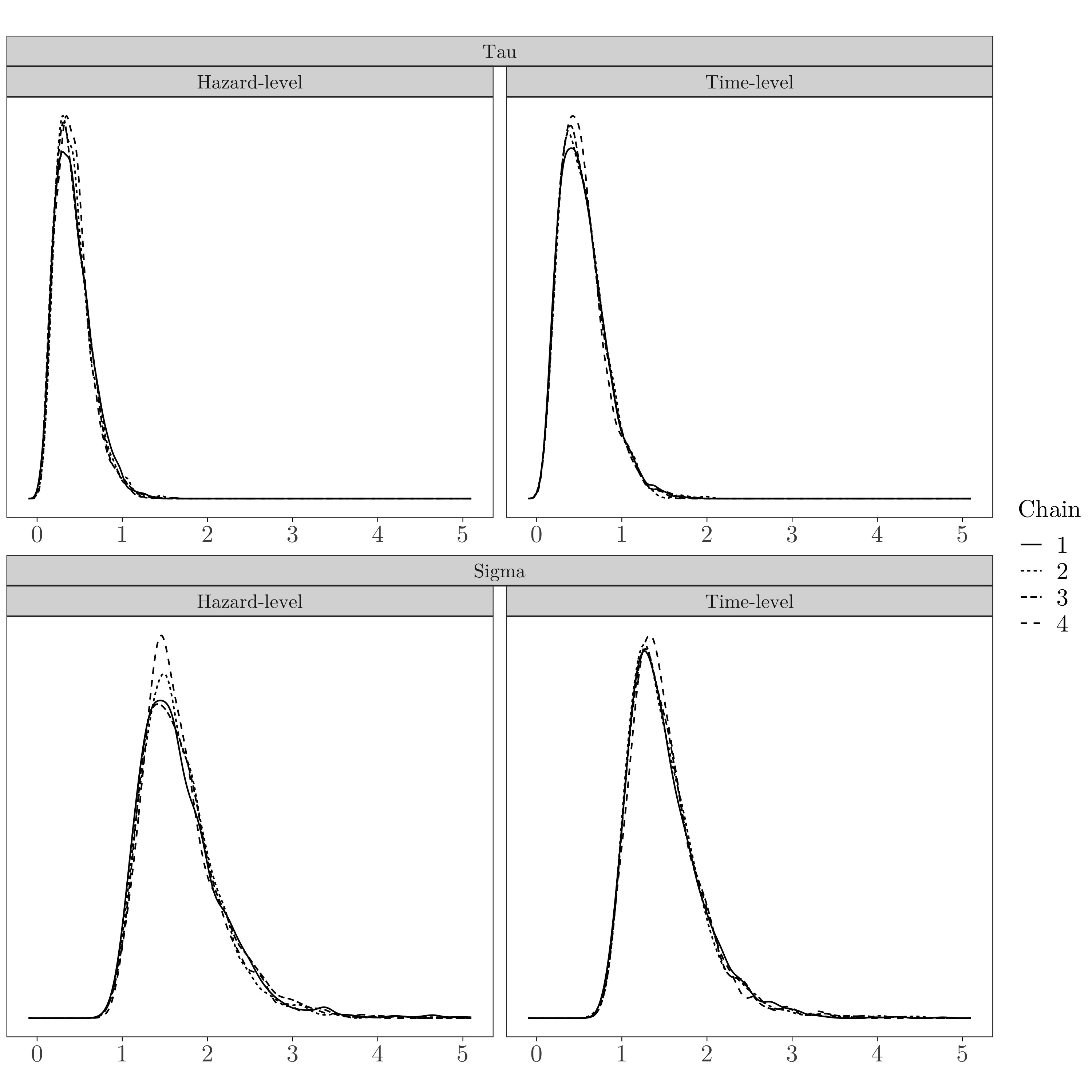}
    \caption{Estimated posterior densities for $\tau_{u}$ (as it appears in the \textbf{hazard-level} component) and $\tau_{\tilde{u}}$ (as it appears in the \textbf{time-level} component) for the ICAR random effects when fitting the RS-SGH LN model. However, aiming to make these results comparable with the estimates showed in Figures \ref{fig:S-posterior_sigma_IID} and \ref{fig:S-posterior_sigma_BYM2} (IID and BYM2 random effects, respectively), we also display the estimated posterior density for $\sigma_{u} = 1 / \sqrt{\tau_{u}}$ and $\sigma_{\tilde{u}} = 1 / \sqrt{\tau_{\tilde{u}}}$. The curves are plotted separately for each posterior chain.}
	\label{fig:S-posterior_sigma_ICAR}
\end{figure}

\begin{figure}[!ht]
	\centering
	\includegraphics[width = 1\textwidth]{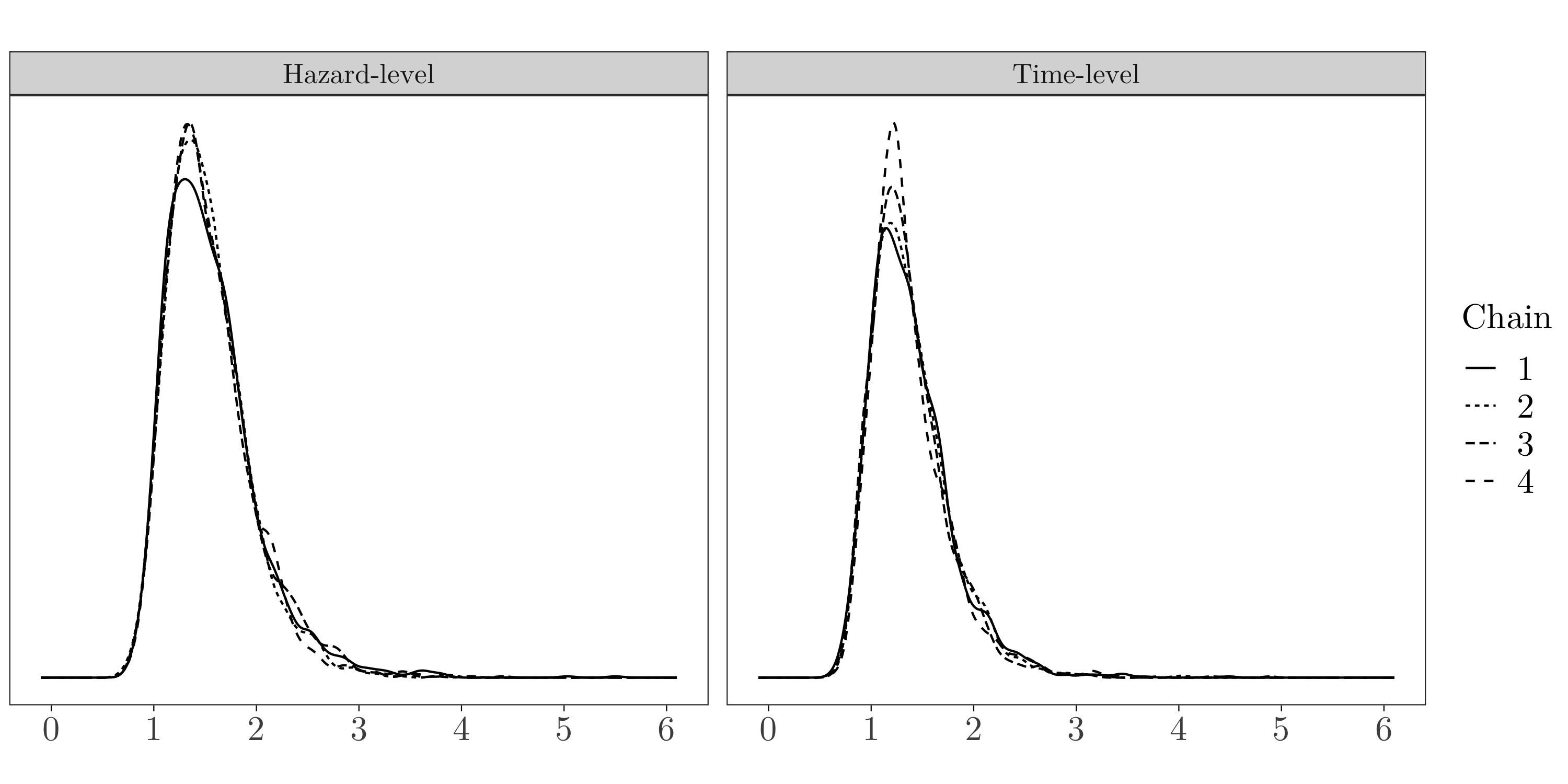}
    \caption{Estimated posterior densities for $\sigma_{u}$ (as it appears in the \textbf{hazard-level} component) and $\sigma_{\tilde{u}}$ (as it appears in the \textbf{time-level} component) for the IID random effects when fitting the RS-SGH LN model. The curves are plotted separately for each posterior chain.}
	\label{fig:S-posterior_sigma_IID}
\end{figure}

\begin{figure}[!ht]
	\centering
	\includegraphics[width = 1\textwidth]{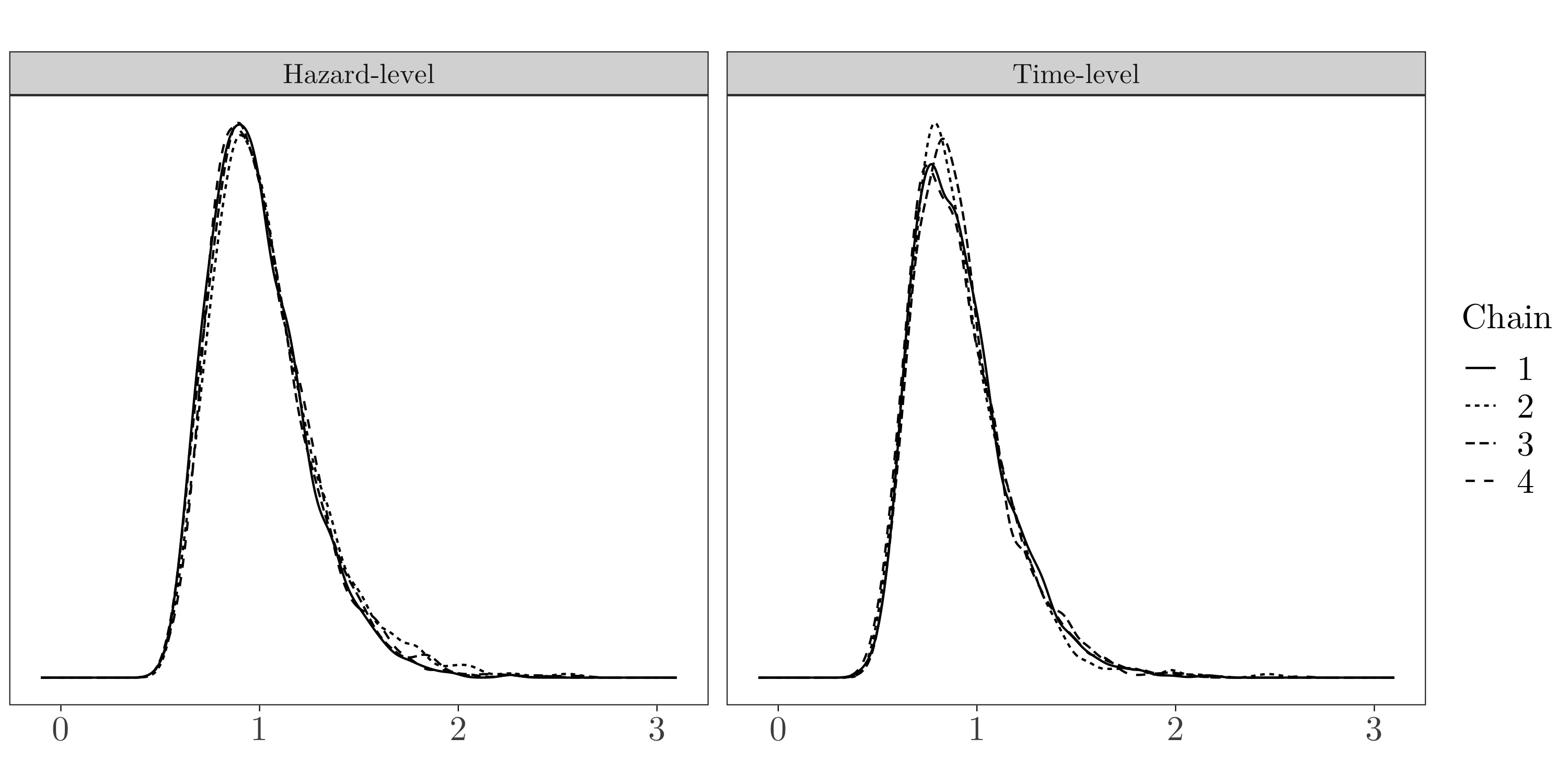}
    \caption{Estimated posterior densities for $\sigma_{u}$ (as it appears in the \textbf{hazard-level} component) and $\sigma_{\tilde{u}}$ (as it appears in the \textbf{time-level} component) for the BYM2 random effects when fitting the RS-SGH LN model. The curves are plotted separately for each posterior chain.}
	\label{fig:S-posterior_sigma_BYM2}
\end{figure}

This section presents complementary results for Section \textcolor{black}{4.3}. In Figures \ref{fig:S-re_LN_ABCD} and \ref{fig:S-re_LN_ABYZ}, we plot the true and estimated spatial effects for models RS-SGH LN IID and RS-SGH LN BYM2, respectively. Similarly, Table \ref{tab:S-estimated-pars-spatial-effects-ALL} show all estimated parameters (with a 95\% equal-tail credible interval) for models RS-SGH LN IID, RS-SGH LN ICAR, and RS-SGH LN BYM2. Moreover, Figures \ref{fig:S-exc_prob_LN_ABST}, \ref{fig:S-exc_prob_LN_ABCD}, and \ref{fig:S-exc_prob_LN_ABYZ} present the estimated relative exceedance probabilities for the same models as before. Lastly, Figures \ref{fig:S-posterior_sigma_ICAR}, \ref{fig:S-posterior_sigma_IID}, and \ref{fig:S-posterior_sigma_BYM2}, show the estimated posterior of $\sigma_{u} = 1 / \sqrt{\tau_{u}}$ (same for $\sigma_{\tilde{u}} = 1 / \sqrt{\tau_{\tilde{u}}}$) for the ICAR, IID, and BYM2 random effects, respectively, when fitting the RS-SGH LN model.

\begin{table}[!ht]
	\caption{Results for the competing models in Section \textcolor{black}{4.3} according to the $\widehat{\text{elpd}}_{\text{PSIS-LOO}}$. The $\widehat{\text{elpd}}_{\text{PSIS-LOO}}$ difference (with standard error) represents the pairwise difference between the others models and the reference model (RS-SGH LN ICAR).}
	\resizebox{\textwidth}{!}{%
	    \centering
		\begin{tabular}{ c | c | c | c | c } 
		Model & RS-SGH LN ICAR & RS-SGH LN IID & RS-SGH LN BYM2 & RS-SGH LN --- \\ \hline
		$\widehat{\text{elpd}}_{\text{PSIS-LOO}}$ difference (SE) & 0.0 (0.0) & -0.1 (1.2) & -0.3 (0.5) & -3187.7 (67.6)
		\end{tabular}%
	}
	\label{tab:S-estimated-elpd}
\end{table}

\begin{table}[!ht]
	\caption{Estimated parameters (with standard deviation and a 95\% equal-tail credible interval) for the RS-SGH LN model with IID, ICAR, and BYM2 random effects, presented in the simulation study (Section \textcolor{black}{4.3}).}
	\resizebox{\textwidth}{!}{%
	    \centering
		\begin{tabular}{ c | c | c | c | c | c | c | c | c } 
		Random Effect & Parameter     & Mean    & SD & 95\% equal-tail CI & Parameter & Mean   & SD & 95\% equal-tail CI \\ \hline\hline
  
		IID~ & $\alpha_1$    & $\phantom{-}0.83$  & 0.03 & ($\phantom{-}0.76$; $\phantom{-}0.90$) & $\beta_5$ & $-0.96$            & 0.04 & ($-1.04$; $-0.87$) \\
		IID~ & $\beta_1$     & $\phantom{-}0.88$  & 0.02 & ($\phantom{-}0.85$; $\phantom{-}0.91$) & $\beta_6$ & $\phantom{-}1.98$  & 0.03 & ($\phantom{-}1.92$; $\phantom{-}2.05$) \\
		IID~ & $\beta_2$     & $-0.90$            & 0.05 & ($-0.93$; $-0.81$)                     & $\mu$     & $\phantom{-}0.55$  & 0.36 & ($-0.14$; $\phantom{-}1.27$) \\
		IID~ & $\beta_3$     & $-0.97$            & 0.05 & ($-1.06$; $-0.88$)                     & $\sigma$  & $\phantom{-}1.15$  & 0.05 & ($\phantom{-}1.06$; $\phantom{-}1.25$) \\
		IID~ & $\beta_4$     & $-0.97$            & 0.04 & ($-1.06$; $-0.88$)                     & ---       & ---                & --- & ---                                    \\
		IID~ & $\tilde{u}_1$ & $\phantom{-}1.94$  & 0.36 & ($\phantom{-}1.26$; $\phantom{-}2.68$) & $u_1$     & $\phantom{-}1.86$  & 0.36 & ($\phantom{-}1.17$; $\phantom{-}2.59$) \\
		IID~ & $\tilde{u}_2$ & $\phantom{-}1.54$  & 0.35 & ($\phantom{-}0.86$; $\phantom{-}2.26$) & $u_2$     & $\phantom{-}1.44$  & 0.36 & ($\phantom{-}0.75$; $\phantom{-}2.17$) \\
		IID~ & $\tilde{u}_3$ & $\phantom{-}0.90$  & 0.36 & ($\phantom{-}0.22$; $\phantom{-}1.63$) & $u_3$     & $\phantom{-}0.86$  & 0.36 & ($\phantom{-}0.17$; $\phantom{-}1.59$) \\
		IID~ & $\tilde{u}_4$ & $\phantom{-}0.52$  & 0.36 & ($-0.17$; $\phantom{-}1.26$)           & $u_4$     & $\phantom{-}0.43$  & 0.36 & ($-0.26$; $\phantom{-}1.16$) \\
		IID~ & $\tilde{u}_5$ & $\phantom{-}0.15$  & 0.36 & ($-0.55$; $\phantom{-}0.89$)           & $u_5$     & $-0.07$            & 0.36 & ($-0.75$; $\phantom{-}0.67$) \\
		IID~ & $\tilde{u}_6$ & $-0.56$            & 0.37 & ($-1.27$; $\phantom{-}0.19$)           & $u_6$     & $-0.56$            & 0.36 & ($-1.26$; $\phantom{-}0.16$) \\
		IID~ & $\tilde{u}_7$ & $-0.78$            & 0.37 & ($-1.50$; $-0.02$)                     & $u_7$     & $-1.12$            & 0.36 & ($-1.81$; $-0.39$) \\
		IID~ & $\tilde{u}_8$ & $-1.23$            & 0.37 & ($-1.93$; $-0.49$)                     & $u_8$     & $-1.54$            & 0.36 & ($-2.24$; $-0.81$) \\
		IID~ & $\tilde{u}_9$ & $-1.51$            & 0.40 & ($-2.31$; $-0.71$)                     & $u_9$     & $-2.27$            & 0.37 & ($-2.97$; $-1.52$) \\ \hline\hline

  		ICAR & $\alpha_1$    & $\phantom{-}0.84$  & 0.03 & ($\phantom{-}0.77$; $\phantom{-}0.91$) & $\beta_5$ & $-0.97$            & 0.04 & ($-1.05$; $-0.88$) \\
		ICAR & $\beta_1$     & $\phantom{-}0.88$  & 0.02 & ($\phantom{-}0.85$; $\phantom{-}0.91$) & $\beta_6$ & $\phantom{-}1.99$  & 0.03 & ($\phantom{-}1.92$; $\phantom{-}2.05$) \\
		ICAR & $\beta_2$     & $-0.91$            & 0.05 & ($-1.00$; $-0.82$)                     & $\mu$     & $\phantom{-}0.65$  & 0.04 & ($\phantom{-}0.57$; $\phantom{-}0.74$) \\
		ICAR & $\beta_3$     & $-0.98$            & 0.04 & ($-1.07$; $-0.89$)                     & $\sigma$  & $\phantom{-}1.21$  & 0.02 & ($\phantom{-}1.18$; $\phantom{-}1.25$) \\
		ICAR & $\beta_4$     & $-0.98$            & 0.04  & ($-1.07$; $-0.89$)                     & ---       & ---                & --- & ---                                    \\
		ICAR & $\tilde{u}_1$ & $\phantom{-}1.86$  & 0.09 & ($\phantom{-}1.69$; $\phantom{-}2.04$) & $u_1$     & $\phantom{-}1.96$  & 0.04 & ($\phantom{-}1.89$; $\phantom{-}2.04$) \\
		ICAR & $\tilde{u}_2$ & $\phantom{-}1.45$  & 0.07 & ($\phantom{-}1.32$; $\phantom{-}1.58$) & $u_2$     & $\phantom{-}1.54$  & 0.03 & ($\phantom{-}1.48$; $\phantom{-}1.60$) \\
		ICAR & $\tilde{u}_3$ & $\phantom{-}0.81$  & 0.08 & ($\phantom{-}0.66$; $\phantom{-}0.97$) & $u_3$     & $\phantom{-}0.97$  & 0.03 & ($\phantom{-}0.90$; $\phantom{-}1.03$) \\
		ICAR & $\tilde{u}_4$ & $\phantom{-}0.41$  & 0.09 & ($\phantom{-}0.23$; $\phantom{-}0.60$) & $u_4$     & $\phantom{-}0.53$  & 0.04 & ($\phantom{-}0.46$; $\phantom{-}0.61$) \\
		ICAR & $\tilde{u}_5$ & $\phantom{-}0.04$  & 0.10 & ($-0.15$; $\phantom{-}0.23$)           & $u_5$     & $\phantom{-}0.04$  & 0.04 & ($-0.04$; $\phantom{-}0.11$) \\
		ICAR & $\tilde{u}_6$ & $-0.69$            & 0.10 & ($-0.88$; $-0.49$)                     & $u_6$     & $-0.45$            & 0.04 & ($-0.53$; $-0.37$) \\
		ICAR & $\tilde{u}_7$ & $-0.90$            & 0.12 & ($-1.14$; $-0.65$)                     & $u_7$     & $-1.01$            & 0.05 & ($-1.10$; $-0.91$) \\
		ICAR & $\tilde{u}_8$ & $-1.33$            & 0.11 & ($-1.55$; $-1.11$)                     & $u_8$     & $-1.43$            & 0.05 & ($-1.52$; $-1.34$) \\
		ICAR & $\tilde{u}_9$ & $-1.65$            & 0.19 & ($-2.01$; $-1.28$)                     & $u_9$     & $-2.15$            & 0.07 & ($-2.29$; $-2.01$) \\ \hline\hline

	      BYM2 & $\alpha_1$    & $\phantom{-}0.83$  & 0.03 & ($\phantom{-}0.77$; $\phantom{-}0.90$) & $\beta_5$ & $-0.96$            & 0.04 & ($-1.04$; $-0.88$) \\
	      BYM2 & $\beta_1$     & $\phantom{-}0.88$  & 0.02 & ($\phantom{-}0.85$; $\phantom{-}0.91$) & $\beta_6$ & $\phantom{-}1.99$  & 0.03 & ($\phantom{-}1.92$; $\phantom{-}2.05$) \\
	      BYM2 & $\beta_2$     & $-0.91$            & 0.05 & ($-1.00$; $-0.82$)                     & $\mu$     & $\phantom{-}0.60$  & 0.12 & ($\phantom{-}0.32$; $\phantom{-}0.81$) \\
	      BYM2 & $\beta_3$     & $-0.97$            & 0.05 & ($-1.06$; $-0.88$)                     & $\sigma$  & $\phantom{-}1.18$  & 0.04 & ($\phantom{-}1.10$; $\phantom{-}1.26$) \\
	      BYM2 & $\beta_4$     & $-0.98$            & 0.04 & ($-1.06$; $-0.89$)                     & ---       & ---                & ---  & ---                                    \\
        BYM2 & $\tilde{\rho}$& $\phantom{-}0.81$  & 0.22 & ($\phantom{-}0.20$; $\phantom{-}1.00$) & $\rho$    & $\phantom{-}0.82$  & 0.21 & ($\phantom{-}0.22$; $\phantom{-}1.00$) \\
	      BYM2 & $\tilde{u}_1$ & $\phantom{-}1.90$  & 0.14 & ($\phantom{-}1.66$; $\phantom{-}2.20$) & $u_1$     & $\phantom{-}1.91$  & 0.12 & ($\phantom{-}1.63$; $\phantom{-}2.12$) \\
	      BYM2 & $\tilde{u}_2$ & $\phantom{-}1.49$  & 0.13 & ($\phantom{-}1.27$; $\phantom{-}1.78$) & $u_2$     & $\phantom{-}1.49$  & 0.12 & ($\phantom{-}1.21$; $\phantom{-}1.70$) \\
	      BYM2 & $\tilde{u}_3$ & $\phantom{-}0.85$  & 0.13 & ($\phantom{-}0.61$; $\phantom{-}1.14$) & $u_3$     & $\phantom{-}0.91$  & 0.12 & ($\phantom{-}0.63$; $\phantom{-}1.12$) \\
	      BYM2 & $\tilde{u}_4$ & $\phantom{-}0.46$  & 0.14 & ($\phantom{-}0.21$; $\phantom{-}0.78$) & $u_4$     & $\phantom{-}0.48$  & 0.12 & ($\phantom{-}0.19$; $\phantom{-}0.69$) \\
	      BYM2 & $\tilde{u}_5$ & $\phantom{-}0.09$  & 0.15 & ($-0.18$; $\phantom{-}0.41$)           & $u_5$     & $\phantom{-}0.01$  & 0.12 & ($-0.30$; $\phantom{-}0.20$)           \\
	      BYM2 & $\tilde{u}_6$ & $-0.63$            & 0.16 &  ($-0.91$; $-0.29$)                    & $u_6$     & $-0.51$            & 0.13 & ($-0.80$; $-0.29$)                     \\
	      BYM2 & $\tilde{u}_7$ & $-0.85$            & 0.17 & ($-1.16$; $-0.48$)                     & $u_7$     & $-1.07$            & 0.13 & ($-1.37$; $-0.85$)                     \\
	      BYM2 & $\tilde{u}_8$ & $-1.29$            & 0.16 & ($-1.57$; $-0.95$)                     & $u_8$     & $-1.49$            & 0.13 & ($-1.79$; $-1.27$)                     \\
	      BYM2 & $\tilde{u}_9$ & $-1.59$            & 0.22 & ($-2.01$; $-1.14$)                     & $u_9$     & $-2.21$            & 0.14 & ($-2.54$; $-1.97$) 
		\end{tabular}%
	}
	\label{tab:S-estimated-pars-spatial-effects-ALL}
\end{table}


\clearpage

\section{Case study} \label{S-sec:supp-applications-cs01}

\begin{figure}[!ht]
	\centering
	\includegraphics[width = 0.525\textwidth]{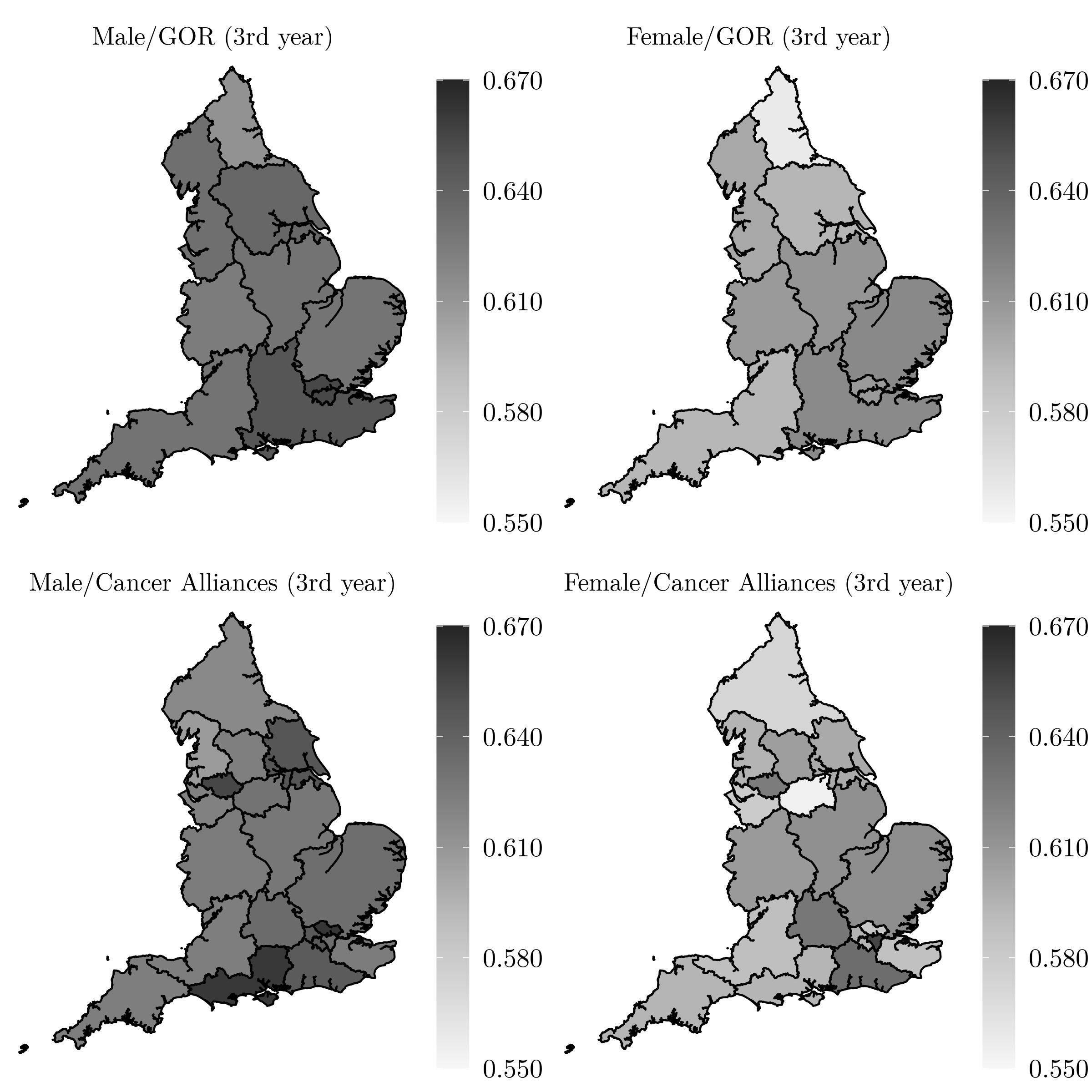}
    \caption{$2.5^{\text{th}}$ net survival percentile for $t = 3$ based on the {(i: top-left panel)} ``Government Office Regions'' (GOR) spatial structure with fitted model RS-SGH LN BYM2 for male patients, {(ii: top-right panel)} ``Government Office Regions'' (GOR) spatial structure with fitted model RS-SGH LN BYM2 for female patients, {(iii: bottom-left panel)} ``Cancer Alliances Regions'' spatial structure with fitted model RS-SGH LN BYM2 for male patients, and {(iv: bottom-right panel)} ``Cancer Alliances Regions'' spatial structure with fitted model RS-SGH LN BYM2 for female patients.}
	\label{fig:S-cs01-t3-0025}
\end{figure}

\begin{figure}[!ht]
	\centering
	\includegraphics[width = 0.525\textwidth]{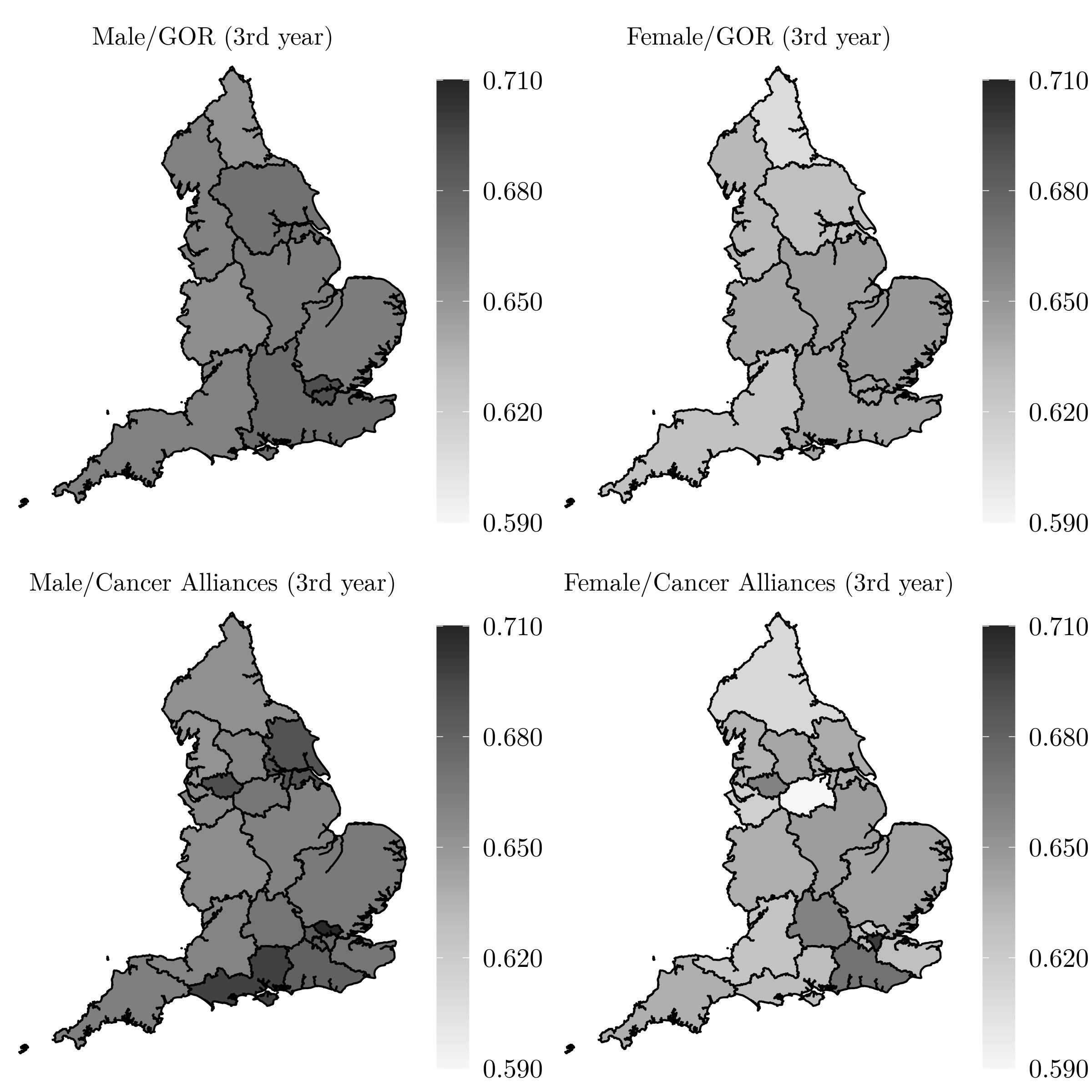}
    \caption{$97.5^{\text{th}}$ net survival percentile for $t = 3$ based on the {(i: top-left panel)} ``Government Office Regions'' (GOR) spatial structure with fitted model RS-SGH LN BYM2 for male patients, {(ii: top-right panel)} ``Government Office Regions'' (GOR) spatial structure with fitted model RS-SGH LN BYM2 for female patients, {(iii: bottom-left panel)} ``Cancer Alliances Regions'' spatial structure with fitted model RS-SGH LN BYM2 for male patients, and {(iv: bottom-right panel)} ``Cancer Alliances Regions'' spatial structure with fitted model RS-SGH LN BYM2 for female patients.}
	\label{fig:S-cs01-t3-0975}
\end{figure}

\begin{figure}[!ht]
	\centering
	\includegraphics[width = 0.525\textwidth]{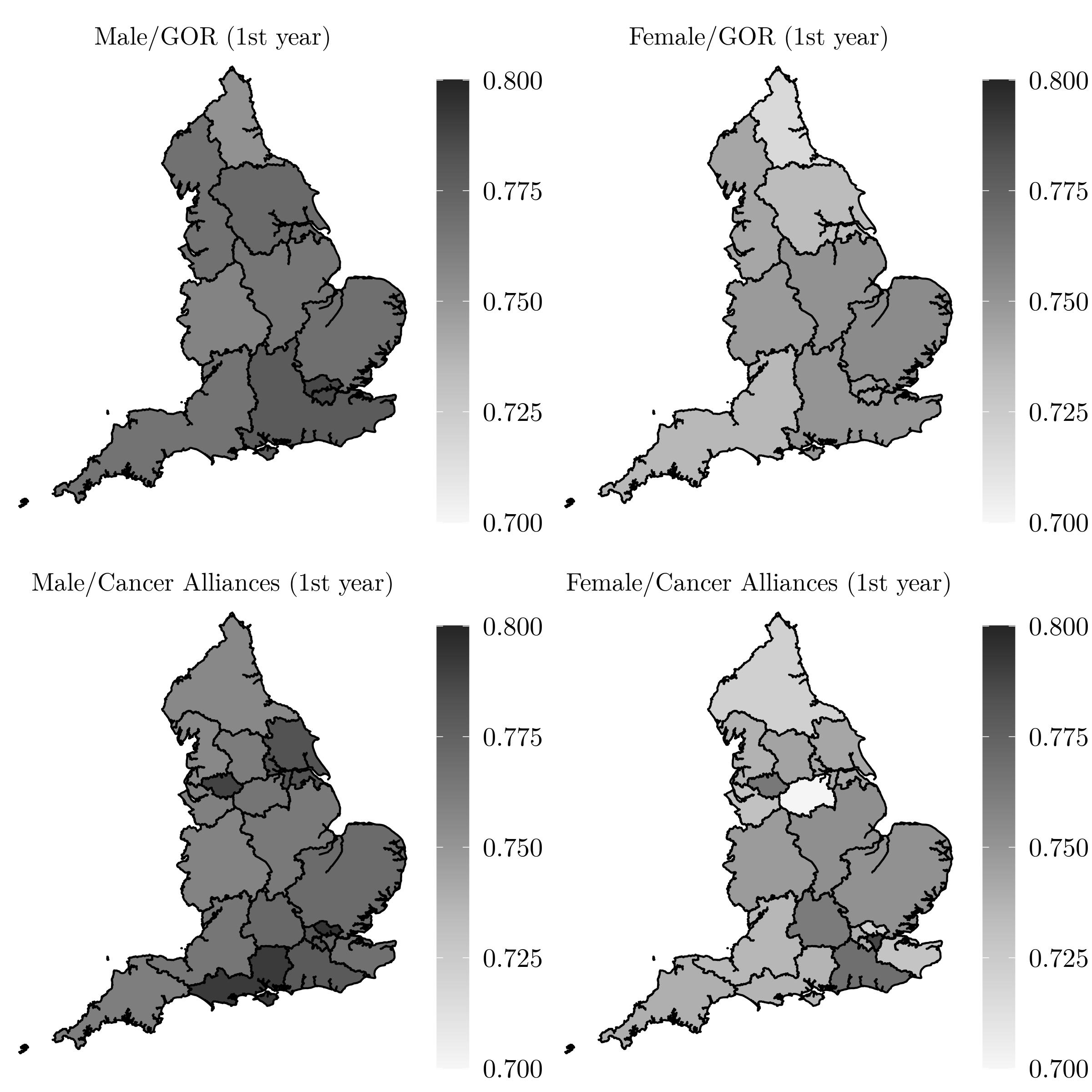}
    \caption{Net survival point estimate for $t = 1$ based on the {(i: top-left panel)} ``Government Office Regions'' (GOR) spatial structure with fitted model RS-SGH LN BYM2 for male patients, {(ii: top-right panel)} ``Government Office Regions'' (GOR) spatial structure with fitted model RS-SGH LN BYM2 for female patients, {(iii: bottom-left panel)} ``Cancer Alliances Regions'' spatial structure with fitted model RS-SGH LN BYM2 for male patients, and {(iv: bottom-right panel)} ``Cancer Alliances Regions'' spatial structure with fitted model RS-SGH LN BYM2 for female patients.}
	\label{fig:S-cs01-t1-mean}
\end{figure}

\begin{figure}[!ht]
	\centering
	\includegraphics[width = 0.525\textwidth]{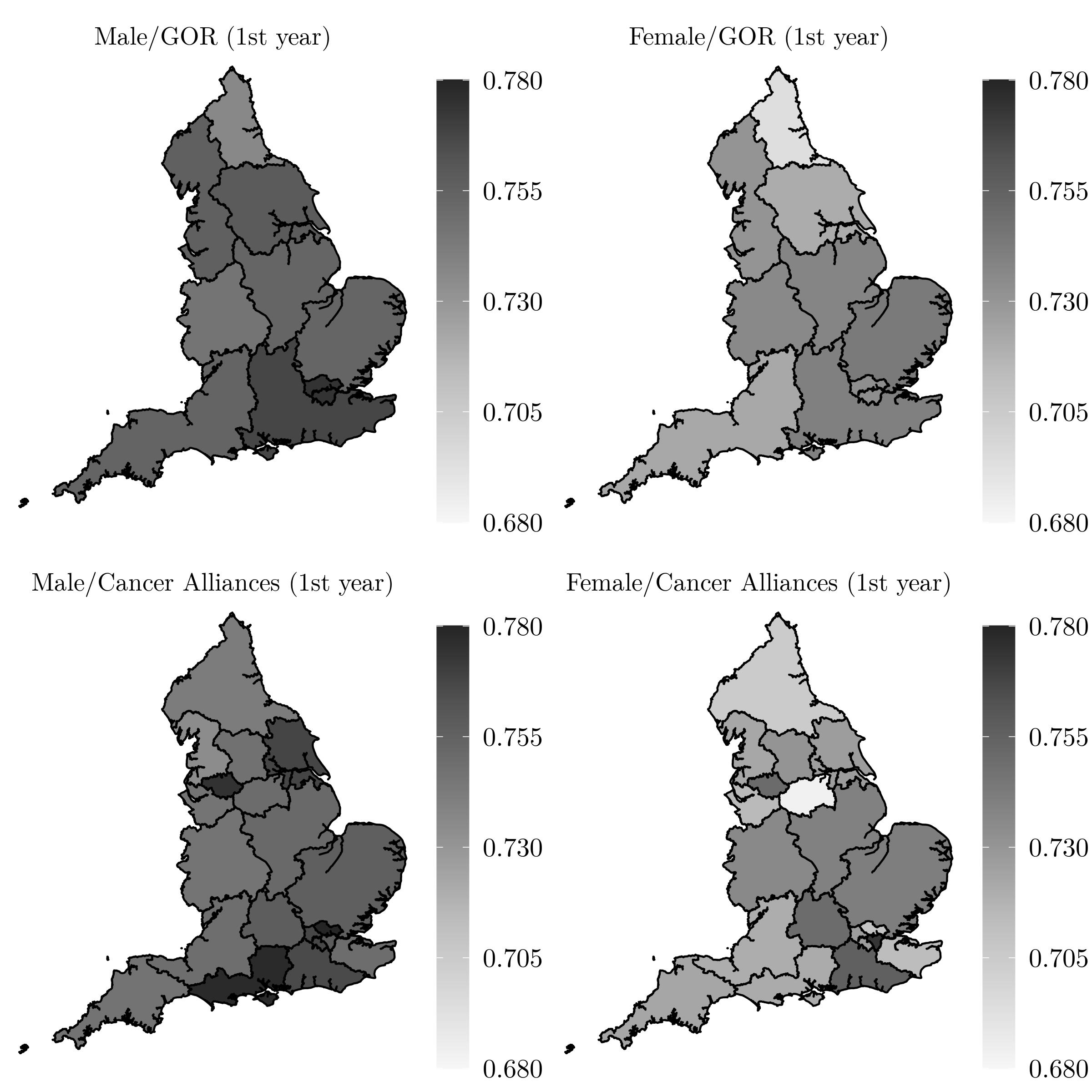}
    \caption{$2.5^{\text{th}}$ net survival percentile for $t = 1$ based on the {(i: top-left panel)} ``Government Office Regions'' (GOR) spatial structure with fitted model RS-SGH LN BYM2 for male patients, {(ii: top-right panel)} ``Government Office Regions'' (GOR) spatial structure with fitted model RS-SGH LN BYM2 for female patients, {(iii: bottom-left panel)} ``Cancer Alliances Regions'' spatial structure with fitted model RS-SGH LN BYM2 for male patients, and {(iv: bottom-right panel)} ``Cancer Alliances Regions'' spatial structure with fitted model RS-SGH LN BYM2 for female patients.}
	\label{fig:S-cs01-t1-0025}
\end{figure}

\begin{figure}[!ht]
	\centering
	\includegraphics[width = 0.525\textwidth]{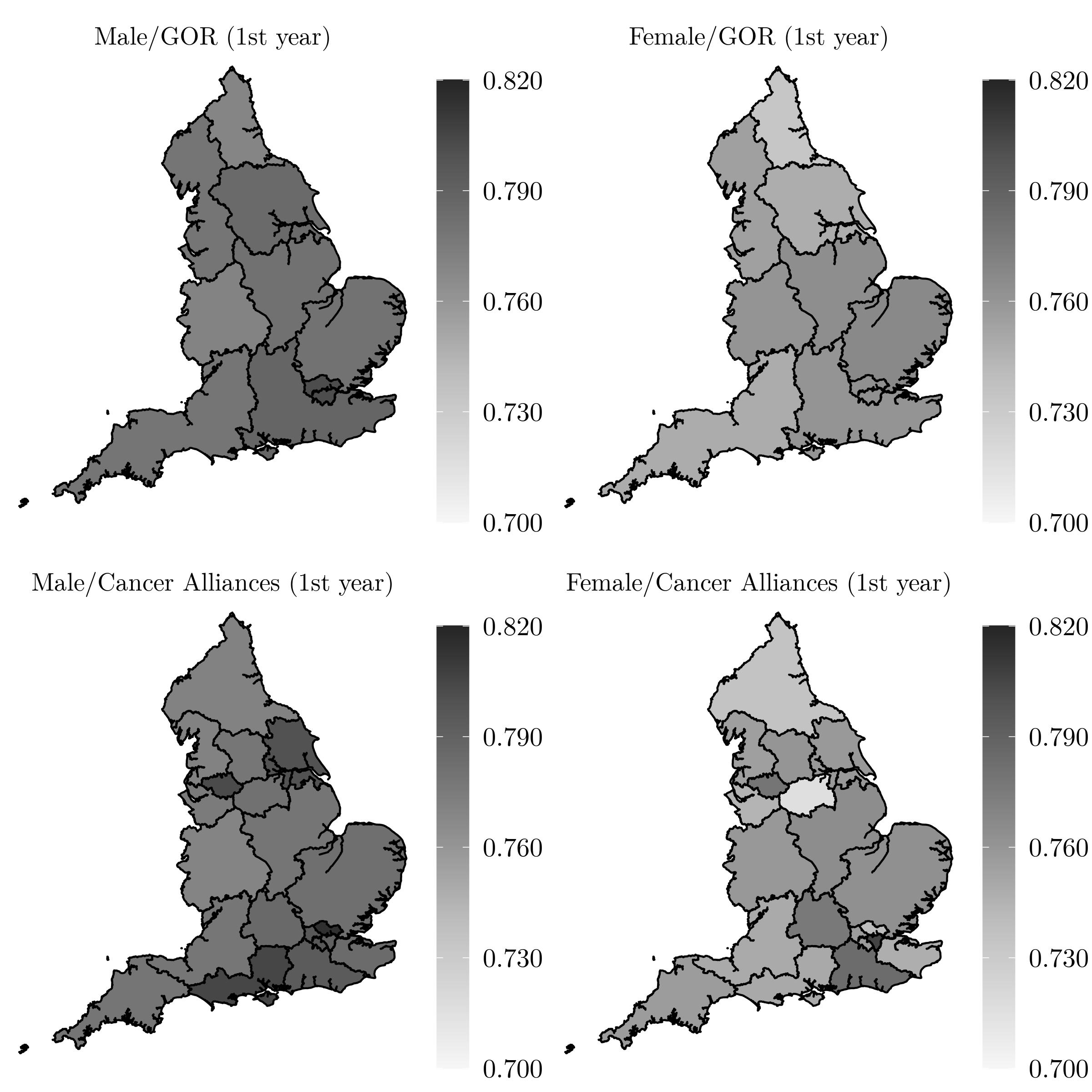}
    \caption{$97.5^{\text{th}}$ net survival percentile for $t = 1$ based on the {(i: top-left panel)} ``Government Office Regions'' (GOR) spatial structure with fitted model RS-SGH LN BYM2 for male patients, {(ii: top-right panel)} ``Government Office Regions'' (GOR) spatial structure with fitted model RS-SGH LN BYM2 for female patients, {(iii: bottom-left panel)} ``Cancer Alliances Regions'' spatial structure with fitted model RS-SGH LN BYM2 for male patients, and {(iv: bottom-right panel)} ``Cancer Alliances Regions'' spatial structure with fitted model RS-SGH LN BYM2 for female patients.}
	\label{fig:S-cs01-t1-0975}
\end{figure}

\begin{figure}[!ht]
	\centering
	\includegraphics[width = 0.5\textwidth]{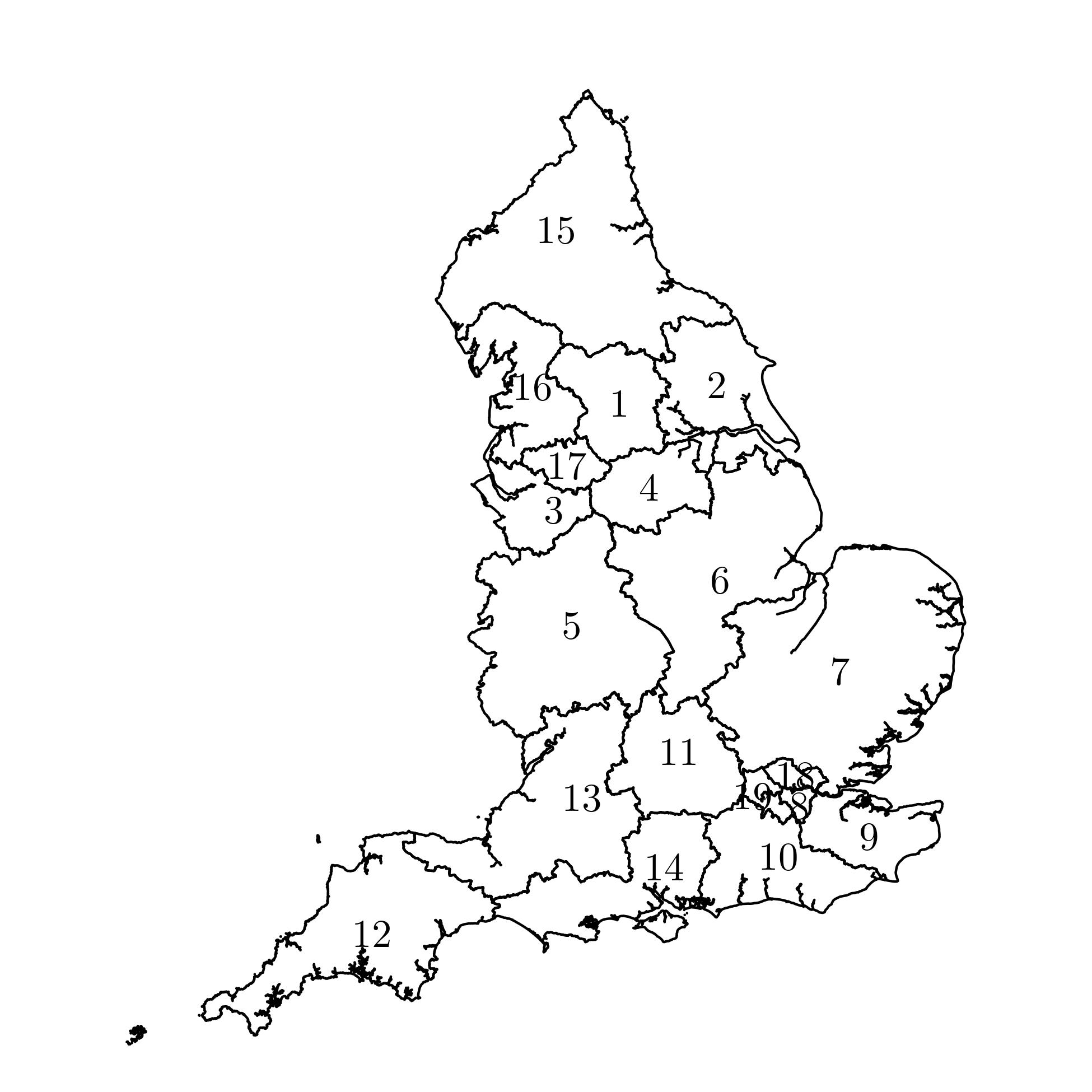}
    \caption{Map of England divided into the 1--19 Cancer Alliances Regions, namely, West Yorkshire, ``Humber, Coast and Vale,'' ``Cheshire and Merseyside,'' ``South Yorkshire, Bassetlaw, North Derbyshire and Hardwick,'' West Midlands, East Midlands, East of England, South East London, ``Kent and Medway,'' ``Surrey and Sussex,'' Thames Valley, Peninsula, ``Somerset, Wiltshire, Avon and Gloucestershire,'' Wessex, ``North East and Cumbria,'' ``Lancashire and South Cumbria,'' ``National Cancer Vanguard: Greater Manchester,''  ``National Cancer Vanguard: North Central and North East London,'' and ``National Cancer Vanguard: North West and South West London,'' respectively.}
	\label{fig:S-map_england_cancer_alliance}
\end{figure}

\begin{figure}[!ht]
	\centering
	\includegraphics[width = 0.60\textwidth]{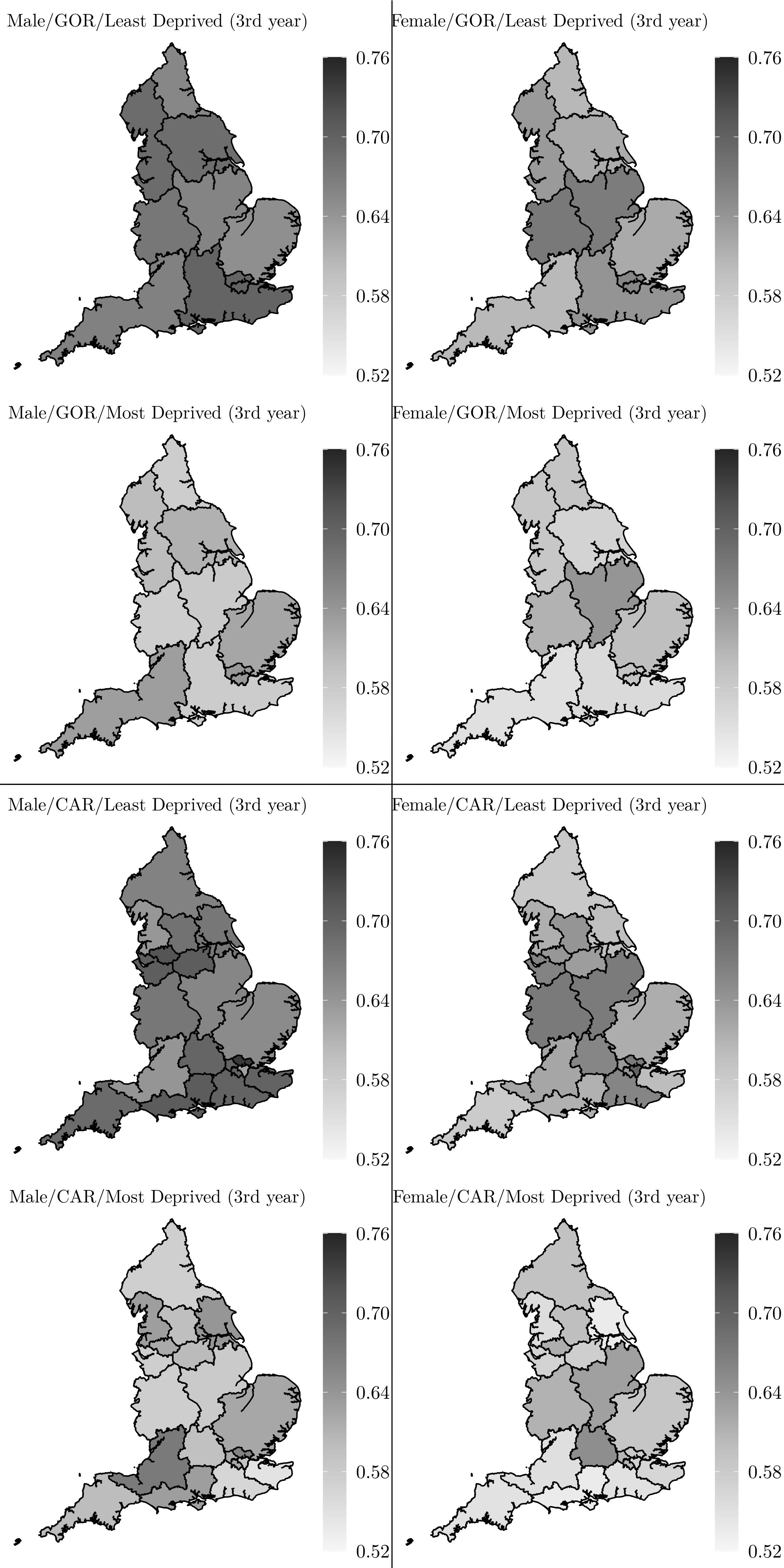}
    \caption{``Deprivation level'' (``1'' being \textit{least deprived} and ``5'' \textit{most deprived}) stratified net survival point estimate for $t = 3$ based on the {(i: top-left maps)} ``Government Office Regions'' (GOR) spatial structure with fitted model RS-SGH LN BYM2 for male patients, {(ii: top-right maps)} ``Government Office Regions'' (GOR) spatial structure with fitted model RS-SGH LN BYM2 for female patients, {(iii: bottom-left maps)} ``Cancer Alliances Regions'' (CAR) spatial structure with fitted model RS-SGH LN BYM2 for male patients, and {(iv: bottom-right maps)} ``Cancer Alliances Regions'' (CAR) spatial structure with fitted model RS-SGH LN BYM2 for female patients. For each panel, the upper map represents the \textit{least deprived} level, and the lower map represents the \textit{most deprived} level.}
	\label{fig:S-cs01-t3-mean-dep}
\end{figure}

\begin{figure}[!ht]
	\centering
	\includegraphics[width = 0.60\textwidth]{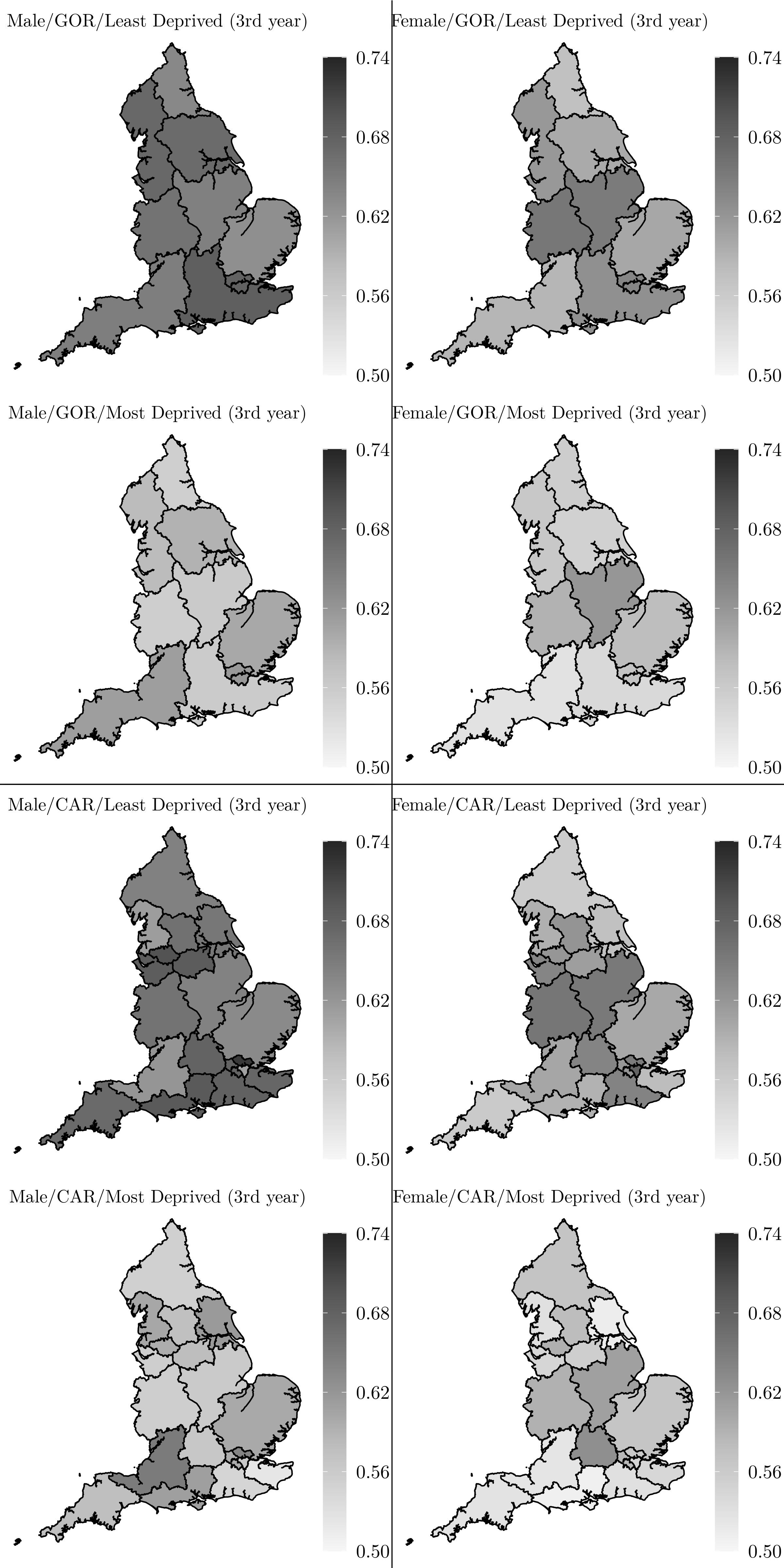}
    \caption{``Deprivation level'' (``1'' being \textit{least deprived} and ``5'' \textit{most deprived}) stratified $2.5^{\text{th}}$ net survival percentile for $t = 3$ based on the {(i: top-left maps)} ``Government Office Regions'' (GOR) spatial structure with fitted model RS-SGH LN BYM2 for male patients, {(ii: top-right maps)} ``Government Office Regions'' (GOR) spatial structure with fitted model RS-SGH LN BYM2 for female patients, {(iii: bottom-left maps)} ``Cancer Alliances Regions'' (CAR) spatial structure with fitted model RS-SGH LN BYM2 for male patients, and {(iv: bottom-right maps)} ``Cancer Alliances Regions'' (CAR) spatial structure with fitted model RS-SGH LN BYM2 for female patients. For each panel, the upper map represents the \textit{least deprived} level, and the lower map represents the \textit{most deprived} level.}
	\label{fig:S-cs01-t3-0025-dep}
\end{figure}

\begin{figure}[!ht]
	\centering
	\includegraphics[width = 0.60\textwidth]{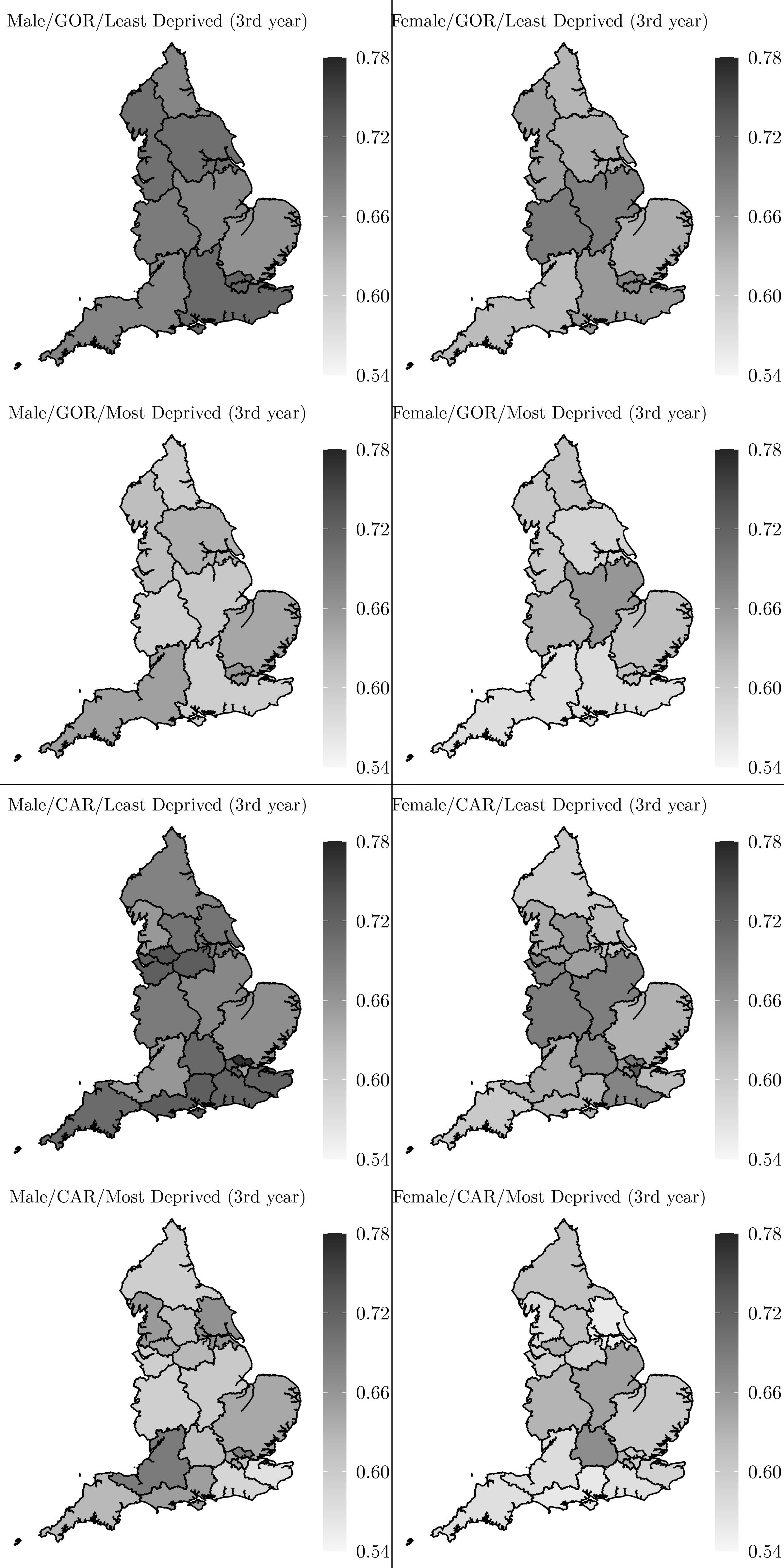}
    \caption{``Deprivation level'' (``1'' being \textit{least deprived} and ``5'' \textit{most deprived}) stratified $97.5^{\text{th}}$ net survival percentile for $t = 3$ based on the {(i: top-left maps)} ``Government Office Regions'' (GOR) spatial structure with fitted model RS-SGH LN BYM2 for male patients, {(ii: top-right maps)} ``Government Office Regions'' (GOR) spatial structure with fitted model RS-SGH LN BYM2 for female patients, {(iii: bottom-left maps)} ``Cancer Alliances Regions'' (CAR) spatial structure with fitted model RS-SGH LN BYM2 for male patients, and {(iv: bottom-right maps)} ``Cancer Alliances Regions'' (CAR) spatial structure with fitted model RS-SGH LN BYM2 for female patients. For each panel, the upper map represents the \textit{least deprived} level, and the lower map represents the \textit{most deprived} level.}
	\label{fig:S-cs01-t3-0975-dep}
\end{figure}

\begin{figure}[!ht]
	\centering
	\includegraphics[width = 0.60\textwidth]{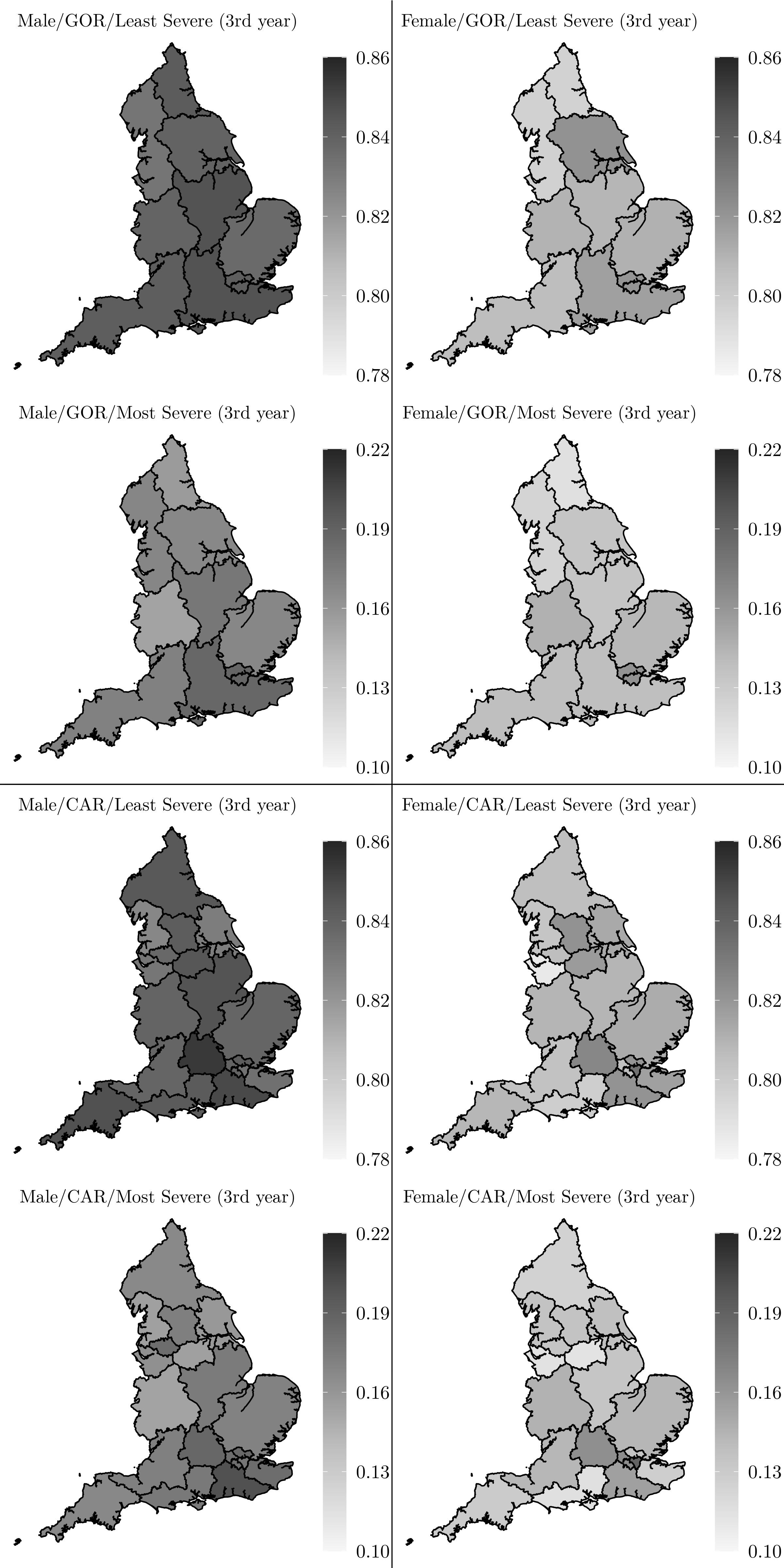}
    \caption{``Cancer stage'' (``1, 2, and 3'' being \textit{least severe} and ``4'' \textit{most severe}) stratified net survival point estimate for $t = 3$ based on the {(i: top-left maps)} ``Government Office Regions'' (GOR) spatial structure with fitted model RS-SGH LN BYM2 for male patients, {(ii: top-right maps)} ``Government Office Regions'' (GOR) spatial structure with fitted model RS-SGH LN BYM2 for female patients, {(iii: bottom-left maps)} ``Cancer Alliances Regions'' (CAR) spatial structure with fitted model RS-SGH LN BYM2 for male patients, and {(iv: bottom-right maps)} ``Cancer Alliances Regions'' (CAR) spatial structure with fitted model RS-SGH LN BYM2 for female patients. For each panel, the upper map represents the \textit{least severe} level, and the lower map represents the \textit{most severe} level. \textbf{Important}: different from all other analyses, notice that the maps for different cancer stages are plotted in \underline{different} scales.}
	\label{fig:S-cs01-t3-mean-cancer}
\end{figure}

\begin{figure}[!ht]
	\centering
	\includegraphics[width = 0.60\textwidth]{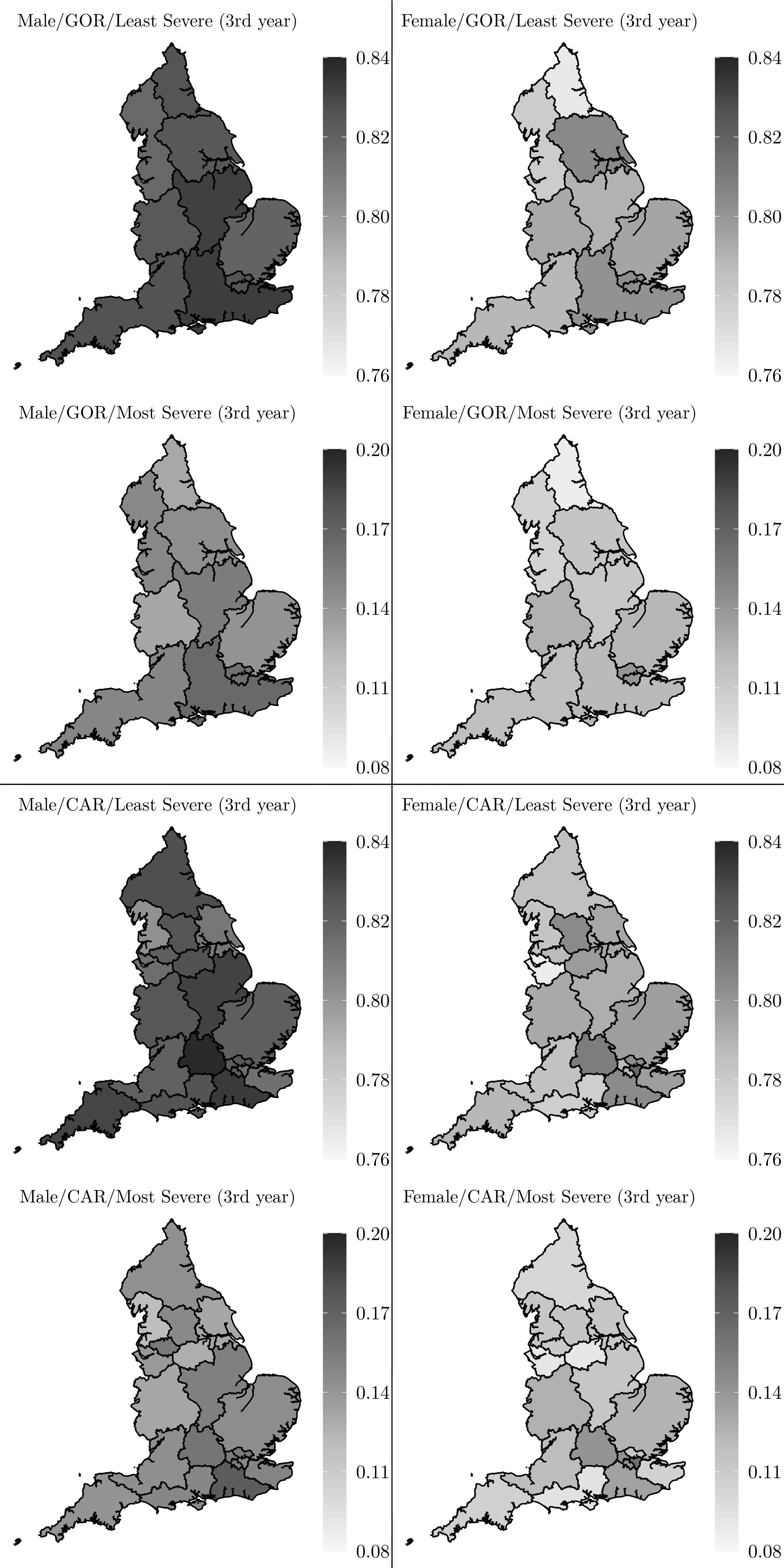}
    \caption{``Cancer stage'' (``1, 2, and 3'' being \textit{least severe} and ``4'' \textit{most severe}) stratified $2.5^{\text{th}}$ net survival percentile for $t = 3$ based on the {(i: top-left maps)} ``Government Office Regions'' (GOR) spatial structure with fitted model RS-SGH LN BYM2 for male patients, {(ii: top-right maps)} ``Government Office Regions'' (GOR) spatial structure with fitted model RS-SGH LN BYM2 for female patients, {(iii: bottom-left maps)} ``Cancer Alliances Regions'' (CAR) spatial structure with fitted model RS-SGH LN BYM2 for male patients, and {(iv: bottom-right maps)} ``Cancer Alliances Regions'' (CAR) spatial structure with fitted model RS-SGH LN BYM2 for female patients. For each panel, the upper map represents the \textit{least severe} level, and the lower map represents the \textit{most severe} level. \textbf{Important}: different from all other analyses, notice that the maps for different cancer stages are plotted in \underline{different} scales.}
	\label{fig:S-cs01-t3-0025-cancer}
\end{figure}

\begin{figure}[!ht]
	\centering
	\includegraphics[width = 0.60\textwidth]{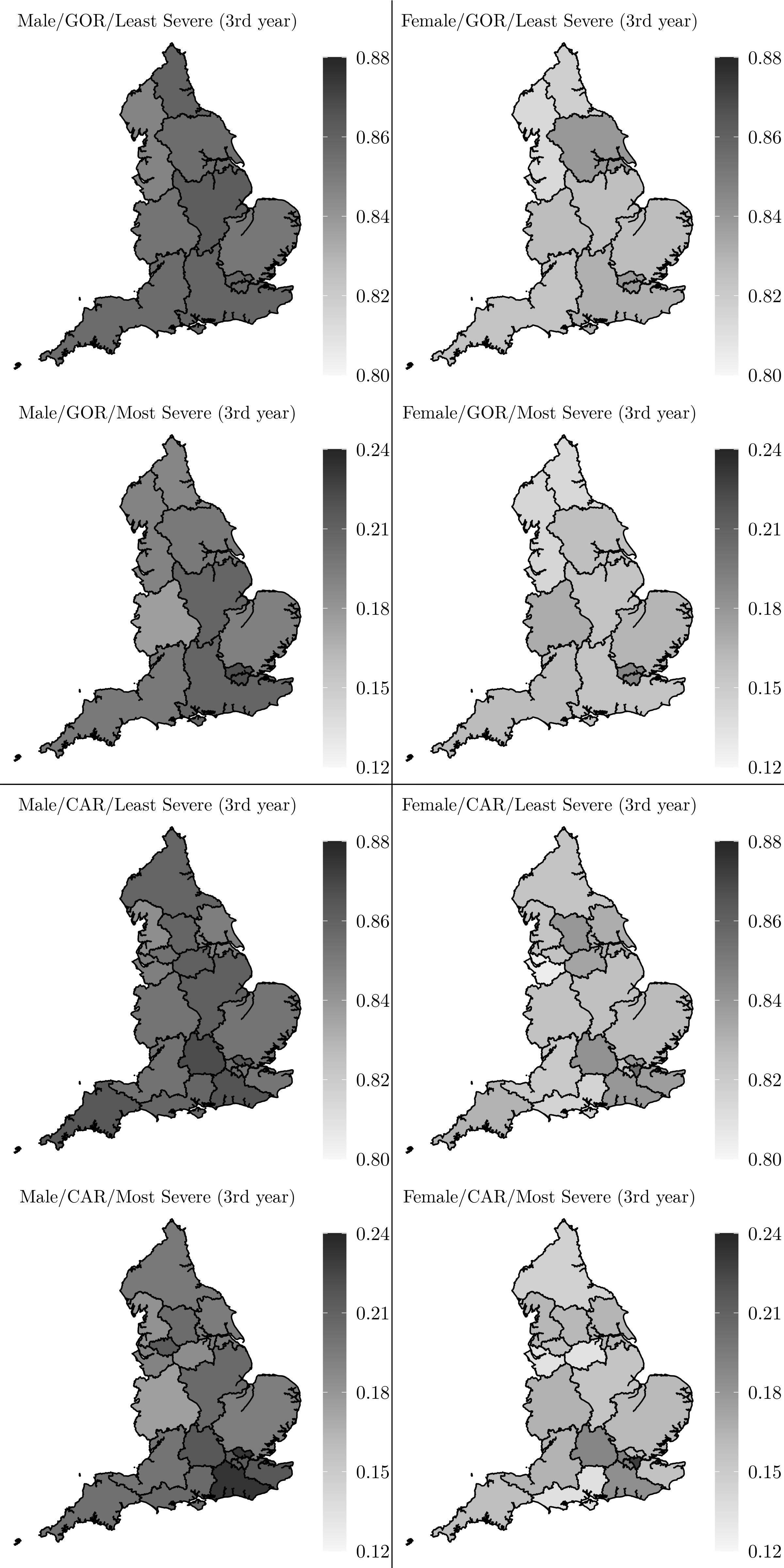}
    \caption{``Cancer stage'' (``1, 2, and 3'' being \textit{least severe} and ``4'' \textit{most severe}) stratified $97.5^{\text{th}}$ net survival percentile for $t = 3$ based on the {(i: top-left maps)} ``Government Office Regions'' (GOR) spatial structure with fitted model RS-SGH LN BYM2 for male patients, {(ii: top-right maps)} ``Government Office Regions'' (GOR) spatial structure with fitted model RS-SGH LN BYM2 for female patients, {(iii: bottom-left maps)} ``Cancer Alliances Regions'' (CAR) spatial structure with fitted model RS-SGH LN BYM2 for male patients, and {(iv: bottom-right maps)} ``Cancer Alliances Regions'' (CAR) spatial structure with fitted model RS-SGH LN BYM2 for female patients. For each panel, the upper map represents the \textit{least severe} level, and the lower map represents the \textit{most severe} level. \textbf{Important}: different from all other analyses, notice that the maps for different cancer stages are plotted in \underline{different} scales.}
	\label{fig:S-cs01-t3-0975-cancer}
\end{figure}

\begin{figure}[!ht]
	\centering
	\includegraphics[width = 0.60\textwidth]{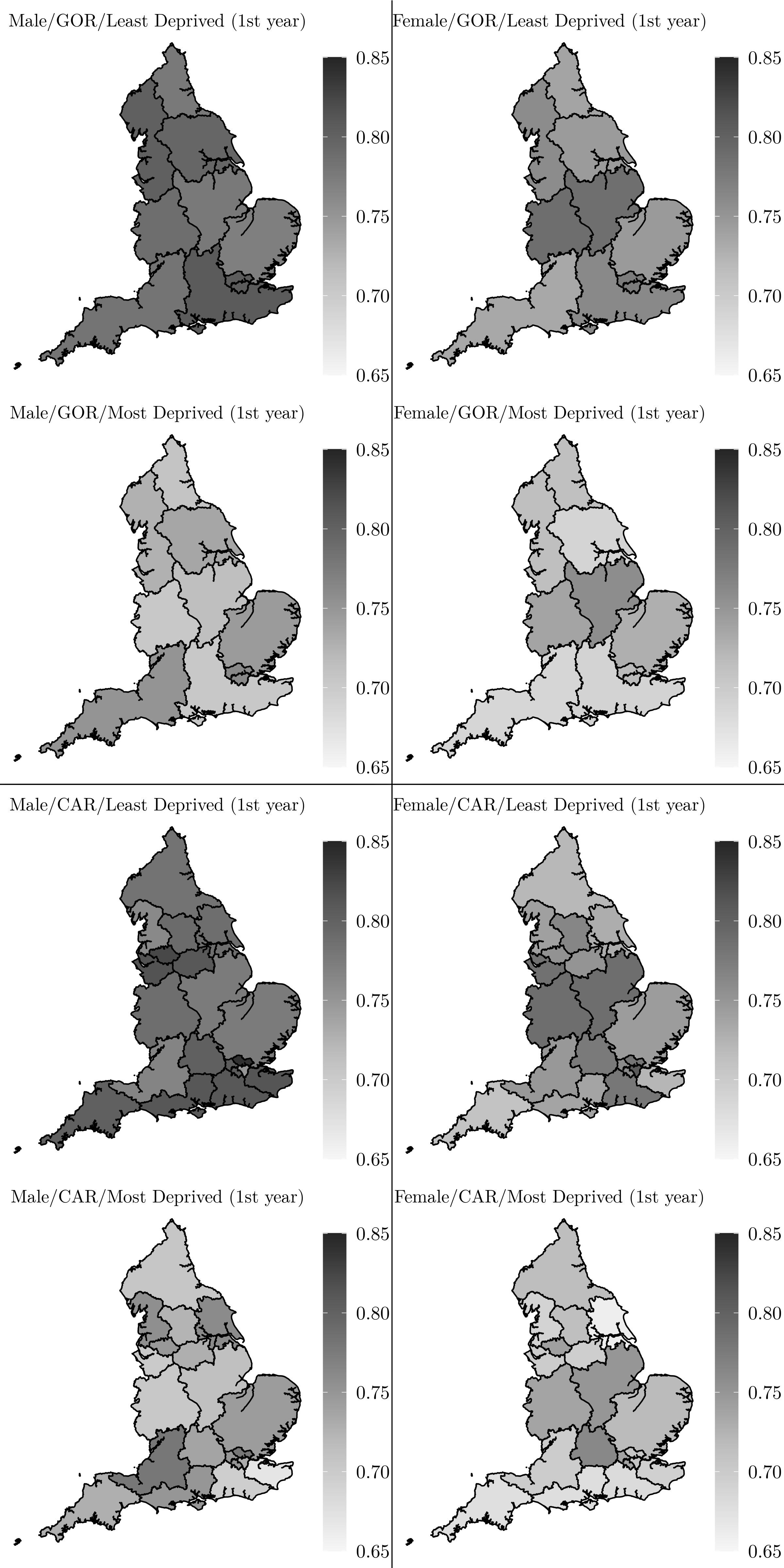}
    \caption{``Deprivation level'' (``1'' being \textit{least deprived} and ``5'' \textit{most deprived}) stratified net survival point estimate for $t = 1$ based on the {(i: top-left maps)} ``Government Office Regions'' (GOR) spatial structure with fitted model RS-SGH LN BYM2 for male patients, {(ii: top-right maps)} ``Government Office Regions'' (GOR) spatial structure with fitted model RS-SGH LN BYM2 for female patients, {(iii: bottom-left maps)} ``Cancer Alliances Regions'' (CAR) spatial structure with fitted model RS-SGH LN BYM2 for male patients, and {(iv: bottom-right maps)} ``Cancer Alliances Regions'' (CAR) spatial structure with fitted model RS-SGH LN BYM2 for female patients. For each panel, the upper map represents the \textit{least deprived} level, and the lower map represents the \textit{most deprived} level.}
	\label{fig:S-cs01-t1-mean-dep}
\end{figure}

\begin{figure}[!ht]
	\centering
	\includegraphics[width = 0.60\textwidth]{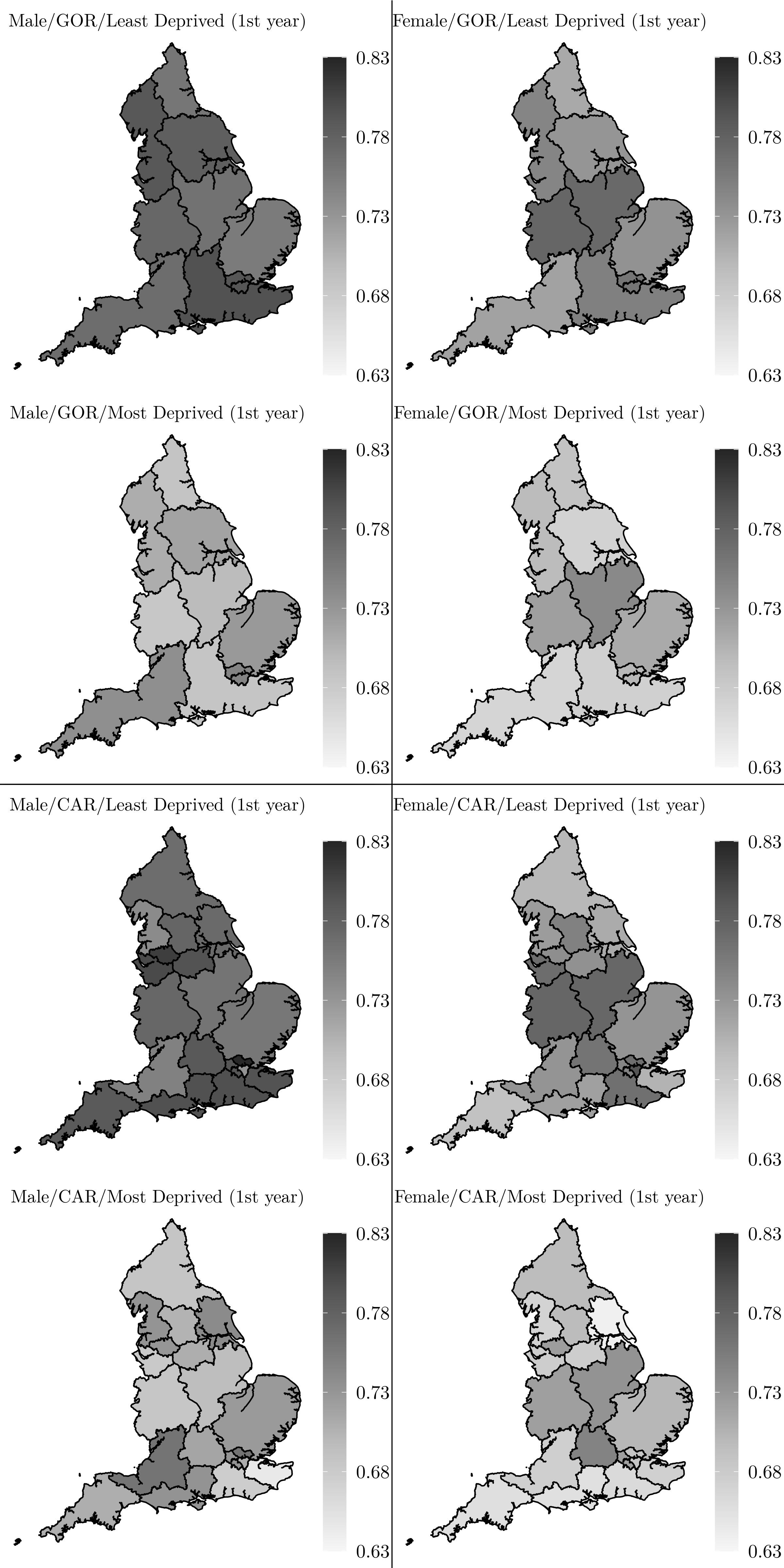}
    \caption{``Deprivation level'' (``1'' being \textit{least deprived} and ``5'' \textit{most deprived}) stratified $2.5^{\text{th}}$ net survival percentile for $t = 1$ based on the {(i: top-left maps)} ``Government Office Regions'' (GOR) spatial structure with fitted model RS-SGH LN BYM2 for male patients, {(ii: top-right maps)} ``Government Office Regions'' (GOR) spatial structure with fitted model RS-SGH LN BYM2 for female patients, {(iii: bottom-left maps)} ``Cancer Alliances Regions'' (CAR) spatial structure with fitted model RS-SGH LN BYM2 for male patients, and {(iv: bottom-right maps)} ``Cancer Alliances Regions'' (CAR) spatial structure with fitted model RS-SGH LN BYM2 for female patients. For each panel, the upper map represents the \textit{least deprived} level, and the lower map represents the \textit{most deprived} level.}
	\label{fig:S-cs01-t1-0025-dep}
\end{figure}

\begin{figure}[!ht]
	\centering
	\includegraphics[width = 0.60\textwidth]{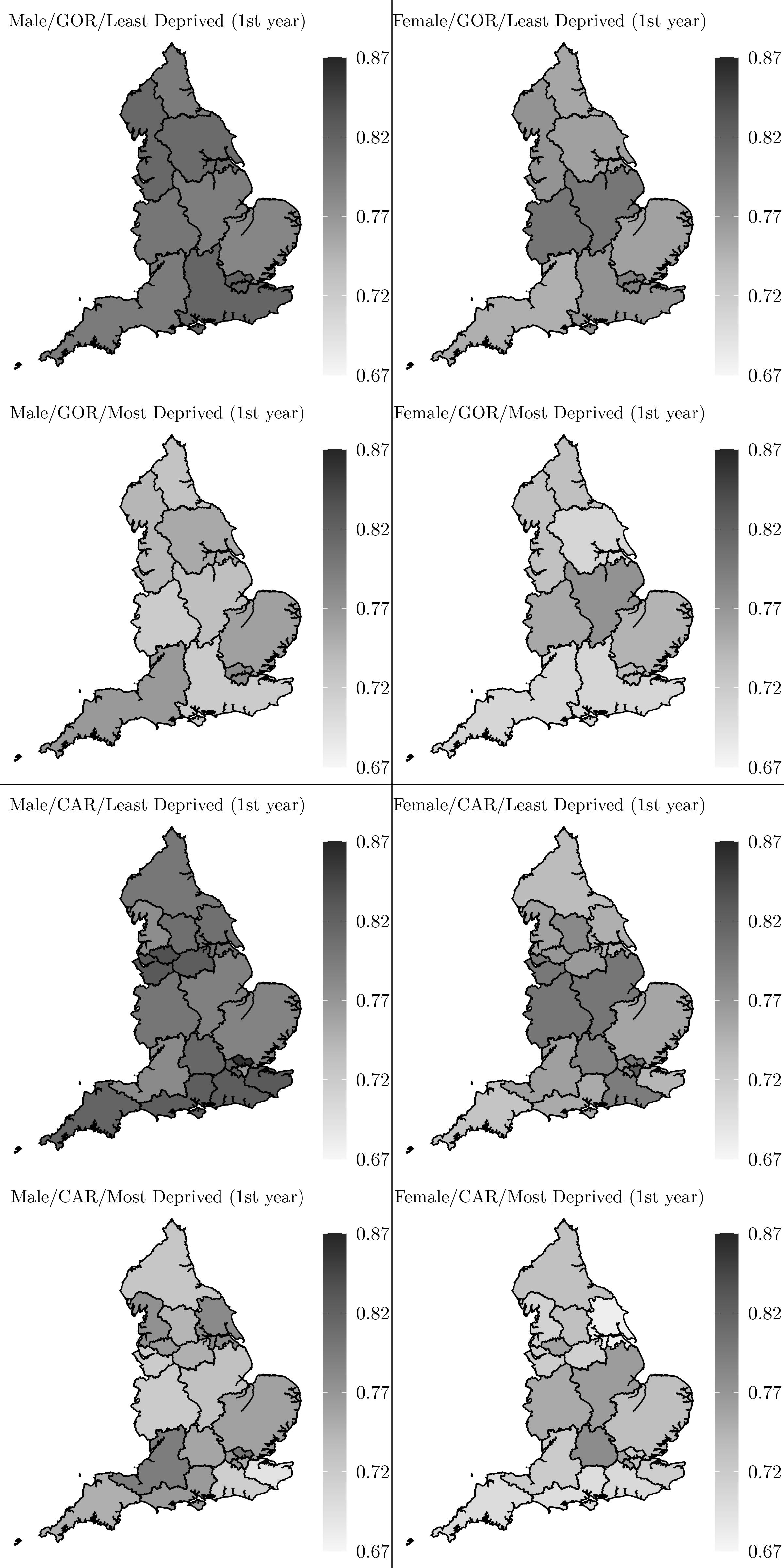}
    \caption{``Deprivation level'' (``1'' being \textit{least deprived} and ``5'' \textit{most deprived}) stratified $97.5^{\text{th}}$ net survival percentile for $t = 1$ based on the {(i: top-left maps)} ``Government Office Regions'' (GOR) spatial structure with fitted model RS-SGH LN BYM2 for male patients, {(ii: top-right maps)} ``Government Office Regions'' (GOR) spatial structure with fitted model RS-SGH LN BYM2 for female patients, {(iii: bottom-left maps)} ``Cancer Alliances Regions'' (CAR) spatial structure with fitted model RS-SGH LN BYM2 for male patients, and {(iv: bottom-right maps)} ``Cancer Alliances Regions'' (CAR) spatial structure with fitted model RS-SGH LN BYM2 for female patients. For each panel, the upper map represents the \textit{least deprived} level, and the lower map represents the \textit{most deprived} level.}
	\label{fig:S-cs01-t1-0975-dep}
\end{figure}

\begin{figure}[!ht]
	\centering
	\includegraphics[width = 0.60\textwidth]{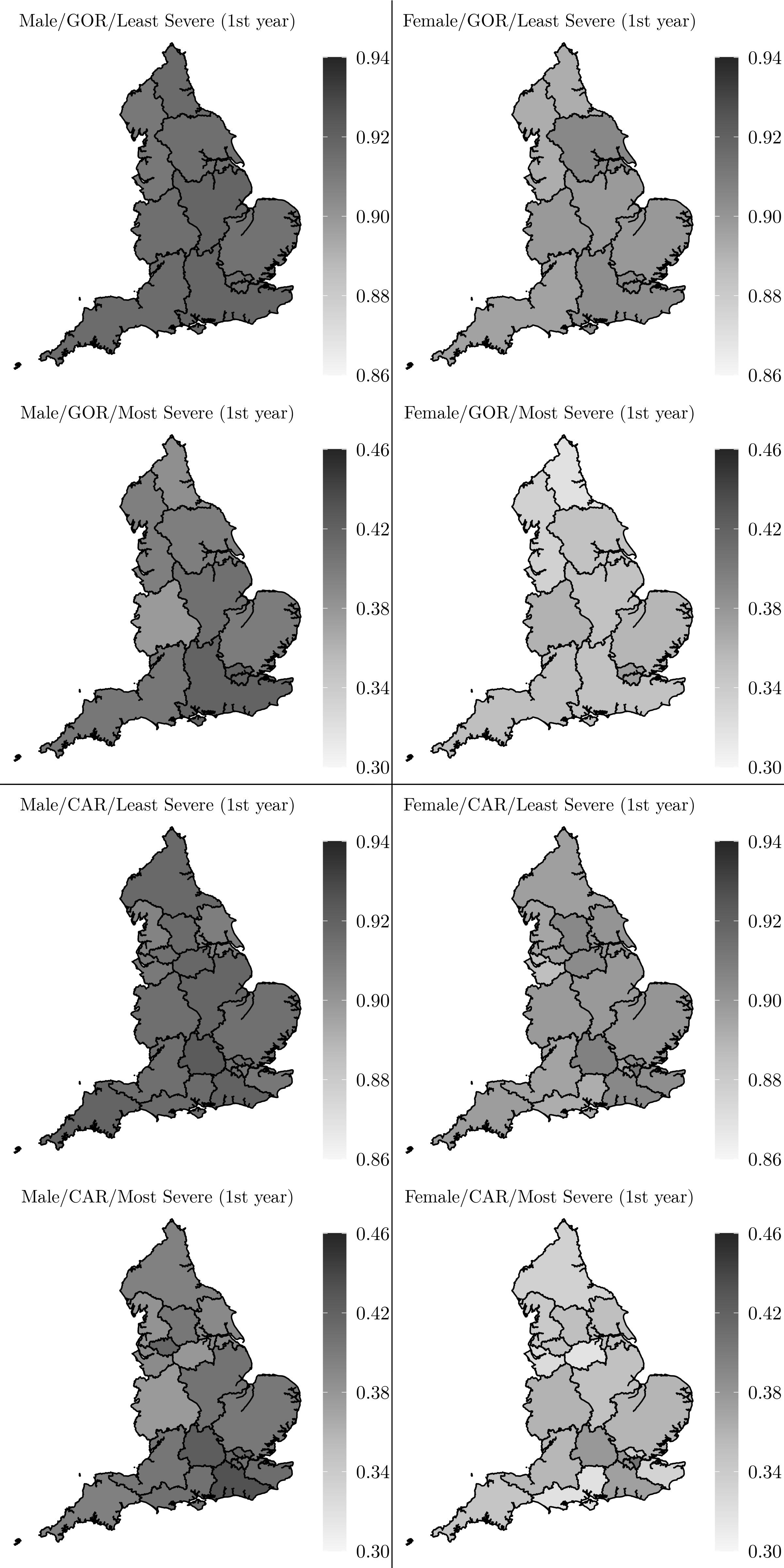}
    \caption{``Cancer stage'' (``1, 2, and 3'' being \textit{least severe} and ``4'' \textit{most severe}) stratified net survival point estimate for $t = 1$ based on the {(i: top-left maps)} ``Government Office Regions'' (GOR) spatial structure with fitted model RS-SGH LN BYM2 for male patients, {(ii: top-right maps)} ``Government Office Regions'' (GOR) spatial structure with fitted model RS-SGH LN BYM2 for female patients, {(iii: bottom-left maps)} ``Cancer Alliances Regions'' (CAR) spatial structure with fitted model RS-SGH LN BYM2 for male patients, and {(iv: bottom-right maps)} ``Cancer Alliances Regions'' (CAR) spatial structure with fitted model RS-SGH LN BYM2 for female patients. For each panel, the upper map represents the \textit{least severe} level, and the lower map represents the \textit{most severe} level. \textbf{Important}: different from all other analyses, notice that the maps for different cancer stages are plotted in \underline{different} scales.}
	\label{fig:S-cs01-t1-mean-cancer}
\end{figure}

\begin{figure}[!ht]
	\centering
	\includegraphics[width = 0.60\textwidth]{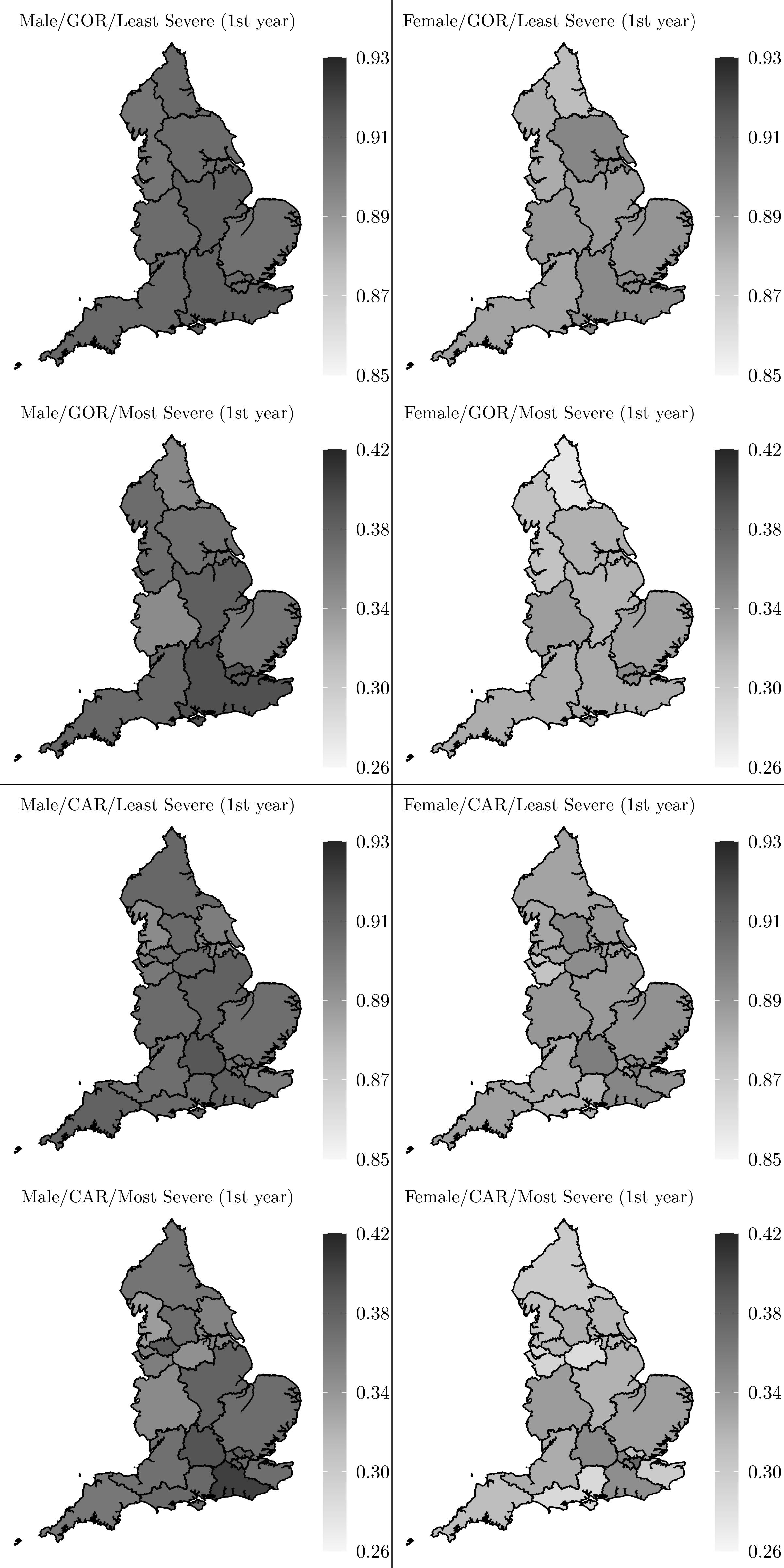}
    \caption{``Cancer stage'' (``1, 2, and 3'' being \textit{least severe} and ``4'' \textit{most severe}) stratified $2.5^{\text{th}}$ net survival percentile for $t = 1$ based on the {(i: top-left maps)} ``Government Office Regions'' (GOR) spatial structure with fitted model RS-SGH LN BYM2 for male patients, {(ii: top-right maps)} ``Government Office Regions'' (GOR) spatial structure with fitted model RS-SGH LN BYM2 for female patients, {(iii: bottom-left maps)} ``Cancer Alliances Regions'' (CAR) spatial structure with fitted model RS-SGH LN BYM2 for male patients, and {(iv: bottom-right maps)} ``Cancer Alliances Regions'' (CAR) spatial structure with fitted model RS-SGH LN BYM2 for female patients. For each panel, the upper map represents the \textit{least severe} level, and the lower map represents the \textit{most severe} level. \textbf{Important}: different from all other analyses, notice that the maps for different cancer stages are plotted in \underline{different} scales.}
	\label{fig:S-cs01-t1-0025-cancer}
\end{figure}

\begin{figure}[!ht]
	\centering
	\includegraphics[width = 0.60\textwidth]{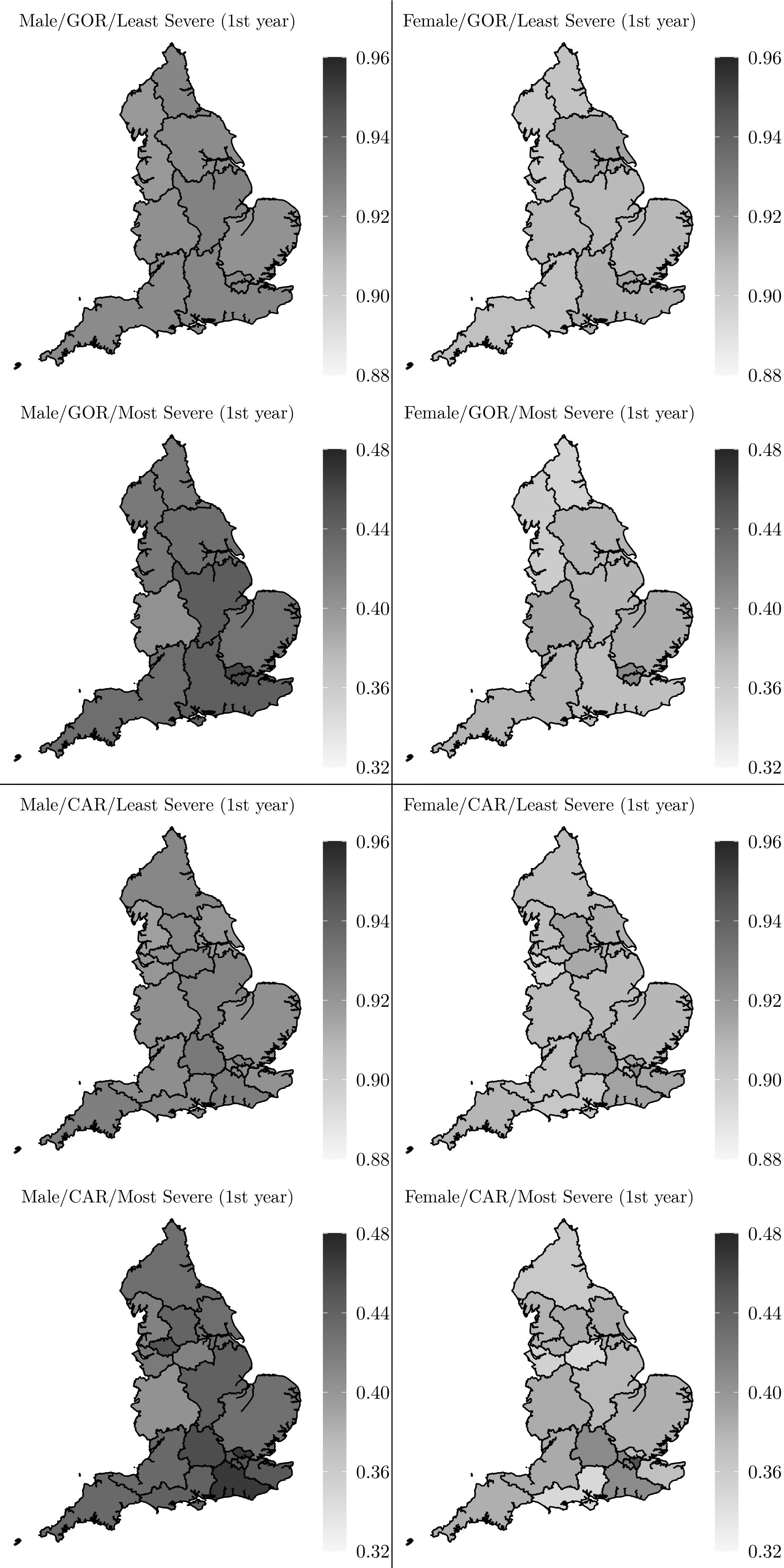}
    \caption{``Cancer stage'' (``1, 2, and 3'' being \textit{least severe} and ``4'' \textit{most severe}) stratified $97.5^{\text{th}}$ net survival percentile for $t = 1$ based on the {(i: top-left maps)} ``Government Office Regions'' (GOR) spatial structure with fitted model RS-SGH LN BYM2 for male patients, {(ii: top-right maps)} ``Government Office Regions'' (GOR) spatial structure with fitted model RS-SGH LN BYM2 for female patients, {(iii: bottom-left maps)} ``Cancer Alliances Regions'' (CAR) spatial structure with fitted model RS-SGH LN BYM2 for male patients, and {(iv: bottom-right maps)} ``Cancer Alliances Regions'' (CAR) spatial structure with fitted model RS-SGH LN BYM2 for female patients. For each panel, the upper map represents the \textit{least severe} level, and the lower map represents the \textit{most severe} level. \textbf{Important}: different from all other analyses, notice that the maps for different cancer stages are plotted in \underline{different} scales.}
	\label{fig:S-cs01-t1-0975-cancer}
\end{figure}

This section presents complementary results for Section \textcolor{black}{5}. Table \ref{tab:S-cs01-model-selection} shows the ranked models (according to the $\widehat{\text{elpd}}_{\text{PSIS-LOO}}$ criterion), Table \ref{tab:S-cs01-fitting-time} displays the fitting time for all models, and Table \ref{tab:S-cs01-model-pars} presents the estimated hyperparameters of the random effects for the highest-ranked BYM2 models in the first case study. Figures \ref{fig:S-cs01-t3-0025} and \ref{fig:S-cs01-t3-0975} report the uncertainty for the estimated net survival, such that $t = 3$ years. Figures \ref{fig:S-cs01-t1-mean}, \ref{fig:S-cs01-t1-0025}, and \ref{fig:S-cs01-t1-0975} present similar results (including the net survival point estimate), however, for $t = 1$ year. Figures \ref{fig:S-cs01-t3-mean-dep}, \ref{fig:S-cs01-t3-0025-dep}, \ref{fig:S-cs01-t3-0975-dep}, \ref{fig:S-cs01-t1-mean-dep}, \ref{fig:S-cs01-t1-0025-dep} and \ref{fig:S-cs01-t1-0975-dep} report the net survival estimates with data stratified by ``deprivation level,'' and Figures \ref{fig:S-cs01-t3-mean-cancer}, \ref{fig:S-cs01-t3-0025-cancer}, \ref{fig:S-cs01-t3-0975-cancer}, \ref{fig:S-cs01-t1-mean-cancer}, \ref{fig:S-cs01-t1-0025-cancer}, and \ref{fig:S-cs01-t1-0975-cancer} report equivalent results with data stratified by ``cancer stage.'' Lastly, Figure \ref{fig:S-map_england_cancer_alliance} shows the England map according to the Cancer Alliance Regions.

\begin{table}[!ht]
	\caption{Results for the competing models in Section \textcolor{black}{5} according to the $\widehat{\text{elpd}}_{\text{PSIS-LOO}}$. The $\widehat{\text{elpd}}_{\text{PSIS-LOO}}$ difference (with standard error) represents the pairwise difference between the others models and the reference model. The ``Geography 01'' refers to the Government Office Regions and the ``Geography 02'' refers to the Cancer Alliances Regions in 2016. The \sout{strikethrough} scenario corresponds to the fitted model for which we did not observe well mixed posterior chains.}
	\resizebox{\textwidth}{!}{%
	    \centering
		\begin{tabular}{ c | c | l | c | c | c | l | c } 
            Sex  & Geography & Model & $\widehat{\text{elpd}}_{\text{PSIS-LOO}}$ diff. (SE) & Sex & Geography & Model & $\widehat{\text{elpd}}_{\text{PSIS-LOO}}$ diff. (SE) \\ \hline
            Male & 01 & RS-SGH LN BYM2  & \phantom{0000-}0.0 (\phantom{00}0.0) & Female & 01 & RS-SGH LN BYM2  & \phantom{0-}0.0 (0.0) \\
            Male & 01 & RS-SGH LN ICAR  & \phantom{0000}-1.4 (\phantom{00}1.4) & Female & 01 & RS-SGH LN ICAR  & \phantom{0}-0.3 (1.4) \\
            Male & 01 & RS-SGH PGW ICAR & \phantom{000}-14.6 (\phantom{00}3.6) & Female & 01 & RS-SGH LL BYM2  & \phantom{0}-9.8 (4.0) \\
            Male & 01 & RS-SGH PGW BYM2 & \phantom{000}-19.7 (\phantom{00}5.4) & Female & 01 & RS-SGH PGW BYM2 & \phantom{0}-9.8 (2.4) \\
            Male & 01 & RS-SGH LL ICAR  & \phantom{000}-24.6 (\phantom{00}6.5) & Female & 01 & RS-SGH PGW ICAR & \phantom{}-20.6 (3.5) \\
            Male & 01 & \sout{RS-SGH LL BYM2}  & \sout{\phantom{}-21284.1 (\phantom{}576.0)} & Female & 01 & RS-SGH LL ICAR & \phantom{}-42.0 (5.8) \\ \hline
            Male & 02 & RS-SGH LN BYM2  & \phantom{0-}0.0 (0.0) & Female & 02 & RS-SGH LN BYM2  & \phantom{0-}0.0 (0.0) \\
            Male & 02 & RS-SGH LN ICAR  & \phantom{0}-2.1 (1.8) & Female & 02 & RS-SGH LN ICAR  & \phantom{0}-2.5 (1.8) \\
            Male & 02 & RS-SGH PGW ICAR & \phantom{}-15.4 (3.8) & Female & 02 & RS-SGH PGW BYM2 & \phantom{0}-9.9 (2.3) \\
            Male & 02 & RS-SGH PGW BYM2 & \phantom{}-16.0 (4.6) & Female & 02 & RS-SGH PGW ICAR & \phantom{}-12.0 (3.0) \\
            Male & 02 & RS-SGH LL BYM2  & \phantom{}-23.3 (6.3) & Female & 02 & RS-SGH LL BYM2  & \phantom{}-41.6 (6.0) \\
            Male & 02 & RS-SGH LN ICAR  & \phantom{}-25.2 (6.6) & Female & 02 & RS-SGH LL ICAR  & \phantom{}-44.2 (6.2)
		\end{tabular}%
	}
	\label{tab:S-cs01-model-selection}
\end{table}

\begin{table}[!ht]
	\caption{Fitting time for all models in Section \textcolor{black}{5}. ``Fitting time (sec.)'' presents the time (in seconds) to fit the corresponding model. The models are ordered as in Table \ref{tab:S-cs01-model-selection}. The models were fitted on a AMD EPYC 7402 CPU at 2.8 Ghz.}
	\resizebox{\textwidth}{!}{%
	    \centering
		\begin{tabular}{ c | c | l | c | c | c | l | c } 
            Sex  & Patients' location & Model & Fitting time (sec.) & Sex & Patients' location & Model & Fitting time (sec.) \\ \hline
            Male & 01 & RS-SGH LN BYM2  & ~70950.50 & Female & 01 & RS-SGH LN BYM2  & ~63060.69 \\
            Male & 01 & RS-SGH LN ICAR  & ~24974.23 & Female & 01 & RS-SGH LN ICAR  & ~27092.05 \\
            Male & 01 & RS-SGH PGW ICAR & ~35546.03 & Female & 01 & RS-SGH LL BYM2  & 143318.70 \\
            Male & 01 & RS-SGH PGW BYM2 & 134614.80 & Female & 01 & RS-SGH PGW BYM2 & ~83139.05 \\
            Male & 01 & RS-SGH LL ICAR  & ~77084.76 & Female & 01 & RS-SGH PGW ICAR & ~44532.04 \\
            Male & 01 & RS-SGH LL BYM2  & 154491.00 & Female & 01 & RS-SGH LL ICAR  & 158862.70 \\ \hline
            Male & 02 & RS-SGH LN BYM2  & ~52695.24 & Female & 02 & RS-SGH LN BYM2  & ~45791.73 \\
            Male & 02 & RS-SGH LN ICAR  & ~15219.16 & Female & 02 & RS-SGH LN ICAR  & ~14379.65 \\
            Male & 02 & RS-SGH PGW ICAR & ~21843.38 & Female & 02 & RS-SGH PGW BYM2 & ~60938.90 \\
            Male & 02 & RS-SGH PGW BYM2 & ~68693.83 & Female & 02 & RS-SGH PGW ICAR & ~17720.56 \\
            Male & 02 & RS-SGH LL BYM2  & 113003.30 & Female & 02 & RS-SGH LL BYM2  & ~99978.17 \\
            Male & 02 & RS-SGH LN ICAR  & ~44572.72 & Female & 02 & RS-SGH LL ICAR  & ~38042.61
		\end{tabular}%
	}
	\label{tab:S-cs01-fitting-time}
\end{table}

\begin{table}[!ht]
	\caption{Estimated parameters (with standard deviation and a 95\% equal-tail credible interval) for the highest-ranked BYM2 models in Section \textcolor{black}{5}. We reported $\sigma_{u} = 1 / \sqrt{\tau_{u}}$ (and $\sigma_{\tilde{u}} = 1 / \sqrt{\tau_{\tilde{u}}}$) and $\rho$ (and $\tilde{\rho}$). The ``Geography 01'' refers to the Government Office Regions and the ``Geography 02'' refers to the Cancer Alliances Regions in 2016.}
	\resizebox{\textwidth}{!}{%
	    \centering
		\begin{tabular}{ l | c | l | c | c | c | c | c | c | c | c } 
            Sex  & Geography & Model & Parameter & Mean & SD & 95\% equal-tail CI & Parameter & Mean & SD & 95\% equal-tail CI \\ \hline
            Male   & 01 & RS-SGH LN BYM2 & $\sigma_{\tilde{u}}$ & 0.09 & 0.07 & (0.00; 0.26) & $\sigma_{u}$ & 0.03 & 0.03 & (0.00; 0.09) \\
            Male   & 01 & RS-SGH LN BYM2 & $\tilde{\rho}$       & 0.53 & 0.35 & (0.00; 1.00) & $\rho$       & 0.52 & 0.36 & (0.00; 1.00) \\
            Male   & 02 & RS-SGH LN BYM2 & $\sigma_{\tilde{u}}$ & 0.08 & 0.06 & (0.00; 0.23) & $\sigma_{u}$ & 0.04 & 0.03 & (0.00; 0.11) \\ 
            Male   & 02 & RS-SGH LN BYM2 & $\tilde{\rho}$       & 0.49 & 0.35 & (0.00; 1.00) & $\rho$       & 0.46 & 0.35 & (0.00; 1.00) \\
            Female & 01 & RS-SGH LN BYM2 & $\sigma_{\tilde{u}}$ & 0.13 & 0.11 & (0.01; 0.41) & $\sigma_{u}$ & 0.04 & 0.03 & (0.00; 0.13) \\
            Female & 01 & RS-SGH LN BYM2 & $\tilde{\rho}$       & 0.52 & 0.35 & (0.00; 1.00) & $\rho$       & 0.52 & 0.35 & (0.00; 1.00) \\ 
            Female & 02 & RS-SGH LN BYM2 & $\sigma_{\tilde{u}}$ & 0.09 & 0.07 & (0.00; 0.26) & $\sigma_{u}$ & 0.03 & 0.03 & (0.00; 0.09) \\
            Female & 02 & RS-SGH LN BYM2 & $\tilde{\rho}$       & 0.53 & 0.35 & (0.00; 1.00) & $\rho$       & 0.52 & 0.36 & (0.00; 1.00) \\
		\end{tabular}%
	}
	\label{tab:S-cs01-model-pars}
\end{table}

\begin{figure}[!ht]
	\centering
	\includegraphics[width = 1\textwidth]{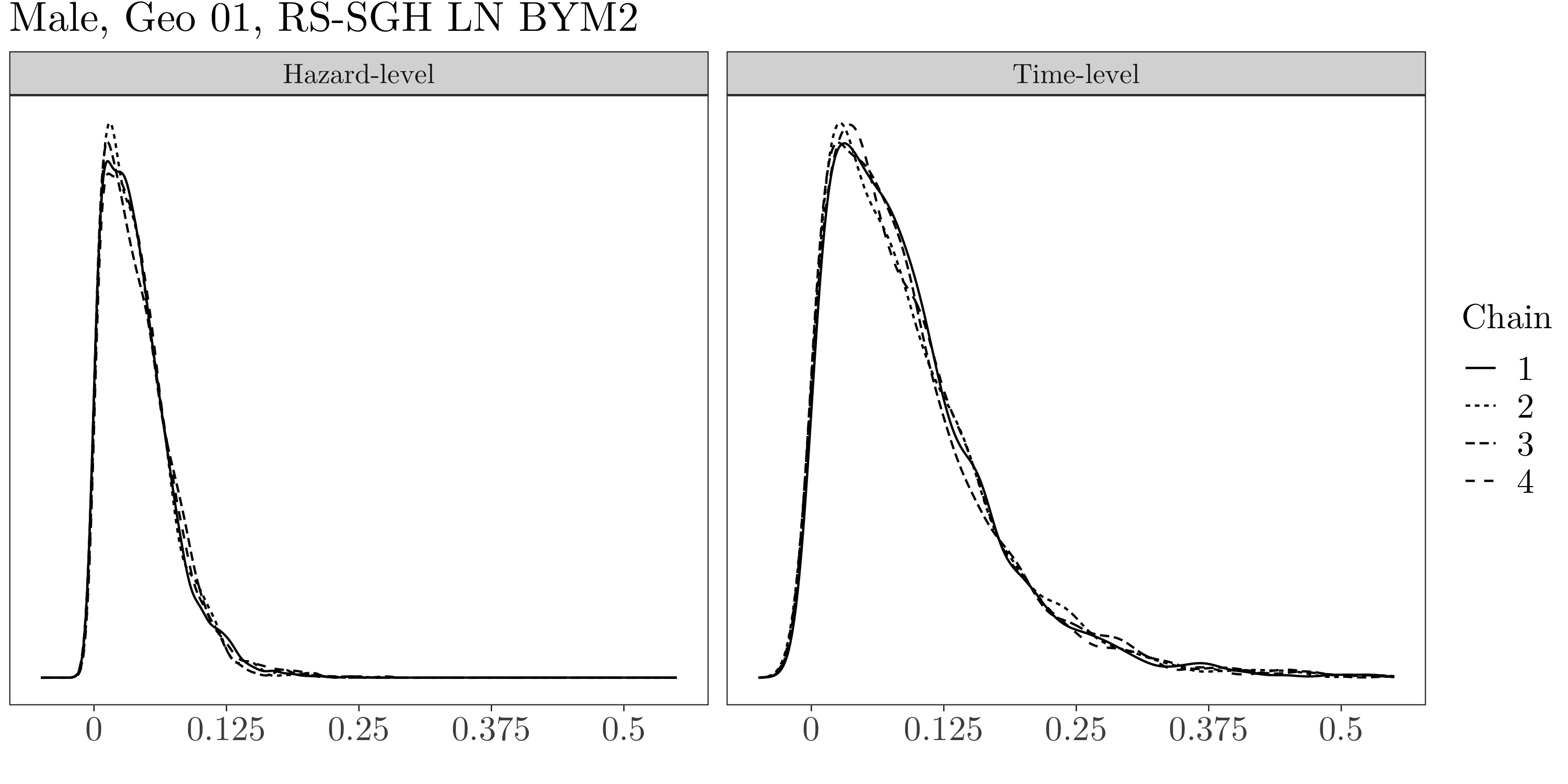}
    \caption{Estimated posterior densities for $\sigma_{u}$ (as it appears in the \textbf{hazard-level} component) and $\sigma_{\tilde{u}}$ (as it appears in the \textbf{time-level} component) for the BYM2 random effects when fitting the RS-SGH LN model for \textit{male} patients based on the ``Government Office Regions'' (``Geo 01''). The curves are plotted separately for each posterior chain.}
	\label{fig:S-cs01-male-geo1-ln-bym2}
\end{figure}

\begin{figure}[!ht]
	\centering
	\includegraphics[width = 1\textwidth]{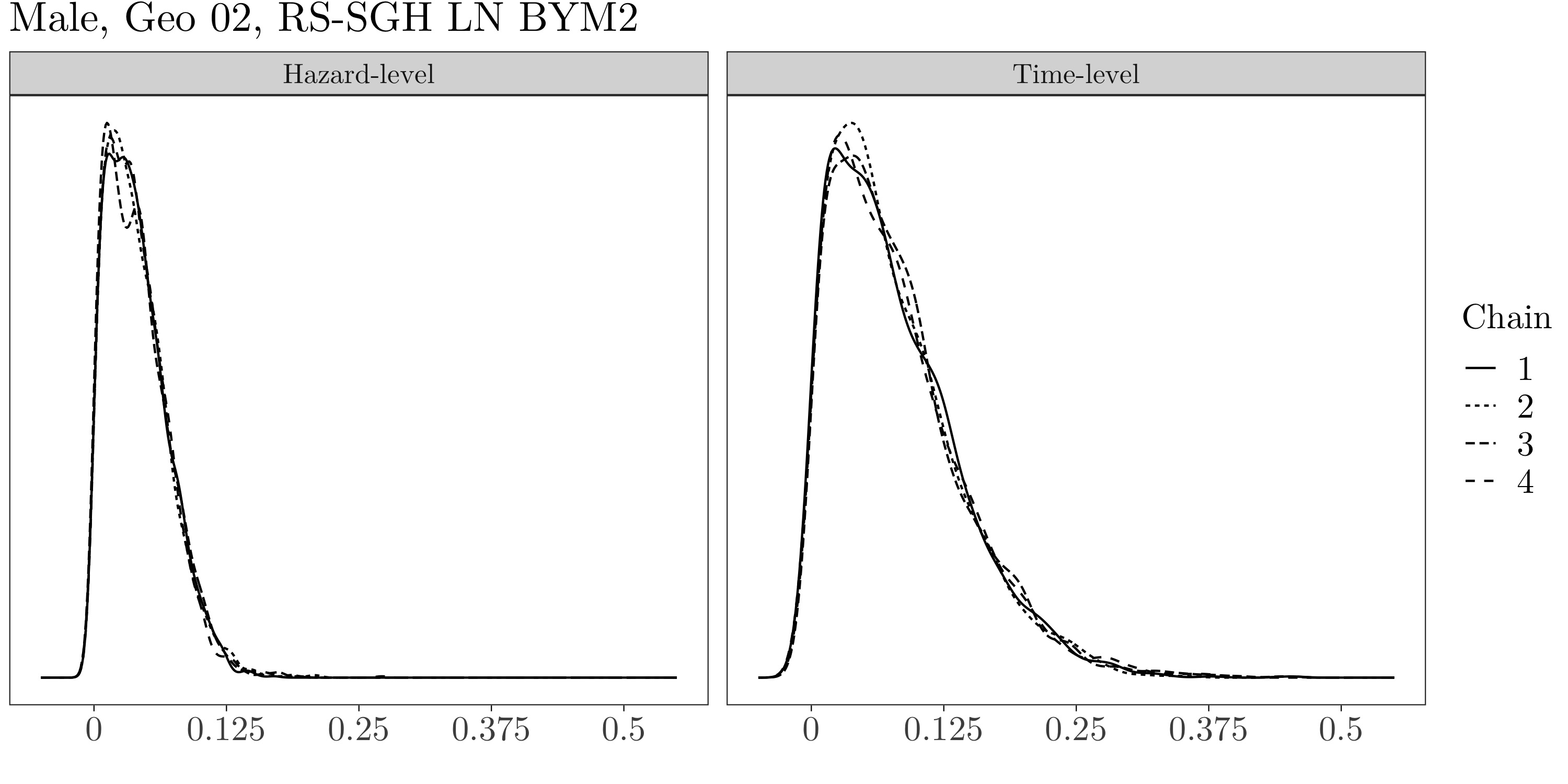}
    \caption{Estimated posterior densities for $\sigma_{u}$ (as it appears in the \textbf{hazard-level} component) and $\sigma_{\tilde{u}}$ (as it appears in the \textbf{time-level} component) for the BYM2 random effects when fitting the RS-SGH LN model for \textit{male} patients based on the ``Cancer Alliances Regions in 2016'' (``Geo 02''). The curves are plotted separately for each posterior chain.}
	\label{fig:S-cs01-male-geo2-ln-bym2}
\end{figure}

\begin{figure}[!ht]
	\centering
	\includegraphics[width = 1\textwidth]{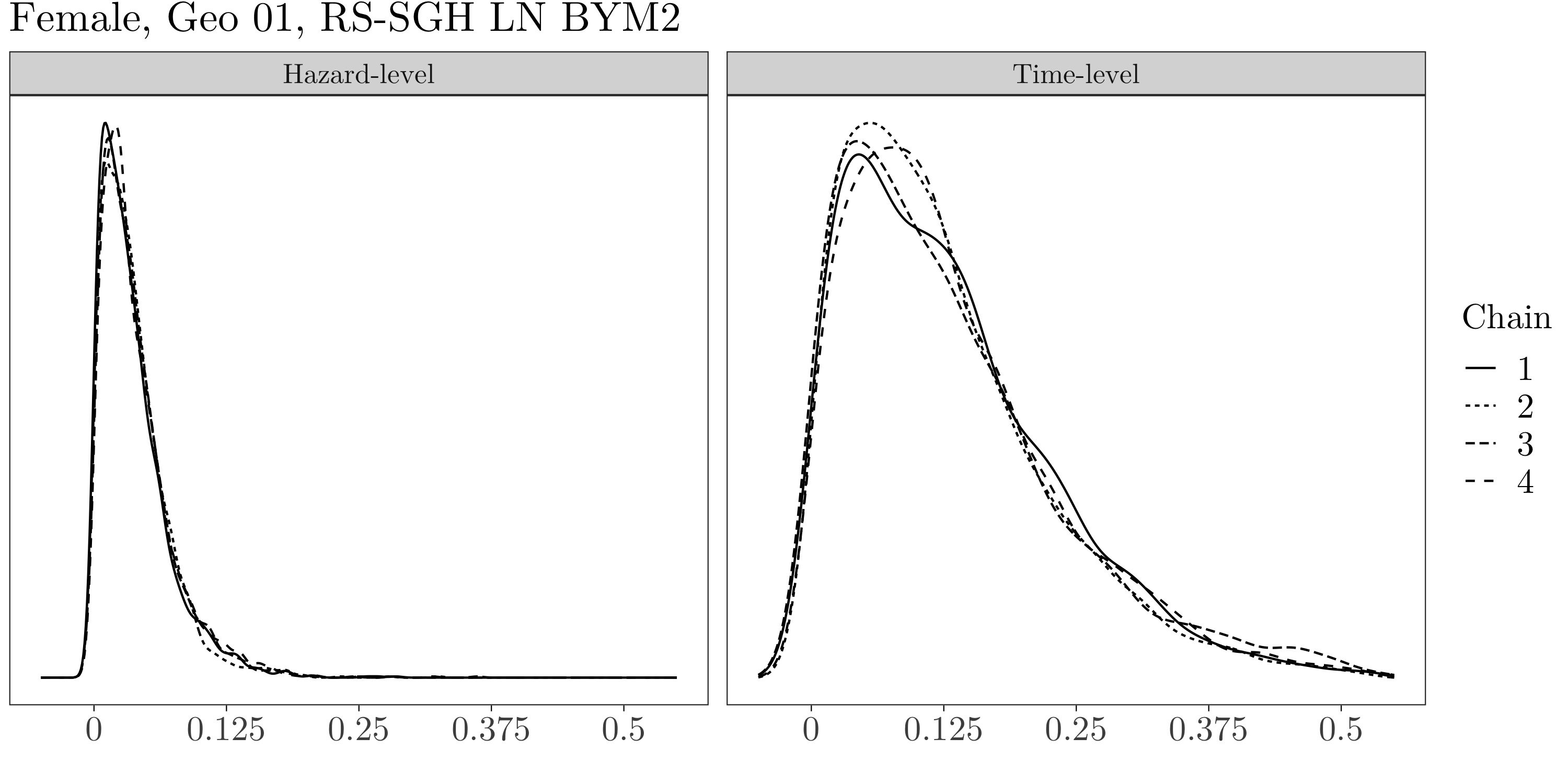}
    \caption{Estimated posterior densities for $\sigma_{u}$ (as it appears in the \textbf{hazard-level} component) and $\sigma_{\tilde{u}}$ (as it appears in the \textbf{time-level} component) for the BYM2 random effects when fitting the RS-SGH LN model for \textit{female} patients based on the ``Government Office Regions'' (``Geo 01''). The curves are plotted separately for each posterior chain.}
	\label{fig:S-cs01-female-geo1-ln-bym2}
\end{figure}

\begin{figure}[!ht]
	\centering
	\includegraphics[width = 1\textwidth]{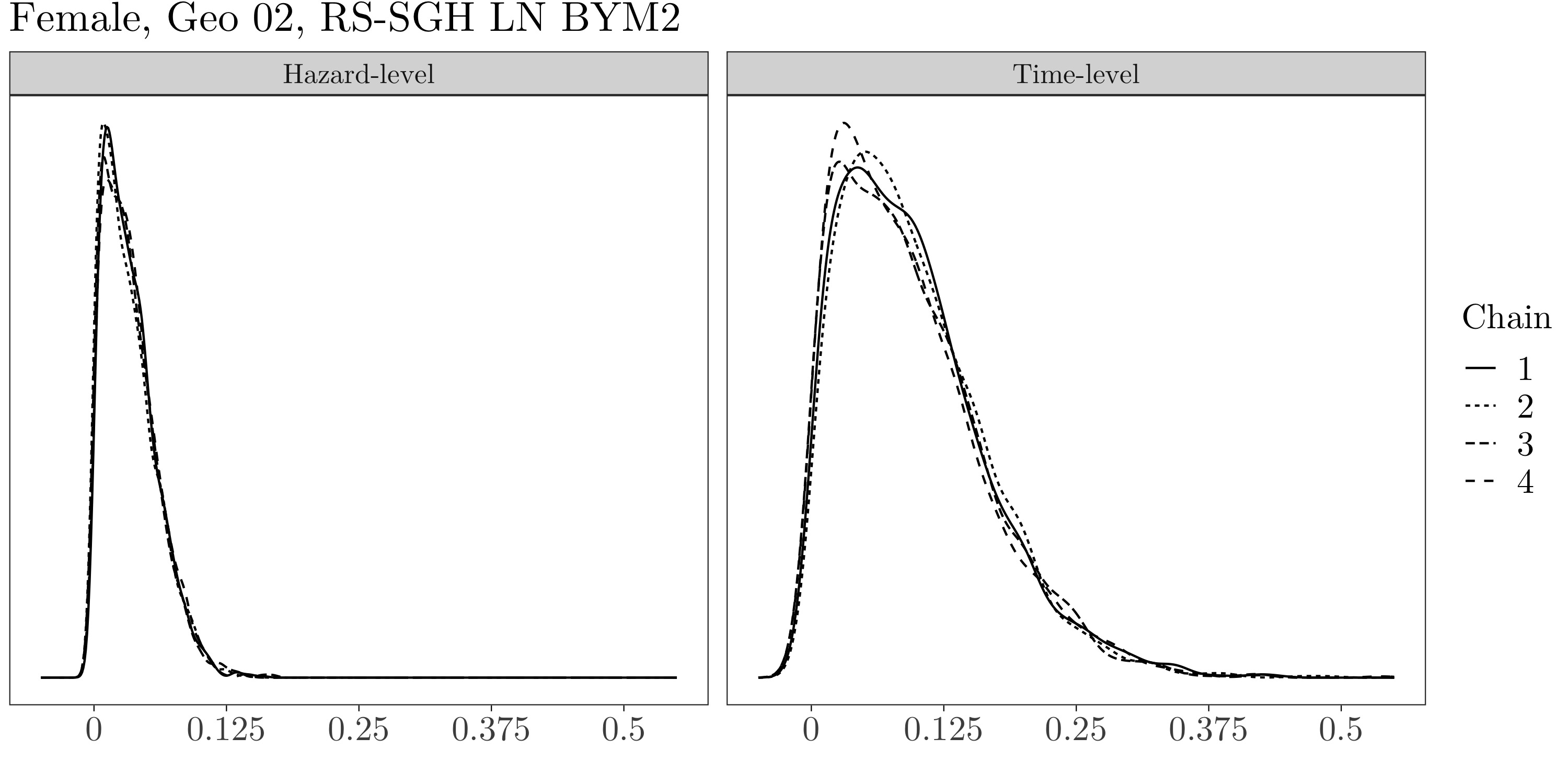}
    \caption{Estimated posterior densities for $\sigma_{u}$ (as it appears in the \textbf{hazard-level} component) and $\sigma_{\tilde{u}}$ (as it appears in the \textbf{time-level} component) for the BYM2 random effects when fitting the RS-SGH LN model for \textit{female} patients based on the ``Cancer Alliances Regions in 2016'' (``Geo 02''). The curves are plotted separately for each posterior chain.}
	\label{fig:S-cs01-female-geo2-ln-bym2}
\end{figure}


\clearpage


\clearpage
\section*{References}
\addcontentsline{toc}{section}{References}
\bibliographystyle{cell}
\bibliography{references}